\documentclass[12pt,prd,aps,floatfix]{revtex4}

\usepackage{graphicx}
\usepackage{amsmath}
\usepackage{amssymb}
\usepackage{ulem}
\usepackage{lscape}
\usepackage{pdflscape}
\usepackage{rotating}

\newcommand{\be}{\begin{equation}}
\newcommand{\ee}{\end{equation}}
\newcommand{\bea}{\begin{eqnarray}}
\newcommand{\eea}{\end{eqnarray}}
\newcommand{\bi}{\begin{itemize}}
\newcommand{\ei}{\end{itemize}}
\newcommand{\ben}{\begin{enumerate}}
\newcommand{\een}{\end{enumerate}}
\newcommand{\bt}{\begin{tabbing}}
\newcommand{\et}{\end{tabbing}}
\newcommand{\nn}{\nonumber}

\newcommand{\bfp}{{\bf p}}
\newcommand{\bfpp}{{{\bf p}^\prime}}

\newcommand{\bfpperp}{{{\bf p}_{\perp}}}
\newcommand{\bfppperp}{{{\bf p}^\prime_{\perp}}}
\newcommand{\bfpppperp}{{{\bf p}^{\prime\prime}_{\perp}}}

\newcommand{\bfpnperp}{{{\bf p}_{\not\perp}}}
\newcommand{\bfppnperp}{{{\bf p}^\prime_{\not\perp}}}

\newcommand{\bfx}{{\bf x}}
\newcommand{\bfxp}{{{\bf x}^\prime}}

\newcommand{\bfz}{{\bf 0}}
\newcommand{\pp}{{p^{\prime}}}
\newcommand{\vp}{{v^{\prime}}}

\newcommand{\epsp}{{\epsilon^{\prime}}}
\newcommand{\epspp}{{\epsilon^{\prime\prime}}}
\newcommand{\epsps}{{\epsilon^{\prime *}}}
\newcommand{\dt}{{\Delta t}}
\newcommand{\dtp}{{\Delta t^\prime}}
\newcommand{\dtmin}{{\dt_{\rm min}}}
\newcommand{\dtpmin}{{\dtp_{\rm min}}}
\newcommand{\tsrc}{{t_{\rm src}}}
\newcommand{\bfxsrc}{{{\bf x}_{\rm src}}}

\newcommand{\epsQ}{{\epsilon_Q}}
\newcommand{\xipi}{{\xi_\pi}}
\newcommand{\xietas}{{\xi_{\eta_s}}}
\newcommand{\xia}{{\xi_a}}

\newcommand{\xiamQ}{{\xi_{am_Q}}}

\newcommand{\gVPpi}{{g_{D^*D\pi}}}
\newcommand{\Vcb}{{|V_{cb}|}}
\newcommand{\mbphys}{{m_{b,\rm phys}}}
\newcommand{\costhl}{{\cos[\theta_\ell]}}
\newcommand{\sinthl}{{\sin[\theta_\ell]}}
\newcommand{\costhv}{{\cos[\theta_v]}}
\newcommand{\sinthv}{{\sin[\theta_v]}}

\newcommand{\sinthlt}{{\sin^2[\theta_\ell]}}
\newcommand{\costhvt}{{\cos^2[\theta_v]}}
\newcommand{\sinthvt}{{\sin^2[\theta_v]}}

\newcommand{\Fo}{{{\mathcal F}_1}}
\newcommand{\Ft}{{{\mathcal F}_2}}
\newcommand{\Fot}{{{\mathcal F}_{1,2}}}
\newcommand{\wref}{{w_{\rm ref}}}
\newcommand{\wmax}{{w_{\rm max}}}

\newcommand{\nuell}{{\nu_\ell}}

\begin{document}

\vspace*{-10mm}
\begin{flushright}
\normalsize
KEK-CP-393   \\[-2mm]
OU-HET-1186
\end{flushright}


\title{ 
$B\to D^*\ell\nuell$ semileptonic form factors
from lattice QCD with M\"obius domain-wall quarks
}

\author{
Y.~Aoki$^{a}$,
B.~Colquhoun$^{b}$,
H.~Fukaya$^{c}$,
S.~Hashimoto$^{d,e}$,
T.~Kaneko$^{d,e,f}$,
R.~Kellermann$^{e}$,
J.~Koponen$^{g}$,
E.~Kou$^{h}$
(JLQCD Collaboration)
}

\affiliation{
$^a$RIKEN Center for Computational Science (R-CCS), Kobe 650-0047, Japan
\\
$^b$SUPA, School of Physics and Astronomy, University of Glasgow,
    Glasgow, G12 8QQ, UK
\\
$^c$Department of Physics, Osaka University, Osaka 560-0043, Japan
\\
$^d$High Energy Accelerator Research Organization (KEK), Ibaraki 305-0801, Japan
\\
$^e$Graduate Institute for Advanced Studies,
    SOKENDAI (The Graduate University for Advanced Studies),
    Ibaraki 305-0801, Japan
\\
$^f$Kobayashi-Maskawa Institute for the Origin of Particles and the Universe,
    Nagoya University, Aichi 464-8602, Japan
\\
$^g$PRISMA+ Cluster of Excellence \& Institute f\"ur Kernphysik,
    Johannes Gutenberg-Universit\"at Mainz, D-55128 Mainz, Germany
\\
$^h$Universit\'e Paris-Saclay, CNRS/IN2P3, IJCLab, 91405 Orsay, France
}

\begin{abstract}
We calculate the form factors for the $B\!\to\!D^*\ell\nuell$ decay
in 2+1 flavor lattice QCD.
For all quark flavors,
we employ the M\"obius domain-wall action,
which preserves chiral symmetry to a good precision.
Our gauge ensembles are generated at three lattice cutoffs
$a^{-1} \sim 2.5$, 3.6 and 4.5 GeV
with pion masses as low as $M_\pi\!\sim\!230$~MeV.
The physical lattice size $L$ satisfies the condition $M_\pi L \geq 4$
to control finite volume effects (FVEs),
while we simulate a smaller size at the smallest $M_\pi$ 
to directly examine FVEs.
The bottom quark masses are chosen in a range
from the physical charm quark mass to 0.7 $a^{-1}$
to control discretization effects.
We extrapolate the form factors to the continuum limit and physical quark masses
based on heavy meson chiral perturbation theory
at next-to-leading order.
Then the recoil parameter dependence
is parametrized using a model independent form
leading to our estimate of the decay rate ratio
between the tau ($\ell\!=\!\tau$) and light lepton ($\ell\!=\!e,\mu$) channels 
$R(D^*)\!=\!0.252(22)$ in the Standard Model.
A simultaneous fit with recent data from the Belle experiment
yields $|V_{cb}|\!=\!39.19(91)\times 10^{-3}$,
which is consistent with previous exclusive determinations,
and shows good consistency
in the kinematical distribution of the differential decay rate
between the lattice and experimental data.

\end{abstract}

\maketitle

\clearpage
\section{Introduction}
\label{sec:intro}


The $B\!\to\!D^*\ell\nuell$ semileptonic decay,
where $\ell\!=\!e,\mu,\tau$ and $\nuell$ represents
the corresponding neutrino,
plays a key role in stringent tests of the Standard Model (SM)
and searches for new physics.
The channels associated with light leptons $\ell\!=\!e,\mu$ provide
a determination of the Cabibbo-Kobayashi-Maskawa (CKM) matrix
element $\Vcb$, which is a fundamental parameter of the SM.
On the other hand,
the $\tau$ channel is expected to be a good probe of new physics,
since, for instance, it is expected to strongly couple
to the charged Higgs boson predicted by supersymmetric models.
Indeed,
there is an intriguing $\gtrsim\!3\,\sigma$ tension
between the SM and experiments on the decay rate ratio
\bea
   R(D^*)
   & = &
   \frac{\Gamma(B\!\to\!D^*\tau\nu_\tau)}{\Gamma(B\!\to\!D^*\ell\nuell)}
   \hspace{5mm}
   (\ell=e,\mu)
   \label{eqn:intro:RD*}
\eea
describing the lepton-flavor universality violation~\cite{HFLAV22}.


However,
there has been a long-standing tension between the $|V_{cb}|$ determinations
from exclusive and inclusive decays
$(B\!\to\!D\ell\nuell+D^*\ell\nuell+D^{**}\ell\nuell+\cdots)$.
The Heavy Flavor Averaging Group reported that
analysis of the inclusive decay yields~\cite{HFLAV22}
\bea
  \Vcb \times 10^3
  & = &
  41.98(45),
  \label{eqn:Vcb:incl}
\eea
which shows a $\lesssim\!4\,\sigma$ (9\,\%) tension with the determination
from the exclusive decays
\bea
  \Vcb \times 10^3
  & = &
  \left\{
  \begin{array}{ll}
    39.14(99) & (B\!\to\!D\ell\nuell), \\
    38.46(68) & (B\!\to\!D^*\ell\nuell).
  \end{array}
  \right.
  \label{eqn:Vcb:excl}
\eea
It has been argued~\cite{Vcb:tension:NP} that   
such a tension can be explained by introducing a tensor interaction beyond the SM,
which, however, largely distorts the $Z\!\to\!\bar{b}b$ decay rate
measured precisely by experiments.
Therefore, it is likely that the $\Vcb$ tension is due to
our incomplete understanding of the theoretical and/or experimental uncertainty.


The largest theoretical uncertainty in the exclusive determination
comes from form factors,
which describe non-perturbative QCD effects to the decay amplitude
through the matrix elements
\bea
   \sqrt{ M_B M_{D^*} }^{-1}
   \langle D^*(\epsp,\pp) | V_\mu | B(p) \rangle
   & = &
   i \varepsilon_{\mu\nu\rho\sigma} \,
   \epsps^\nu v^{\prime\rho} v^\sigma \, h_V(w),
   \label{eqn:intro:ff:V:hqet}
   \\[2mm]
   \sqrt{ M_B M_{D^*} }^{-1}
   \langle D^*(\epsp,\pp) | A_\mu | B(p) \rangle
   & = &
   (w+1) \, \epsilon_\mu^{\prime *} \,   h_{A_1}(w)
 - (\epsps v) \left\{ v_\mu \, h_{A_2}(w) + \vp_\mu \, h_{A_3}(w) \right\},
   \hspace{10mm}
   \label{eqn:intro:ff:A:hqet}
\eea
where    
$w=v\vp$ is the recoil parameter
defined by four velocities $v\!=\!p/M_B$ and $\vp\!=\!\pp/M_{D^{(*)}}$,
and $\epsp$ is the polarization vector of $D^*$,
which satisfies $\pp\epsp\!=\!0$.
Note that, here and in the following,
kinematical variables (momentum, velocity, polarization vector,
but not space-time coordinates) with and without the symbol ``$\prime$'' represent
those for $D^*$ and $B$, respectively.
%
%

Lattice QCD is a powerful method to provide a first-principles calculation of the form factors
with systematically improvable accuracy~\cite{review:B2Dstar:lat22:V,review:B2Dstar:lat22:K}.
However, until recently, only $h_{A_1}(1)$ at the zero-recoil limit $w=1$
had been calculated in unquenched lattice QCD~\cite{B2Dstar:Nf3:hA11:Fermilab/MILC1,B2Dstar:Nf3:hA11:Fermilab/MILC2,B2Dstar:Nf3:hA11:HPQCD}.
In the previous determination of $\Vcb$, therefore,
other information of the form factors in addition to $\Vcb$
had to be determined by fitting experimental data of the differential decay rate
to a theoretical expression~\cite{HFLAV22,Vcb:anly:BLPR17,Vcb:anly:BGS17,Vcb:anly:GK17}.
While a model independent parametrization of the form factors 
by Boyd, Grinstein and Lebed (the BGL parametrization) is available~\cite{BGL},
its large number of parameters due to the poor knowledge on the form factors
makes the fit unstable~\cite{Vcb:HQS:BGS17,Vcb:BGLvsCLN:BLR19,Vcb:BGL:GJS19,Vcb:BGLvsCLN:FUW21}
even with experimental data of differential distribution
with respect to all kinematical variables,
namely the recoil parameter and three decay angles,
from the KEKB/Belle experiment~\cite{B2Dstar:Belle:untag,B2Dstar:Belle:tag}~\footnote{
We note that a more recent data on the normalized distribution
from the full Belle data set~\cite{B2Dstar:Belle:tag:full}
as well as a recent report from
the succeeding SuperKEKB/Belle II experiment~\cite{B2Dstar:BelleII:tag}
are available in a different event-reconstruction approach.}.


One of the largest problems in the lattice study of $B$ meson physics
is that the simulation cost rapidly increases as the lattice spacing decreases.
In spite of recent progress in computer performance and
development of simulation algorithms,
it is still difficult to simulate lattice cutoffs $a^{-1}$ sufficiently larger
than the physical bottom quark mass $\mbphys$
to suppress $O((am_b)^n)$ discretization errors to, say, the 1\,\% level. 
For the time being, a practical approach to $B$ physics on the lattice
is to employ a heavy quark action based on an effective field theory,
such as the heavy quark effective theory (HQET), to directly simulate $\mbphys$
at currently available cutoffs $a^{-1}\!\lesssim\!\mbphys$.
Another straightforward approach is the so-called relativistic approach,
where a lattice QCD action is used also for heavy quarks
but with their masses sufficiently smaller than the lattice cutoff
to suppress discretization effects.
The Fermilab/MILC reported the first lattice calculation of
all form factors at zero and non-zero recoils~\cite{B2Dstar:Nf3:Fermilab/MILC}
using the Fermilab approach for heavy quarks,
namely a HQET interpretation of the anisotropic Wilson-clover quark action~\cite{FermilabAppr},
and staggered light quarks.
It was recently followed by the HPQCD Collaboration
with a fully relativistic approach~\cite{B2Dstar:Nf4:HPQCD},
where the highly improved staggered quark action was used
both for light and heavy flavors.


In this paper,
we present an independent study with a fully relativistic approach 
using the M\"obius domain-wall quark action~\cite{MoebiusDWF}
for all relevant quark flavors.
This preserves chiral symmetry to good precision
at finite lattice spacing,
reducing the leading discretization errors to second order, {\it i.e.}  $O((am_b)^2)$.
We can also use chiral perturbation theory
in the continuum limit ($a\!=\!0$) to guide the chiral extrapolation
to the physical pion mass.
Note also that
we do not need explicit renormalization of the weak axial and vector currents
on the lattice to calculate relevant form factors,
because 
the renormalization constants of these currents coincide with each other
and are canceled by taking appropriate ratios of relevant
correlation functions~\cite{ratio}.
Our preliminary analyses and discussions on these features
can be found in our earlier reports~\cite{B2Dstar:Nf3:JLQCD:lat18,B2Dstar:Nf3:JLQCD:lat19,B2Dstar:Nf3:JLQCD:lat21}.


This paper is organized as follows.
After a brief introduction to our choice of simulation parameters,
the form factors are extracted from correlation functions
at the simulation points in Sec.~\ref{sec:ff}.
These results are extrapolated to the continuum limit
and physical quark masses in Sec.~\ref{sec:cont+chiral_fit}.
In Sec.~\ref{sec:q2-param}, we generate synthetic data of the form factors
and fit them to a model-independent parametrization
in terms of the recoil parameter to make a comparison
with the previous lattice and phenomenological studies
as well as to estimate the SM value of $R(D^*)$.
We also perform a simultaneous fit
with experimental data to estimate $|V_{cb}|$.
Our conclusions are given in Sec.~\ref{sec:concl}.

\section{Form factors at simulation points}
\label{sec:ff}


\subsection{Gauge ensembles}
\label{subsec:ff:ensemble}


Our gauge ensembles are generated for 2+1-flavor lattice QCD,
where dynamical up and down quarks are degenerate
and dynamical strange quark has a distinct mass.
We take three lattice spacings of
$a\!\simeq\!0.044, 0.055$ and 0.080~fm~\cite{B2pi:Nf3:JLQCD},
which correspond to the lattice cutoffs
$a^{-1} \sim 2.453(4)$, 3.610(9) and 4.496(9) GeV, respectively.
They are determined using the Yang-Mills gradient flow~\cite{WilsonFlow}.
The tree-level improved Symanzik gauge action~\cite{Symanzik:W,Symanzik:WW,Symanzik:LW}
and the M\"obius domain-wall quark action~\cite{MoebiusDWF}
are used to control discretization errors
and to preserve chiral symmetry to a good precision.
We refer the interested reader to Refs.~\cite{MoebiusDWF,B2pi:Nf3:JLQCD,Boyle:MDWF:Lat14}
for the five dimensional formulation of the quark action and
our implementation.
Its four-dimensional effective Dirac operator is given by
\bea
  \frac{1+m_q}{2} 
+ \frac{1-m_q}{2} \gamma_5\, \epsilon\left[ H_M \right],
\label{eqn:ff:moebius:4d}
\eea
where $m_q$ represents the quark mass.
We employ the kernel operator $H_M\!=\!2 \gamma_5 D_W(-M) / \left\{2+D_W(-M)\right\}$,
where $D_W(-M)$ is the Wilson-Dirac operator with a negative mass with $M\!=\!1$
and with stout smearing~\cite{StoutSmr} applied three successive times
to the gauge links.
With our choice of the staple weight ($\rho\!=\!0.1$),
the smearing radius $\sim \! 1.5\,a$ is at the scale of the lattice spacing
and vanishes in the limit of $a\!\to\!0$.
We therefore expect that the link smearing does not change the continuum limit
of the $B\!\to\!D^*$ form factors.
It may induce additional discretization effects, which, however, do not
turn out to be large in our continuum and chiral extrapolation
in Sec.~\ref{sec:cont+chiral_fit}.
The polar decomposition approximation in Eq.~(\ref{eqn:ff:moebius:4d})
is realized for the sign function $\epsilon\left[H_M\right]$
in the five-dimensional domain-wall implementation. 
With this choice,
the residual quark mass, which is a measure of the chiral symmetry violation,
is suppressed to the level of 1~MeV at $a\!\simeq\!0.080$~fm,
and 0.2~MeV or less on finer lattices,
which are significantly smaller
than the physical up and down quark masses~\cite{B2pi:Nf3:JLQCD}.
With reasonably small values of the fifth-dimensional size $N_5\!=\!8$\,--\,12,
the computational cost is largely reduced
from that of our previous large-scale simulation~\cite{ConfGene:Nf2:ovr:JLQCD,ConfGene:Nf3:ovr:JLQCD}
using the overlap formulation
that also preserves chiral symmetry~\cite{Compara:JLQCD}.

\begin{table}[b]
\centering
\small
\caption{
  Parameters of gauge ensembles.
  We denote the five dimensional lattice size as $N_s^3 \times N_t \times N_5$.
  Quark masses are bare value in lattice units.
}
\label{tbl:ffs:param}
\begin{tabular}{l|lllllll}
   \hline 
   lattice parameters 
   & $m_{ud}$ & $m_s$ & $M_\pi$[MeV] & $M_K$[MeV] & $\dt + \dtp$
   & $N_{\rm conf}$ & $N_\tsrc$
   \\ \hline
   $\beta\!=\!4.17$, \ 
   $a^{-1}\!=\!2.453(44)$~GeV, \ 
   $32^3\!\times\!64\!\times\!12$
   & 0.0190 & 0.0400 & 499(1) & 618(1) & 12, 16, 24, 28  & 100 & 1 
   \\
   & 0.0120 & 0.0400 & 399(1) & 577(1) & 12, 16, 22, 26  & 100 & 1 
   \\
   & 0.0070 & 0.0400 & 309(1) & 547(1) & 12, 16, 22, 26  & 100 & 2
   \\
   & 0.0035 & 0.0400 & 230(1) & 527(1) & 8, 12, 16, 20   & 100 & 2
   \\ \hline
   $\beta\!=\!4.17$, \ 
   $48^3\!\times\!96\!\times\!12$
   & 0.0035 & 0.0400 & 226(1) & 525(1) & 8, 12, 16, 20   & 100 & 2
   \\ \hline
   $\beta\!=\!4.35$, \ 
   $a^{-1}\!=\!3.610(65)$~GeV, \ 
   $48^3\!\times\!96\!\times\!8$
   & 0.0120 & 0.0250 & 501(2) & 620(2) & 18, 24, 36, 42  & 50 & 1
   \\
   & 0.0080 & 0.0250 & 408(2) & 582(2) & 18, 24, 33, 39  & 50 & 1
   \\
   & 0.0042 & 0.0250 & 300(1) & 547(2) & 18, 24, 33, 39  & 50 & 2
   \\ \hline
   $\beta\!=\!4.47$, \ 
   $a^{-1}\!=\!4.496(80)$~GeV, \ 
   $64^3\!\times\!128\!\times\!8$
   & 0.0030 & 0.0150 & 284(1) & 486(1) & 16, 24, 32, 40  & 50 & 2
   \\ \hline
\end{tabular}
\end{table}

\begin{table}[t]
\centering
\small
\caption{
  Bare heavy quark masses in lattice units
  used to calculate relevant two- and three-point functions.
  The smallest value corresponds to the physical charm mass
  fixed from the spin averaged charmonium mass.
}
\label{tbl:ffs:param:mQ}
\begin{tabular}{l|l}
   \hline 
   $\beta$ & $m_Q$
   \\ \hline
   4.17    & 0.44037, 0.55046, 0.68808
   \\ \hline
   4.35    & 0.27287, 0.34109, 0.42636, 0.53295, 0.66619
   \\ \hline
   4.47    & 0.210476, 0.263095, 0.328869, 0.4110859, 0.5138574, 0.6423218
   \\ \hline
\end{tabular}
\end{table}


We employ the M\"obius domain-wall action for all relevant quark flavors.
Our choices of the degenerate up and down quark mass $m_{ud}$
correspond to pion masses from $M_\pi\!\sim\!500$\,MeV down to 230\,MeV.
Chiral symmetry allows us to use 
heavy meson chiral perturbation theory (HMChPT)~\cite{HMChPT:RW,HMChPT:S}
for our chiral extrapolation of simulation results to the physical pion mass
without introducing terms to describe discretization effects.
We take a strange quark mass $m_s$ close to its physical value
fixed from $M_{\eta_s}^2\!=\!2M_K^2 - M_\pi^2$.
The charm quark mass $m_c$ is set to its physical value
fixed from the spin-averaged charmonium mass $(M_{\eta_c}+3M_{J/\Psi})/4$~\cite{mc:Nf3:JLQCD}.
Depending on the lattice spacing $a$, 
we take three to six bottom quark masses $m_Q\!=\!1.25^n m_c$
$(n\!=\!0,1,\cdots)$ satisfying the condition $m_Q\!\leq\!0.7 a^{-1}$
to avoid large discretization errors.
We note that, with chiral symmetry,
the leading discretization error is reduced to $O((am_Q)^2)$.
As we will see in the next section,
$a$ and $m_Q$ dependences of the form factors are mild,
and the extrapolation to the continuum limit and physical quark masses
is under good control.
Our simulation parameters are summarized in Tables~\ref{tbl:ffs:param}
and \ref{tbl:ffs:param:mQ}.

The spatial lattice size $L=N_s a$ is chosen to satisfy the condition
$M_\pi L \! \gtrsim \! 4$ to suppress finite volume effects (FVEs).
At our smallest $M_\pi$ on the coarsest lattice,
we also simulate a smaller physical lattice size $M_\pi L\!\sim\!3$
to directly examine the FVEs.


The statistics at the lattice cutoffs $a^{-1}\simeq\!2.5$, 3.6 and 4.5~GeV,
are 5,000, 2,500 and 2,500 Hybrid Monte Carlo trajectories
with a unit trajectory length of 1, 2 and 4
times the molecular dynamics time unit, respectively,
which is increased to take account of the longer auto correlation
toward the continuum limit.
On each ensemble, 
we take $N_{\rm conf}$ configurations in an equal trajectory interval
to calculate correlation functions relevant to the $B\!\to\!D^*\ell\nu$ decay.
Our choice of $N_{\rm conf}$ is listed in Table~\ref{tbl:ffs:param}.
More details on our simulation setup can be found in Ref.~\cite{B2pi:Nf3:JLQCD}.


\subsection{Ratio method}
\label{subsec:ff:ratio}


The $B\!\to\!D^{*}$ matrix elements can be extracted from the
three- and two-point functions
\bea
   C_{\mathcal O_\Gamma}^{BD^*}(\dt,\dtp;\bfp,\bfpp,\epsp)
   & = & 
   \frac{1}{N_s^3 N_{\tsrc}}\sum_{\tsrc, \bfxsrc}
   \sum_{\bfx, \bfxp}
   \left\langle 
       {\mathcal O}_{D^*}(\bfxp,\tsrc+\dt+\dtp;\epsp)
   \right.
   \nn \\
   &&
   \hspace{10mm}
   \times
   \left.
      {\mathcal O}_\Gamma(\bfx,\tsrc+\dt)
      {\mathcal O}_{B}(\bfxsrc,\tsrc)^\dagger
   \right\rangle
   e^{-i\bfp(\bfx-\bfxsrc)-i\bfpp(\bfxp-\bfx)},
   \hspace{10mm}
   \label{eqn:ff:corr_3pt}
   \\
   C^P(\dt;\bfp)
   & = & 
   \frac{1}{N_s^3 N_{\tsrc}}\sum_{\tsrc,\bfxsrc}
   \sum_{\bfx}
   \langle 
      {\mathcal O}_P(\bfx,\tsrc+\dt)
      {\mathcal O}_P(\bfxsrc,\tsrc)^\dagger
   \rangle
   e^{-i\bfp(\bfx-\bfxsrc)}
   \label{eqn:ff:corr_2pt}
\eea
through their ground state contribution   
\bea
   C_{\mathcal O_\Gamma}^{BD^*}(\dt,\dtp;\bfp,\bfpp,\epsp)
   & \xrightarrow[\dt,\dtp \to \infty ]{} &
   \frac{Z_{D^*}^*(\bfpp,\epsp)\,Z_B(\bfp)}{4E_{D^{*}}(\bfpp)E_B(\bfp)}
   \langle D^{*}(\pp,\epsp) | {\mathcal O}_\Gamma | B(p) \rangle
   e^{-E_{D^{*}}(\bfpp)\dtp -E_B(\bfp)\dt },
   \hspace{10mm}
   \label{eqn:ff:corr_3pt:ground}
   \nn \\
   C^P(\dt;\bfp)
   & \xrightarrow[\dt \to \infty ]{} &
   \frac{Z_P^*(\bfp)\,Z_P(\bfp)}{2E_P(\bfp)}
   e^{-E_P(\bfp)\dt},
   \label{eqn:ff:corr_2pt:ground}
\eea
where
${\mathcal O}_\Gamma$ is the weak vector ($V_\mu$) or axial ($A_\mu$) current,
and we omit the symbol ``$\prime$'' on the momentum variable and 
the argument $\epsp$ 
for the $D^*$ two-point function $C^{D^*}(\dt;\bfpp,\epsp)$ for simplicity.
The meson interpolating field is denoted by 
${\mathcal O}_P$ ($P\!=\!B, D^*$),
where Gaussian smearing is applied
to enhance their overlap with the ground state
$Z_P(\bfp)\!=\!\langle P | {\mathcal O}_P^\dagger \rangle$.
The $B$ meson is at rest ($\bfp\!=\!\bfz$) throughout this paper.
The $w$ dependence of the form factors
is studied by varying the three-momentum of $D^{*}$
as $|\bfpp|^2\!=\!0,1,2,3,4$
(in this paper, we denote the momentum on the lattice in units of $2\pi/L$).
To this end, we also calculate $D^*$ two-point functions
with the local sink operator ${\mathcal O}_{D^*,\rm lcl}^\dagger$
\bea
   C^{D^*}_{\rm sl}(\dt;\bfpp,\epsp)
   & = & 
   \frac{1}{N_s^3 N_{\tsrc}}\sum_{\tsrc,\bfxsrc}
   \sum_{\bfx}
   \langle 
      {\mathcal O}_{D^*,\rm lcl}(\bfx,\tsrc+\dt;\epsp)
      {\mathcal O}_{D^*}(\bfxsrc,\tsrc;\epsp)^\dagger
   \rangle
   e^{-i\bfpp(\bfx-\bfxsrc)},
   \hspace{8mm}
   \label{eqn:ff:corr_2pt:lcl}
\eea
which are used in a correlator ratio~(\ref{eqn:ff:ratio:r1p}) below.


The three-point functions are calculated by using the sequential source method,
where the total temporal separation $\dt+\dtp$ between the source and sink operators 
is fixed
and the temporal location of the weak current is varied.
In order to control the contamination from the excited states,
we repeat our measurement
for four values of the source-sink separation $\dt+\dtp$
listed in Table~\ref{tbl:ffs:param}.


The statistical accuracy of the two- and three-point functions are improved
by averaging over the spatial location of the source operator $\bfxsrc$
as indicated in Eqs.~(\ref{eqn:ff:corr_3pt}) and (\ref{eqn:ff:corr_2pt}).
To this end, we employ a volume source with $Z_2$ noise.
At $M_\pi\!\lesssim\!300$~MeV,
we repeat our measurement
over two values of the source time-slices $\tsrc\!=\!0$ and $N_t/2$
to take the average of the correlators.
Namely $N_\tsrc$ in Eqs.~(\ref{eqn:ff:corr_3pt}) and (\ref{eqn:ff:corr_2pt})
is 2 at $M_\pi\!\lesssim\!300$~MeV, and 1 otherwise.
The correlators with non-zero momentum $\bfp$ are also averaged
over all possible $\bfp$'s based on parity and rotational symmetries on the lattice.


\begin{figure}[t]
\begin{center}
  \includegraphics[angle=0,width=0.48\linewidth,clip]{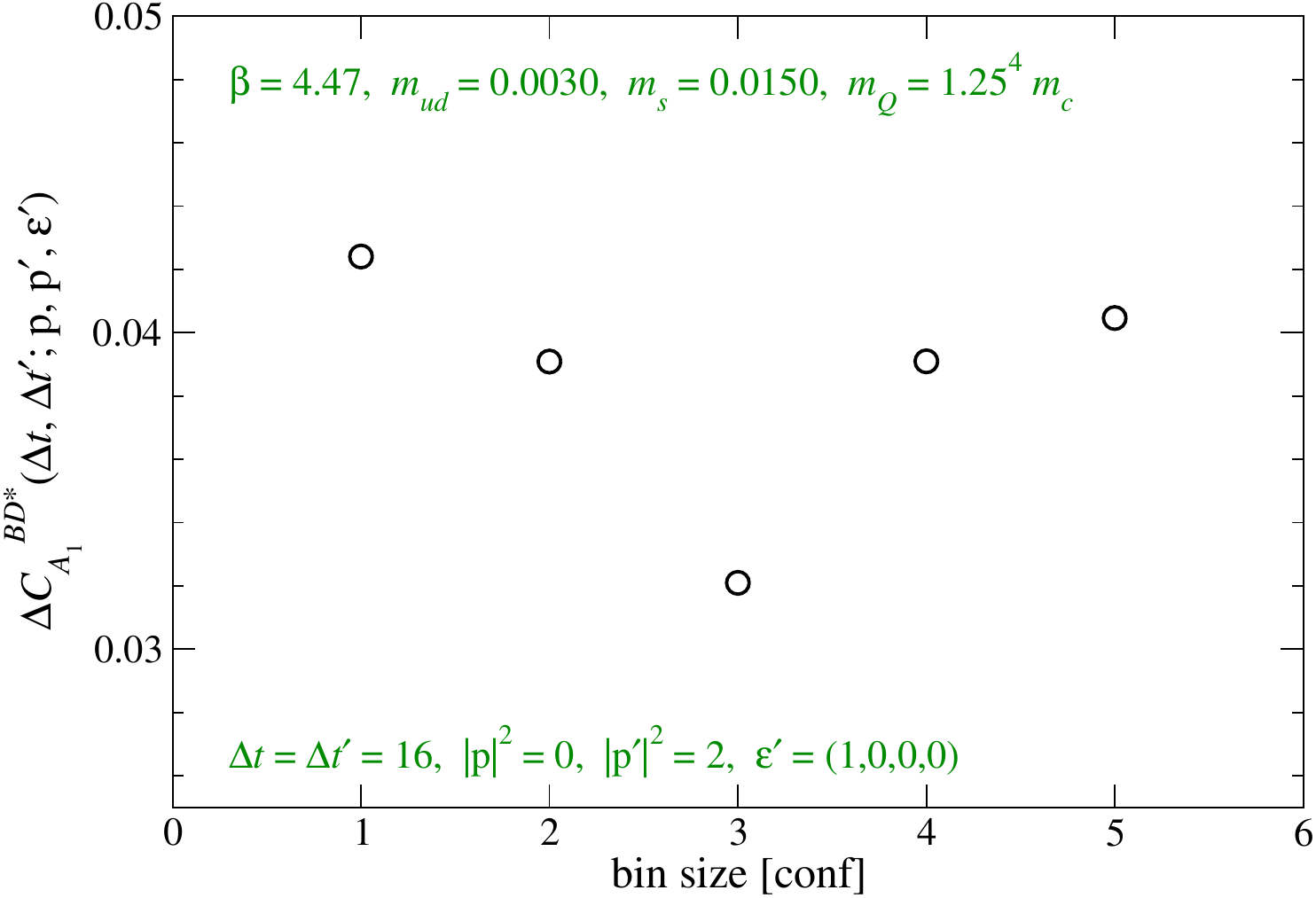}
  \vspace{-3mm}
  \caption{
    Bin size dependence of relative statistical error of 
    three-point function $C_{\mathcal O_\Gamma}^{BD^*}(\dt,\dtp;\bfp,\bfpp,\epsp)$.
    We plot data at the largest cutoff $a^{-1}\!\simeq\!4.5$~GeV,
    $m_Q\!=\!1.25^4 m_c$,
    $\dt\!=\!\dtp\!=\!16$,
    $|\bfp|^2=0$, $|\bfpp|^2=2$
    and $\epsp\!=\!(1,0,0,0)$.
  }
  \label{fig:ff:vs_bins}
\end{center}
\end{figure}

The number of measurements on each ensemble 
is given as $N_{\rm conf} N_\tsrc$ in Table~\ref{tbl:ffs:param}.
As mentioned above,
correlation functions for each configuration are averaged
over $N_\tsrc$ values of the source time-slice $\tsrc$.
Then, simulation data of $N_{\rm conf}$ configurations are divided into 50 bins:
namely, the bin size is two (one) configurations for $\beta\!=\!4.17$
(4.35 and 4.47).
The statistical error is estimated by the bootstrap method with 5,000 replicas.
Figure~\ref{fig:ff:vs_bins} shows the bin size dependence 
of the relative statistical error of a three-point function
at our largest cutoff, where the topological charge $Q$
changes much less frequently leading to larger auto-correlation
compared to the coarser lattices~\cite{B2pi:Nf3:JLQCD}.
Our choice of the bin size on this finest lattice is one configuration,
and we do not observe significant increase of the statistical error
toward larger bin sizes.
This suggests that our bin size is sufficiently large
partly due to our choice of larger unit trajectory length towards the continuum limit.
We also note that the error estimate is stable 
when we vary the number of bootstrap replicas
as well as when we employ the jackknife method instead.


\begin{figure}[t]
\begin{center}
  \includegraphics[angle=0,width=0.48\linewidth,clip]{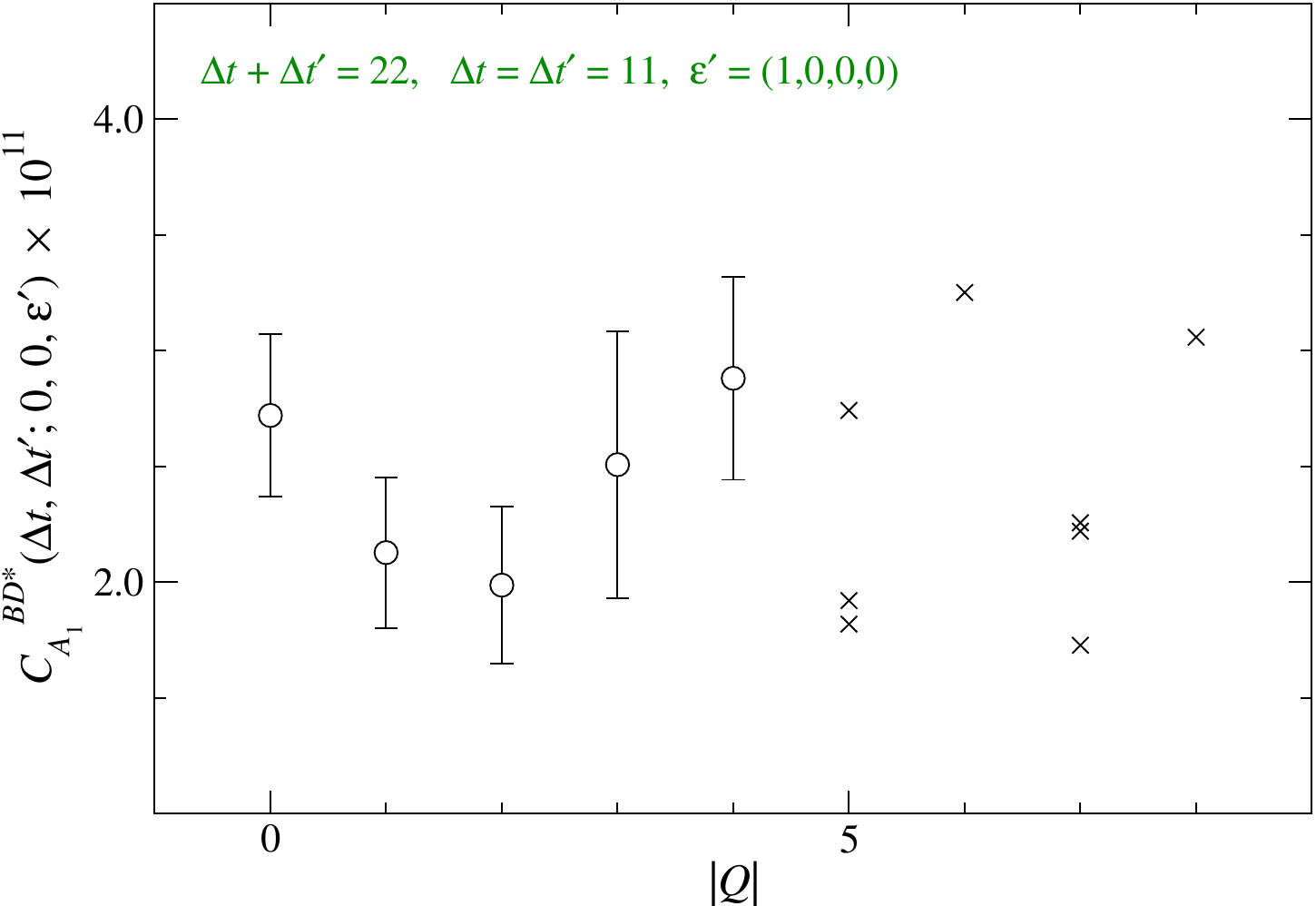}
  \vspace{-3mm}
  \caption{
    Topological charge dependence of 
    three-point function $C_{\mathcal O_\Gamma}^{BD^*}(\dt,\dtp;\bfp,\bfpp,\epsp)$
    at $a^{-1}\!\simeq\!2.5$~GeV and $M_\pi\!\simeq\!300$~MeV.
    We plot data with $m_Q\!=\!1.25^2 m_c$,
    $\dt\!=\!\dtp\!=\!11$,
    zero momenta $|\bfp|^2=|\bfpp|^2=0$
    and $\epsp\!=\!(1,0,0,0)$.
    The error is estimated by the jackknife method.
    At $|Q|\!\geq\!5$, the number of data are smaller than 10,
    and it is difficult to reliably estimate the statistical error.
    We therefore plot individual data by crosses
    rather than their statistical average.
  }
  \label{fig:ff:Qdep}
\end{center}
\end{figure}

As observed in our previous simulation
in the trivial topological sector~\cite{ConfGene:Nf2:ovr:JLQCD},
the local topological fluctuation are active
even if we fix the ``global'' topological charge $Q$,
namely the whole-volume integral of the topological charge density.
In Ref.~\cite{Topology:Nf2:ovr:JLQCD},
we demonstrated that the topological susceptibility $\chi_t$ is calculable
within the fixed topology setup.
Our result for $\chi_t$ shows quark mass dependence consistent with ChPT,
and its value extrapolated to the chiral limit is in good agreement
with that from our simulation
in the $\epsilon$-regime~\cite{TopSucep:Nf2:ovr:JLQCD:Let,TopSucep:Nf2:ovr:JLQCD}.
References~\cite{FixedQ:BCNW03,FixedQ:AFHO07} argued that 
the bias due to the freezing of the global topology can be considered as
FVEs suppressed by the inverse space-time volume.
In our study of the $B\!\to\!\pi\ell\nu$ decay~\cite{B2pi:Nf3:JLQCD},
we confirmed that such FVEs on the pion effective mass are well suppressed,
so that its $|Q|$ dependence is not significant.
Figure~\ref{fig:ff:Qdep}
shows the statistical average of a $B\!\to\!D^*$ three-point function
calculated for each $|Q|$.
We do not observe any significant $|Q|$ dependence of the average
at $|Q|\!\leq\!4$.
At larger $|Q|$, we do not have enough data for a reliable error estimate,
but individual data plotted by crosses are consistent with
the averages at $|Q|\!\leq\!4$.
We therefore conclude that the topology freezing effect is insignificant
within our statistical accuracy.


\begin{figure}[t]
\begin{center}
  \includegraphics[angle=0,width=0.48\linewidth,clip]{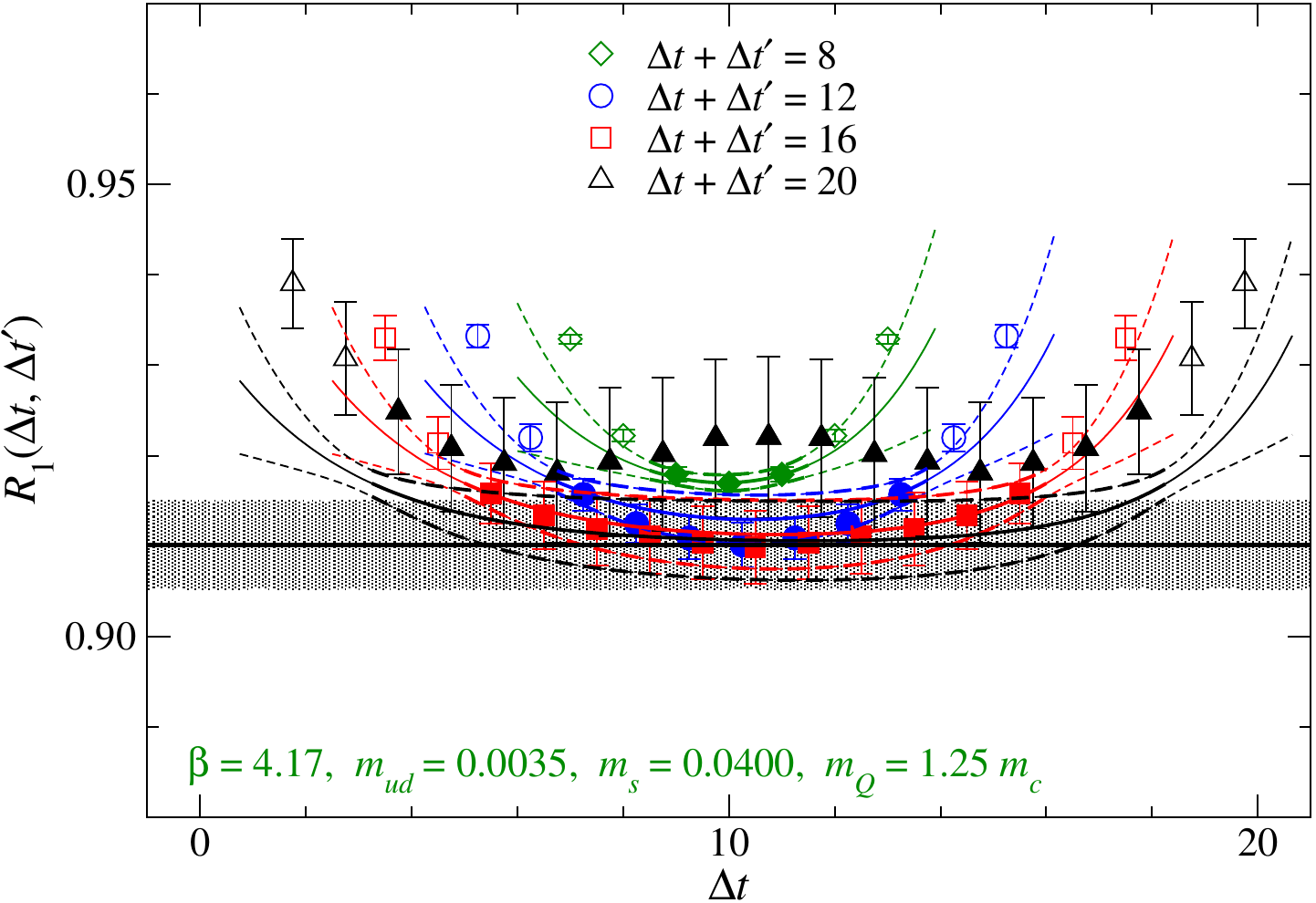}
  \hspace{3mm}
  \includegraphics[angle=0,width=0.48\linewidth,clip]{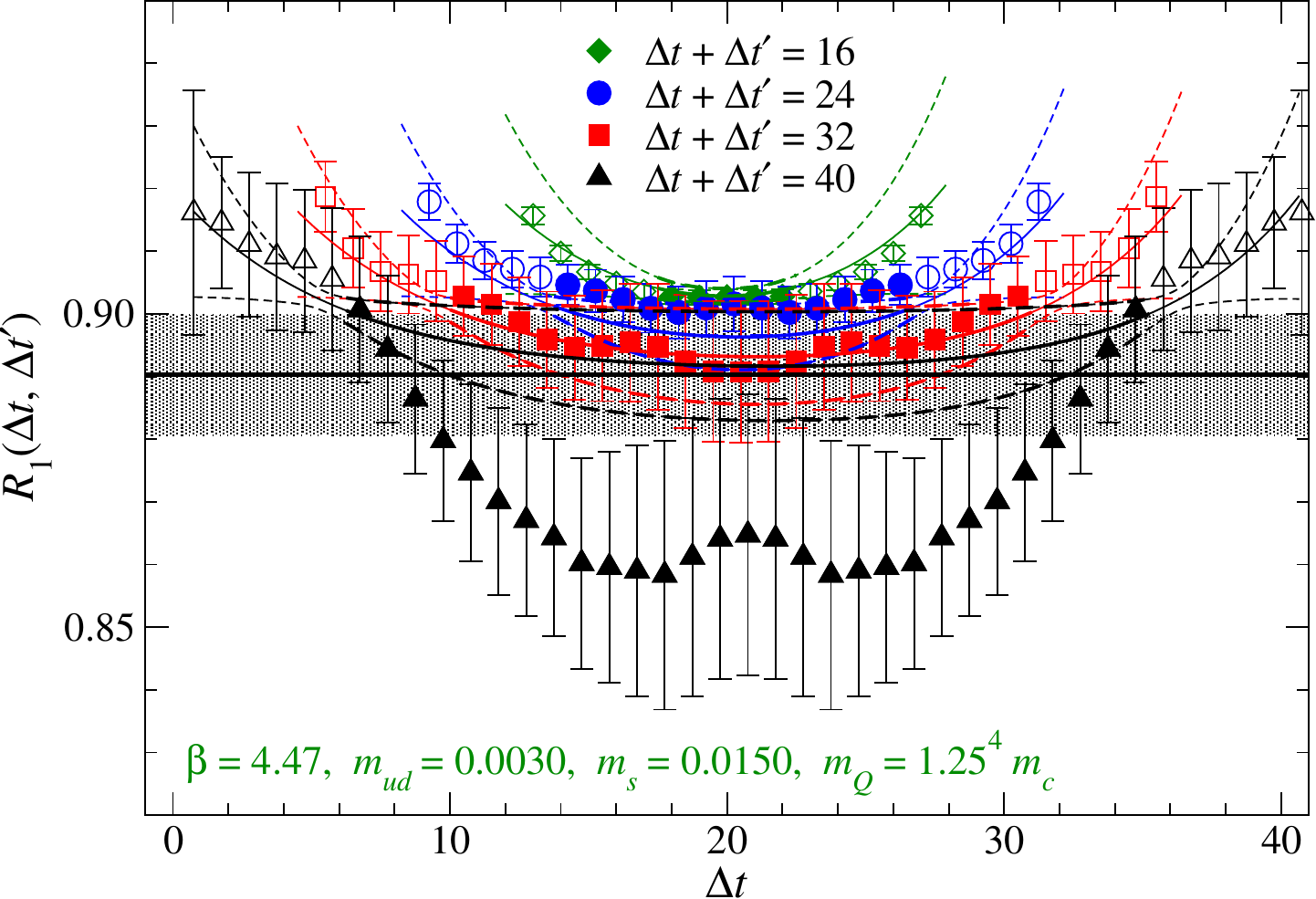}
  \vspace{-3mm}
  \caption{
    Double ratio~(\ref{eqn:ff:ratio:r1}) to estimate $h_{A_1}(1)$
    as a function of temporal location of weak currents $\dt$.
    The open symbols show all data at different choices of
    the source-sink separations $\dt+\dtp$,
    whereas the filled symbols are data used in the fit (\ref{eqn:ff:ratio:r1:fit})
    to extract the ground state matrix element shown by the horizontal black band.
    The thick (thin) curves are fit curves inside (outside) the fit range. 
    The left panel shows the data at 
    our smallest pion mass $M_\pi\!\simeq230$~MeV
    ($a^{-1}\!\simeq\!2.5$~GeV, $m_Q\!=\!1.25 m_c$),
    whereas the right panel is for 
    our largest cutoff $a^{-1}\!\simeq\!4.5$~GeV
    ($M_\pi\!\simeq\!300$~GeV, $m_Q\!=\!1.25^4 m_c$).
    All data are shifted along the horizontal axis 
    so that the mid-point $\dt\!=\!\dtp$ is located
    at the center of the panel
    ($\dt=10$ and 20 in the left and right panels, respectively).
    The symbols are further but slightly shifted along the horizontal axis
    for clarity.
  }
  \label{fig:ff:drat14}
\end{center}
\end{figure}

To precisely extract the form factors,
we construct ratios of the correlation functions, in which
unnecessary overlap factors and exponential damping factors cancel
for the ground-state contribution~\cite{ratio}.
The statistical fluctuation is also expected to partly cancel in such a ratio.
With $B$ and $D^*$ mesons at rest,
the vector current matrix element (\ref{eqn:intro:ff:V:hqet}) vanishes,
and the axial vector one (\ref{eqn:intro:ff:A:hqet})
is sensitive only to the axial vector form factor $h_{A_1}$ at zero recoil,
which is the fundamental input to determine $|V_{cb}|$.
In order to precisely determine $h_{A_1}(1)$,
we employ a double ratio
\bea
   R_1(\dt,\dtp)
   & = &
   \frac{ C_{A_1}^{BD^*}(\dt,\dtp;\bfz,\bfz,\epsp)\,
          C_{A_1}^{D^*B}(\dt,\dtp;\bfz,\bfz,\epsp) }
        { C_{V_4}^{BB}(\dt,\dtp;\bfz,\bfz)\,
          C_{V_4}^{D^*D^*}(\dt,\dtp;\bfz,\bfz,\epsp,\epsp) }
   \xrightarrow[\dt,\dtp\to\infty]{}
   |h_{A_1}(1)|^2,
   \label{eqn:ff:ratio:r1}
\eea
where the $D^*$ polarization is chosen as $\epsp\!=\!(1,0,0,0)$
along the polarization of the current inserted, {\it i.e.} $A_1$.
The left panel of Fig.~\ref{fig:ff:drat14} shows our results for this ratio
at our smallest $M_\pi$ and $a^{-1}$ with $m_Q\!=\!1.25\,m_c$.
Note that
the ratio is symmetrized as 
$R_1(\dt,\dtp)\!\to\!\left\{ R_1(\dt,\dtp) + R_1(\dtp,\dt) \right\}/2$,
since we use the same smearing function for both the source and sink operators.
We observe a reasonable consistency among data at intermediate values of
the source-sink separation $\dt+\dtp\!=\!12$ (circles) and 16 (squares).
The excited state contribution could potentially be significant but not large
($\lesssim\,1$\,\%) for the smallest $\dt+\dtp\!=\!8$ (diamonds).
Data at the largest $\dt+\dtp\!=\!20$ (triangles) show a long plateau
and are consistent with those at smaller $\dt+\dtp$
within the large statistical errors.
These observations suggest that
the data at $\dt+\dtp\!=\!12$, which is roughly 1~fm,
are dominated by the ground state contribution
at least at the midpoint $\dt\!\sim\!\dtp$.
The situation is similar in the right panel of the same figure
for our largest $a^{-1}$ and with $m_Q\!=\!1.25^4 m_c$,
where data around midpoint $\dt\!\sim\!\dtp$ are consistent
among all simulated source-sink separations within 2\,$\sigma$.


\begin{figure}[t]
\begin{center}
  \includegraphics[angle=0,width=0.48\linewidth,clip]{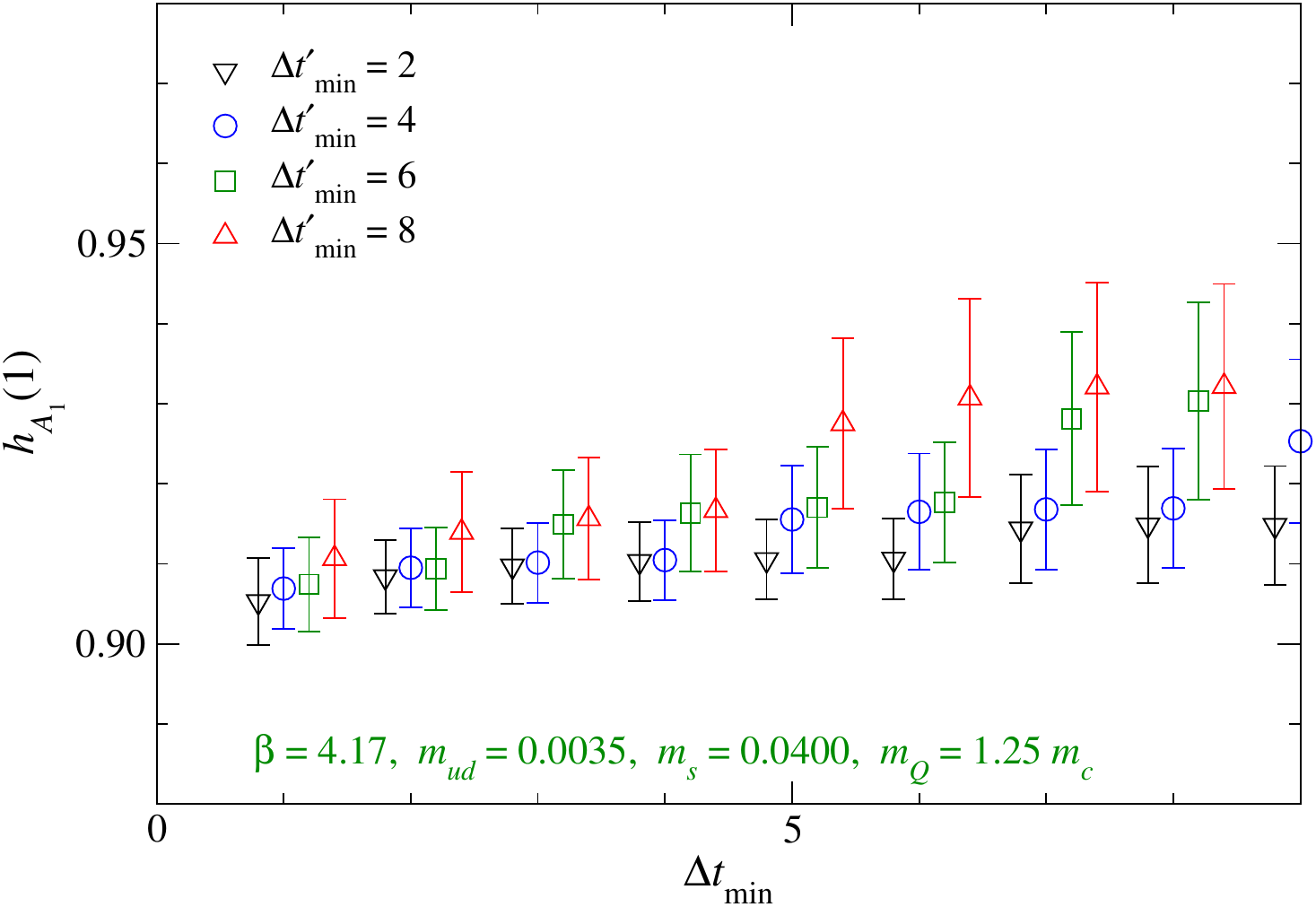}
  \hspace{3mm}
  \includegraphics[angle=0,width=0.48\linewidth,clip]{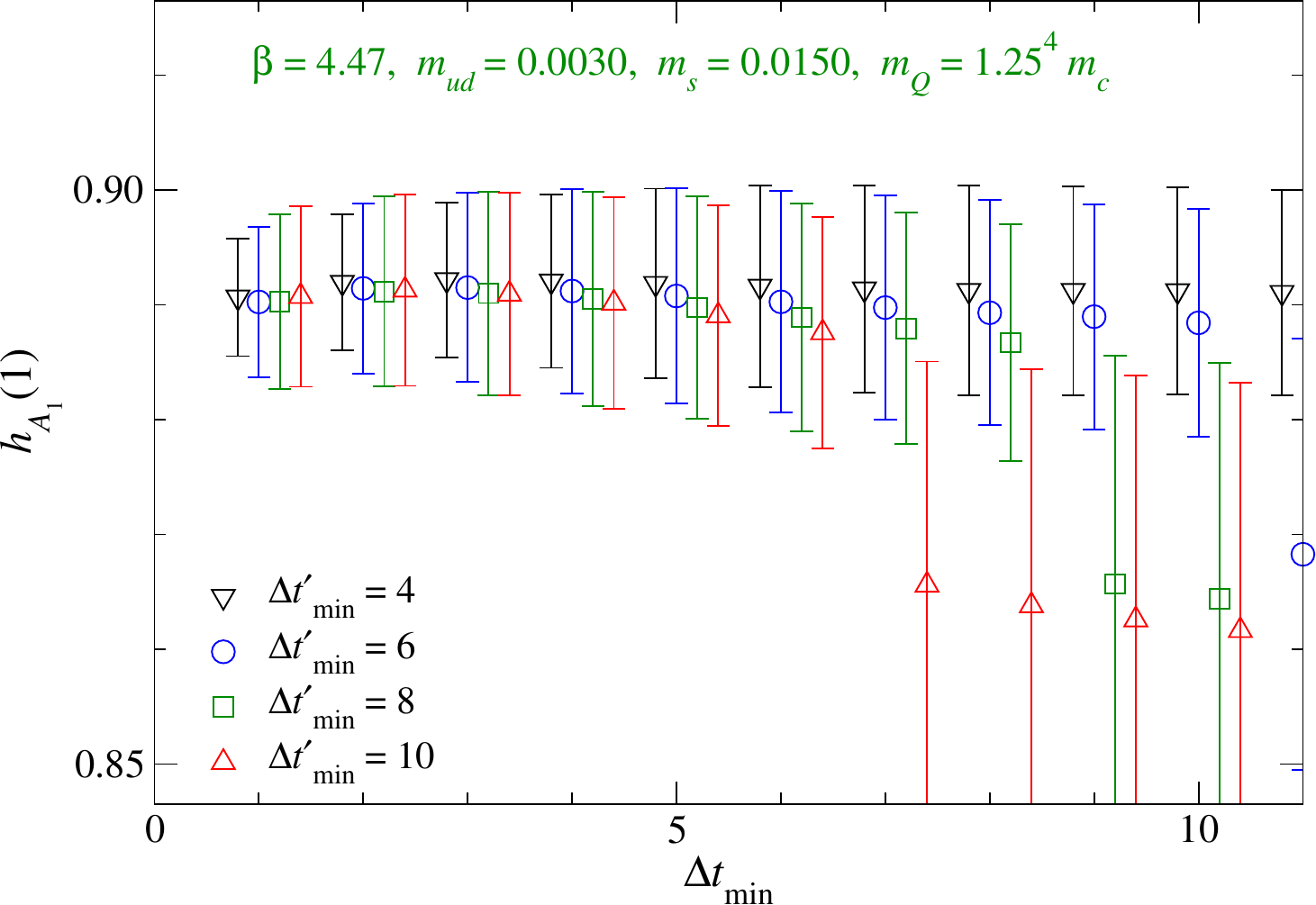}
  \vspace{-3mm}
  \caption{
    Axial form factor $h_{A_1}(1)$ from fit (\ref{eqn:ff:ratio:r1:fit})
    as a function of lower cut $\dtmin$.
    Left and right panels show data at
    $(a^{-1},M_\pi)\!\simeq\!(2.5~{\rm GeV},230~{\rm MeV})$
    and $(4.5~{\rm GeV},300~{\rm MeV})$, respectively.
    Different symbols show data with different values for $\dtpmin$.
    Symbols are slightly shifted along the horizontal axis for clarity.
  }
  \label{fig:ff:drat14:fit_range}
\end{center}
\end{figure}

We find that all data are well described by the following fitting form
including effects from the first excited states 
\bea
   R_1(\dt,\dtp)
   & = &
   |h_{A_1}(1)|^2
   \left(
      1 + a e^{-\Delta M_B \dt}  + b e^{-\Delta M_{D^*} \dt}
        + a e^{-\Delta M_B \dtp} + b e^{-\Delta M_{D^*} \dtp}
   \right)
   \label{eqn:ff:ratio:r1:fit}
\eea
with $\chi^2/{\rm d.o.f.}\!\lesssim\!1$
for a wide range of the values of $\dt$ and $\dtp$.
Here $\Delta M_{B(D^*)}$ represents the energy difference
between the $B(D^*)$ meson ground state 
and the first excited state of the same quantum numbers,
and is set to the value estimated from
a two-exponential fit to the two-point function $C^{B(D^*)}(\dt;\bfz)$.
The fit range is chosen such that
all the data of $R_1(\dt,\dtp)$
with the source-to-current (current-to-sink) separation $\dt^{(\prime)}$
equal to or larger than a lower cut $\dt^{(\prime)}_{\rm min}$
are included irrespective of the source-sink separation $\dt+\dtp$.
As shown in the left panel of Fig.~\ref{fig:ff:drat14:fit_range},
our result for $h_{A_1}(1)$ is stable against the choice of the lower cuts
$\dtmin$ and $\dtpmin$.


The right panels of
Figs.~\ref{fig:ff:drat14} and \ref{fig:ff:drat14:fit_range} for our largest $a^{-1}$
also show ground state dominance
with $\dt+\dtp\!\gtrsim\!23a\!\sim\!1$~fm
and stability of $h_{A_1}(1)$ against the choice of the fit range,
respectively.
The situation is similar for other simulation parameters.


\begin{figure}[t]
\begin{center}
  \includegraphics[angle=0,width=0.48\linewidth,clip]{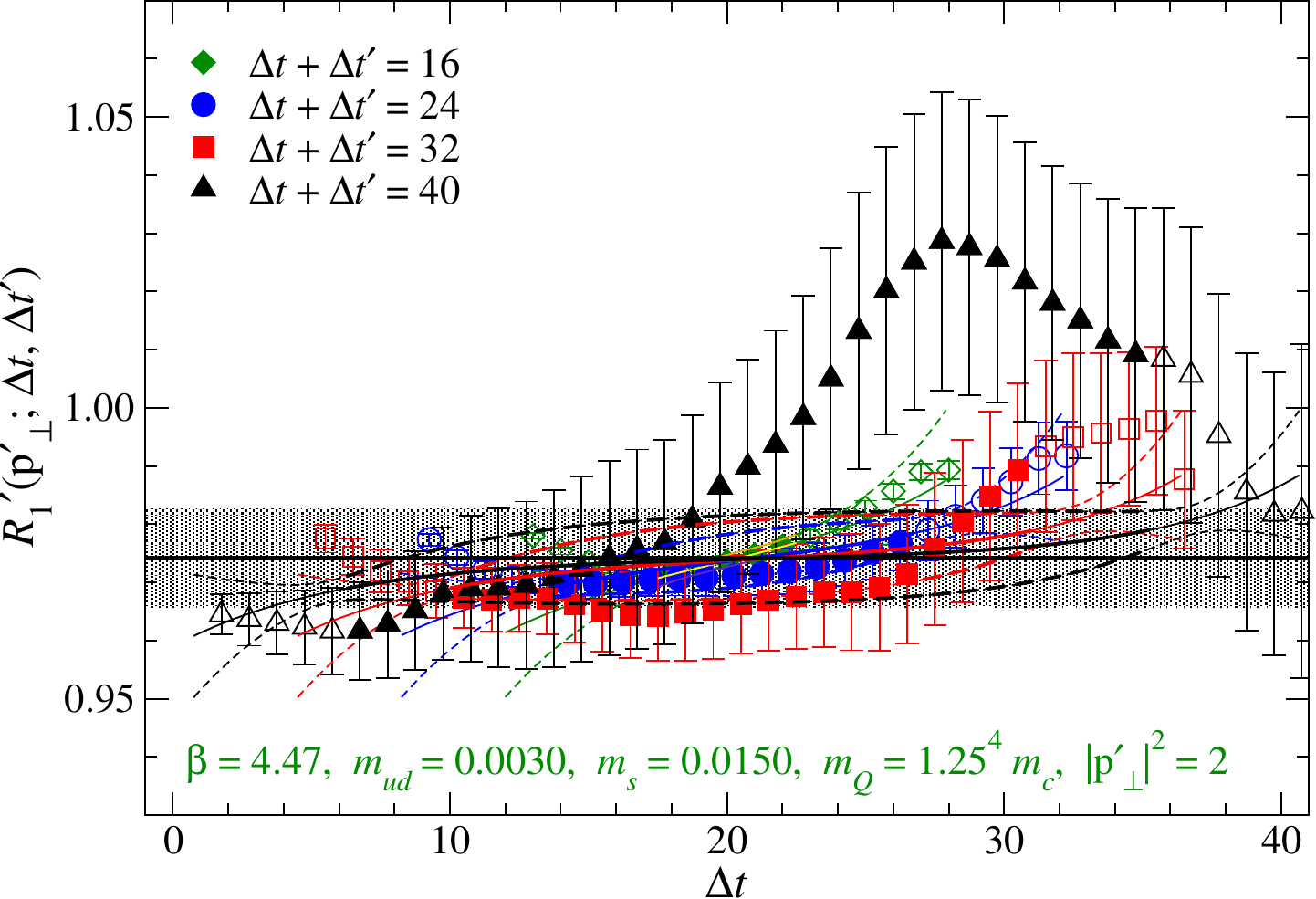}
  \hspace{3mm}
  \includegraphics[angle=0,width=0.48\linewidth,clip]{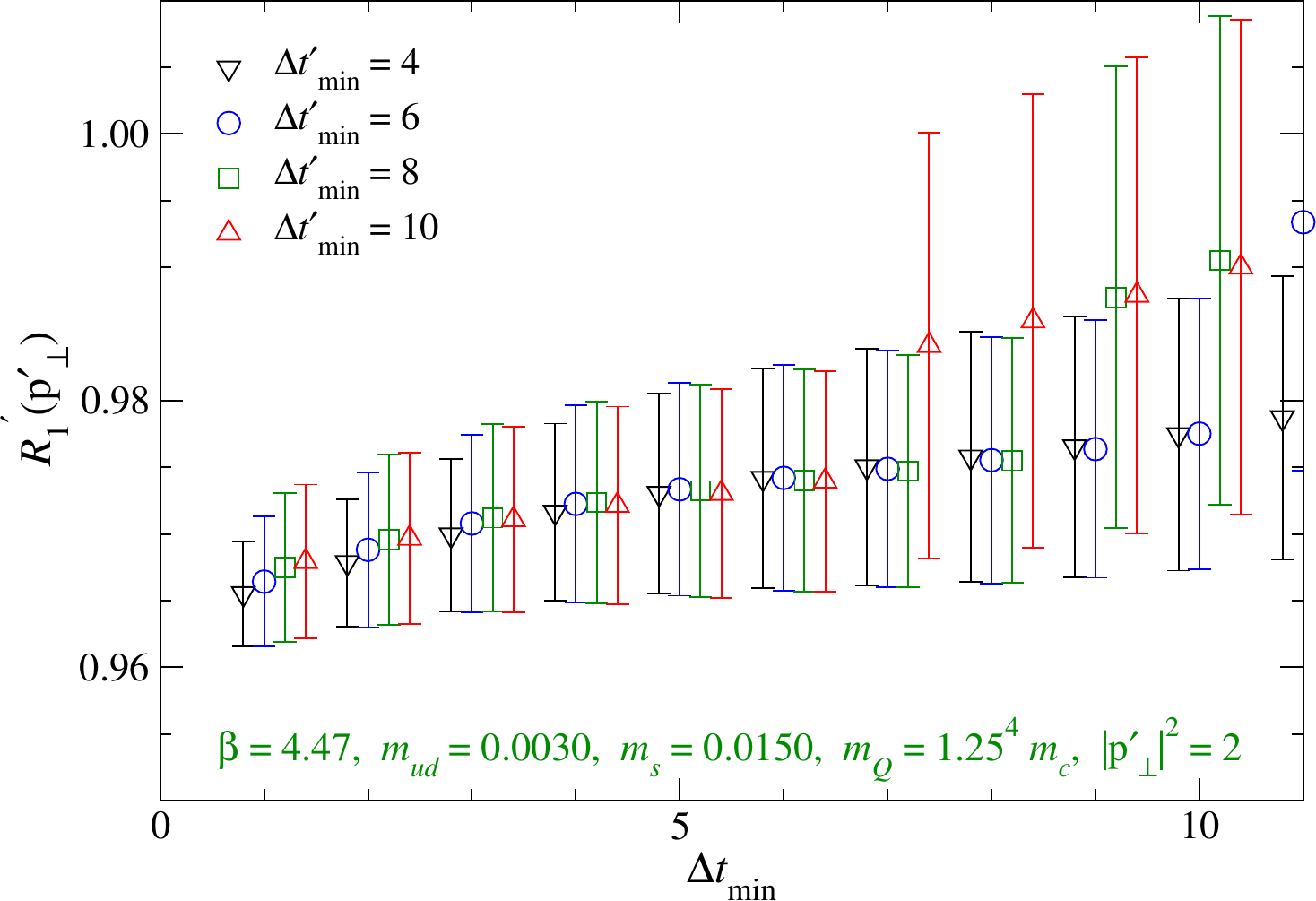}
  \vspace{-3mm}
  \caption{
    Left: ratio $R_1^\prime(\bfppperp; \dt, \dtp)$~(\ref{eqn:ff:ratio:r1p})
    to study $w$-dependence of $h_{A_1}$.
    Right: fitted value of $R_1^\prime(\bfppperp; \dt, \dtp)$
    with different choices of the lower cuts $\dtmin$ and $\dtpmin$.
    Both panels show data at 
    $a^{-1}\!\simeq\!4.5$\,GeV and $M_\pi\!\simeq\!300$\,MeV
    with the $D^*$ momentum $|\bfpperp|=\sqrt{2}$.
  }
  \label{fig:ff:drat24}
\end{center}
\end{figure}

In order to extract form factors at non-zero recoils,
the $D^*$ momentum and polarization vector need to be chosen appropriately.
The axial vector matrix element (\ref{eqn:intro:ff:A:hqet})
is sensitive only to $h_{A_1}$
with a polarization vector ${\boldsymbol \epsilon}^\prime\!=\!(1,0,0)$
and a momentum $\bfppperp$ that satisfies $\epsps v\!=\!0$.
We note that, here and in the following,
three dimensional polarization vector ${\boldsymbol \epsilon}$
is accompanied by its temporal component to be fixed from the convention
$\epsps p^\prime_\perp\!=\!0$, which is assumed
for Eqs.~(\ref{eqn:intro:ff:V:hqet}) and (\ref{eqn:intro:ff:A:hqet}).
Then,
the recoil parameter dependence of $h_{A_1}$ may be studied from the following ratio
\bea
   R_1^\prime(\bfppperp; \dt, \dtp) 
   =
   \frac{C_{A_1}^{B D^*}(\dt, \dtp; \bfz, \bfppperp, \epsp) \,
         C^{D^*}_{sl}(\dtp; \bfz, \epsp)}
        {C_{A_1}^{B D^*}(\dt, \dtp; \bfz, \bfz,      \epsp) \,
         C^{D^*}_{sl}(\dtp, \bfppperp, \epsp)}
   \xrightarrow[\dt,\dtp\to\infty]{}
   \frac{w+1}{2} \frac{h_{A_1}(w)}{h_{A_1}(1)}.
   \label{eqn:ff:ratio:r1p}
\eea
Here we use the two-point function with the local sink
$C^{D^*}_{\rm sl}$~(\ref{eqn:ff:corr_2pt:lcl})
as in our study of the $K\!\to\!\pi\ell\nuell$ decay~\cite{Kl3:Nf3:ovr:JLQCD}.
The vector form factor $h_V$ can be extracted from the following ratio
through the vector current matrix element (\ref{eqn:intro:ff:V:hqet})
\bea
   R_V(\bfpppperp, \bfppperp; \dt, \dtp)
   & = &
   \frac{ C_{V_1}^{BD^*}(\dt,\dtp;\bfz,\bfpppperp, \epspp) }
        { C_{A_1}^{BD^*}(\dt,\dtp;\bfz,\bfppperp,  \epsp) }
   \xrightarrow[\dt,\dtp\to\infty]{}
   \frac{ i \varepsilon_{1ij} \epsilon^{\prime\prime *}_i v_{\perp j}^{\prime\prime}}{w+1}
   \frac{h_V(w)}{h_{A_1}(w)}.
   \label{eqn:ff:ratio:rv}
\eea
Here we use two different polarization vectors,
${\boldsymbol \epsilon}^\prime\!=\!(1,0,0)$ and
${\boldsymbol \epsilon}^{\prime\prime}\!=\!(0,1,0)$,
and momenta $\bfp^\prime_\perp$ and $\bfp^{\prime\prime}_\perp$
that satisfy
$\bfp^{\prime(\prime\prime)}_\perp\!\perp\!{\boldsymbol \epsilon}^{\prime(\prime\prime)}$.
Note that 
$|\bfpppperp|\!=\!|\bfppperp|$ to share the same recoil parameter $w$,
and $v_{\perp j}^{\prime\prime}\!=\!p_{\perp j}^{\prime\prime}/M_{D^*}$.
We emphasize that 
renormalization factors of the weak vector and axial currents
cancel thanks to chiral symmetry preserved in our simulations.

\begin{figure}[t]
\begin{center}
  \includegraphics[angle=0,width=0.48\linewidth,clip]{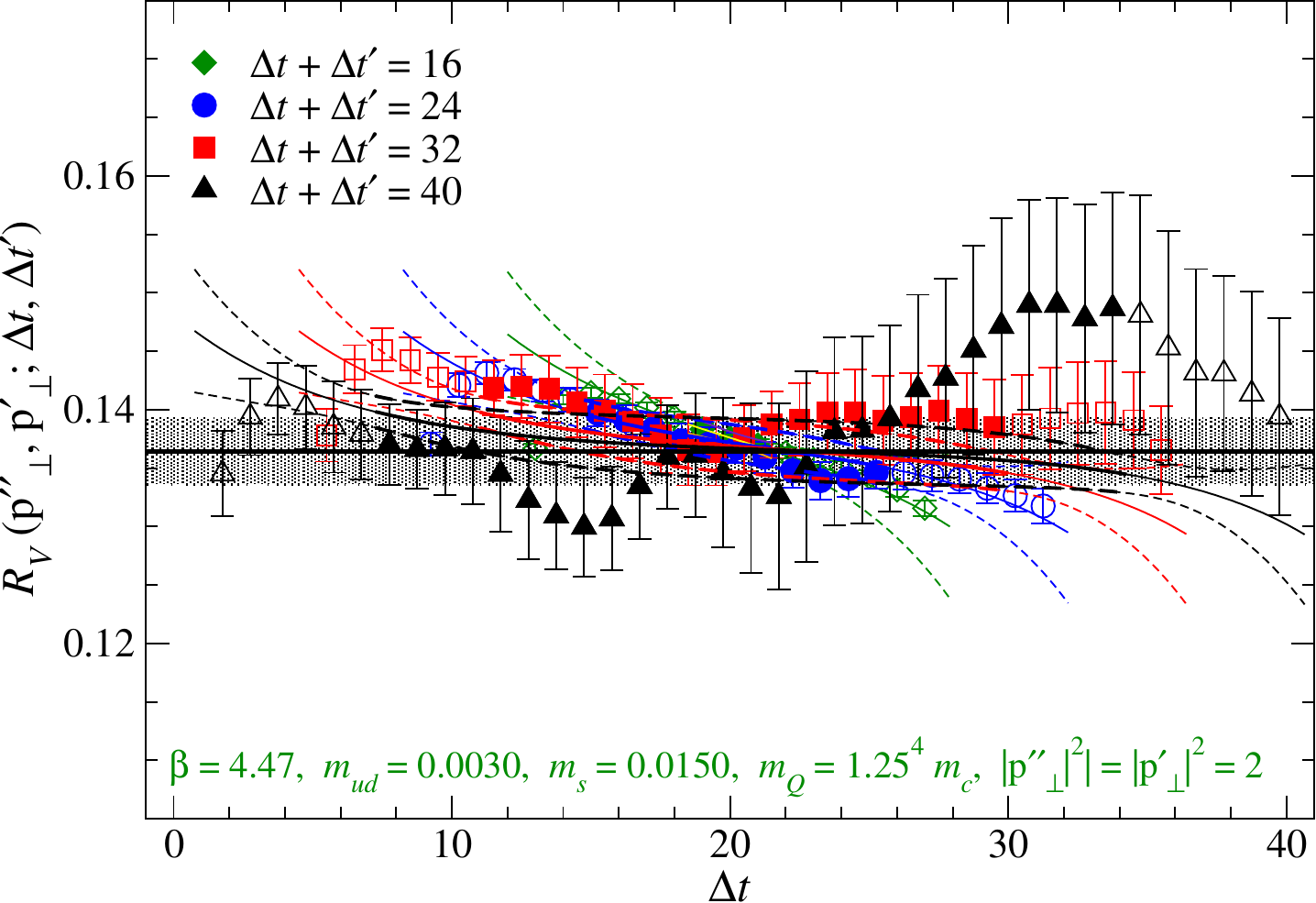}
  \hspace{3mm}
  \includegraphics[angle=0,width=0.48\linewidth,clip]{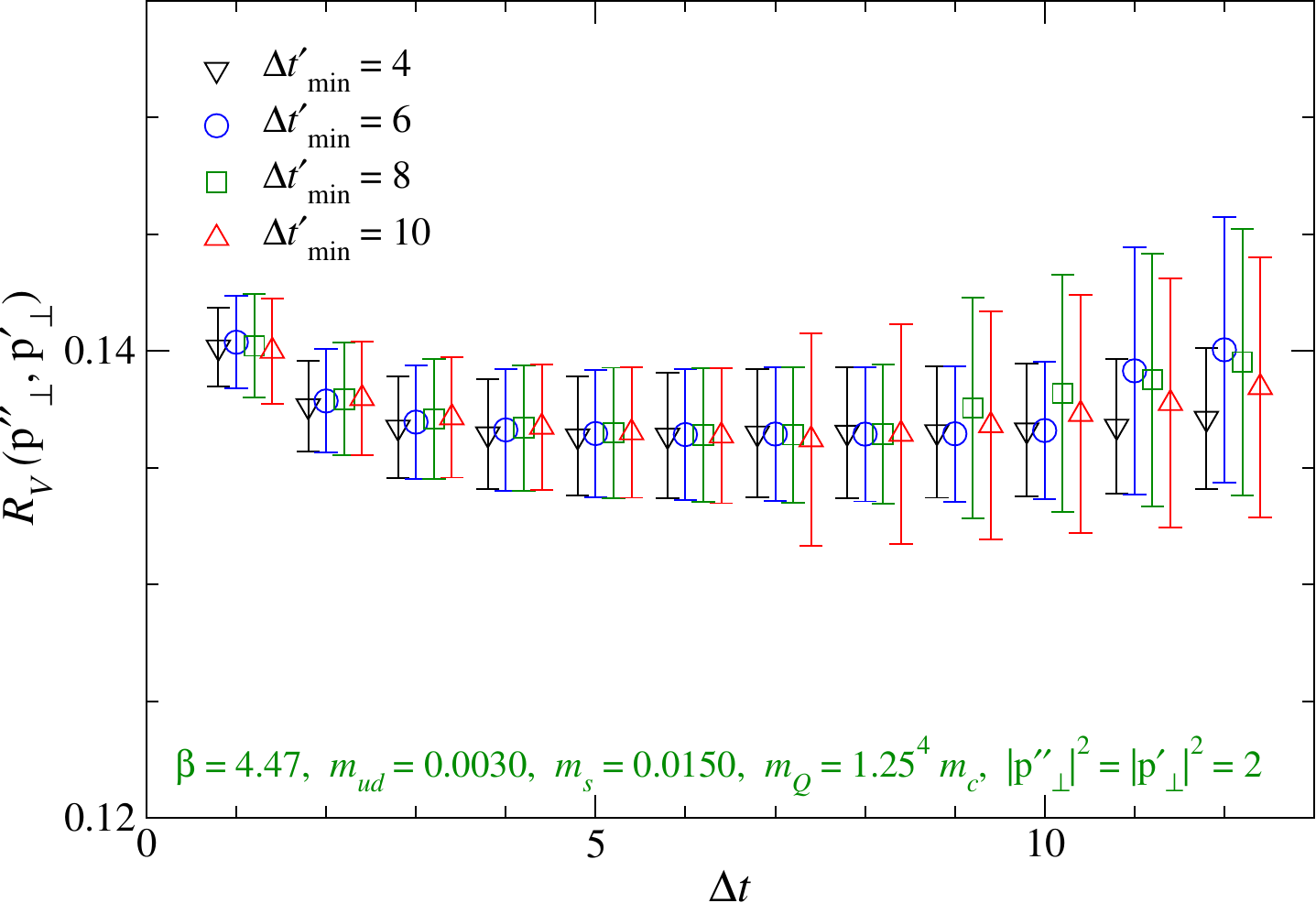}
  \vspace{-3mm}
  \caption{
    Same as Fig.~\protect\ref{fig:ff:drat24}
    but for ratio $R_V(\bfpppperp, \bfppperp; \dt, \dtp)$ (\ref{eqn:ff:ratio:rv})
    to study $h_V(w)$.
  }
  \label{fig:ff:ratv}
\end{center}
\end{figure}


Figures~\ref{fig:ff:drat24} and \ref{fig:ff:ratv} show
an example of our results for $R_1^\prime$ and $R_V$
at our largest cutoff $a^{-1}\!\simeq\!4.5$\,GeV and $m_Q\!=\!1.25^4 m_c$.
Around the mid-point $\dt\!\sim\!\dtp$,
we observe good consistency in these ratios
among all simulated values of the source-sink separation $\dt+\dtp\!\gtrsim\!1$~fm.
A similar ground state dominance is also observed at other simulation points.
We carry out a simultaneous fit using a fitting form
that takes account of the first excited state contribution as
\bea
   R_1^\prime(\bfppperp; \dt,\dtp)
   & = &
   R_1^\prime(\bfppperp)
   \left(
      1 + a e^{-\Delta M_B \dt}  + b e^{-\Delta M_{D^*} \dtp}
   \right)
   \label{eqn:ff:ratio:relse:fit}
\eea
to extract the form factor ratio $h_{A_1}(w)/h_{A_1}(1)$
from the ground state contribution $R_1^\prime(\bfppperp)$,
and a similar form for $R_V(\bfpppperp, \bfppperp; \dt, \dtp)$
to extract $h_V(w)/h_{A_1}(w)$.
Our data are well described by this fitting form
with $\chi^2/{\rm d.o.f.}\!\lesssim\!1$.
From these form factor ratios and $h_{A_1}(1)$ from $R_1$,
we calculate $h_{A_1}(w)$ and $h_V(w)$ at simulated values of $w$.


\begin{figure}[t]
\begin{center}
  \includegraphics[angle=0,width=0.48\linewidth,clip]{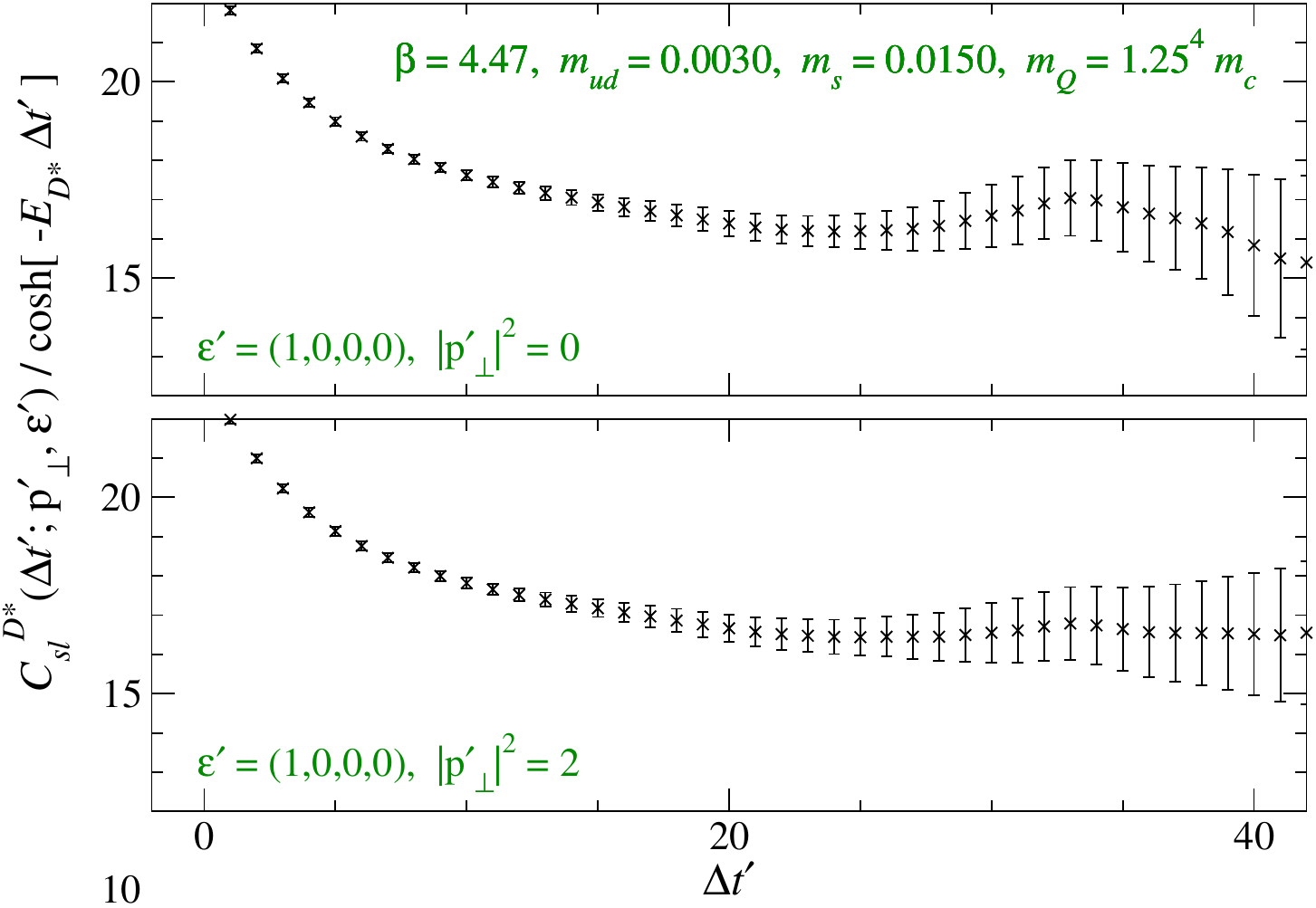}
  \hspace{3mm}
  \includegraphics[angle=0,width=0.48\linewidth,clip]{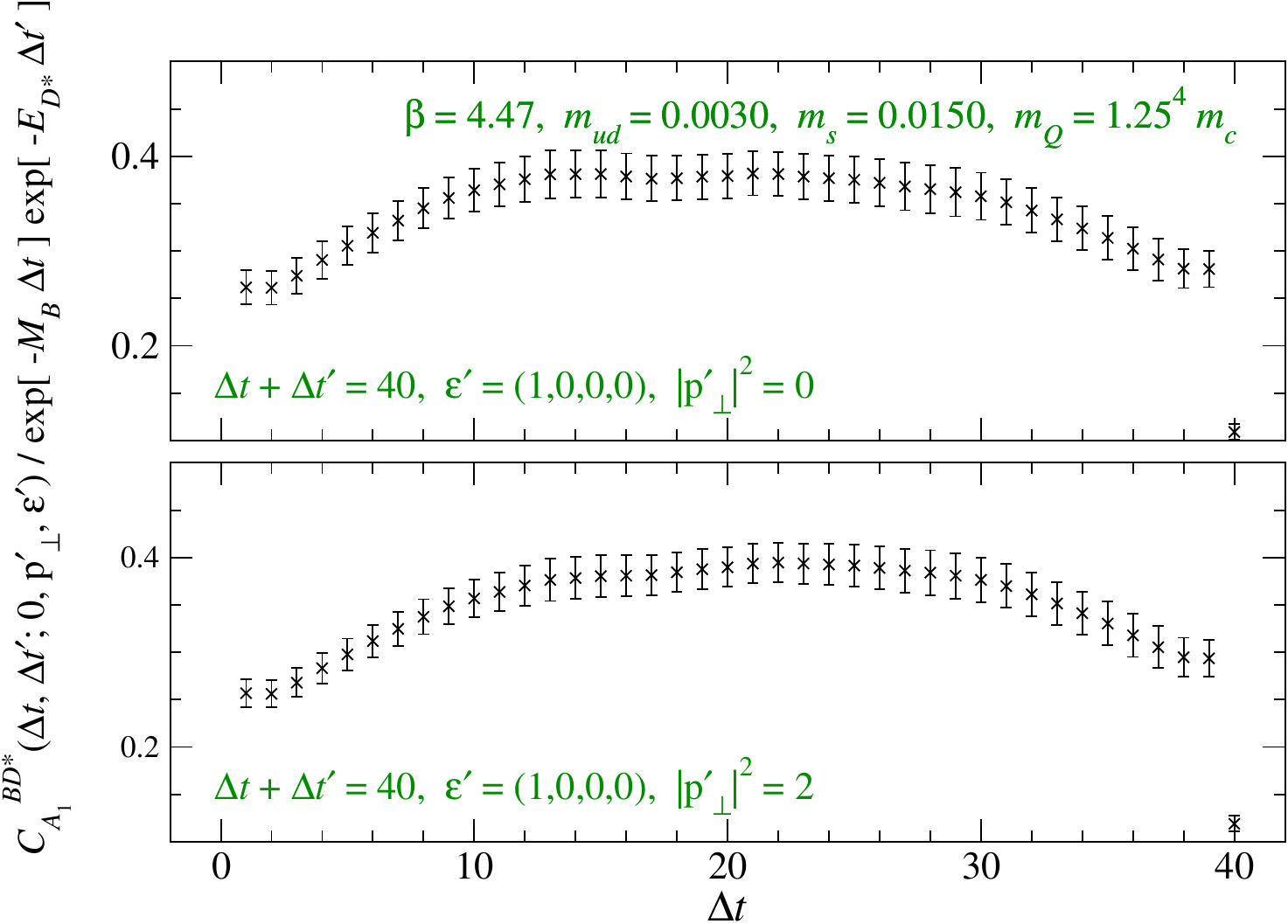}
  \vspace{-3mm}
  \caption{
    Effective plot of two- and three-point functions,
    $C^{D^*}_{sl}(\dtp, \bfppperp, \epsp)$ (left panels) and
    $C_{A_1}^{B D^*}(\dt, \dtp; \bfz, \bfppperp, \epsp)$ (right panels).
    Data divided by the exponential damping factor(s) for the ground state(s)
    are plotted as a function of temporal separations $\dtp$ and $\dt$, respectively.
    These are used to calculate $R_1^\prime(\bfppperp; \dt, \dtp)$
    for $\dt+\dtp\!=\!40$ shown in Fig.\protect\ref{fig:ff:drat24}.
    The top and bottom panels show data with the $D^*$ momentum
    $|\bfppperp|^2\!=\!0$ and 2, respectively.
  }
  \label{fig:ff:bump}
\end{center}
\end{figure}

\begin{figure}[t]
\begin{center}
  \includegraphics[angle=0,width=0.48\linewidth,clip]{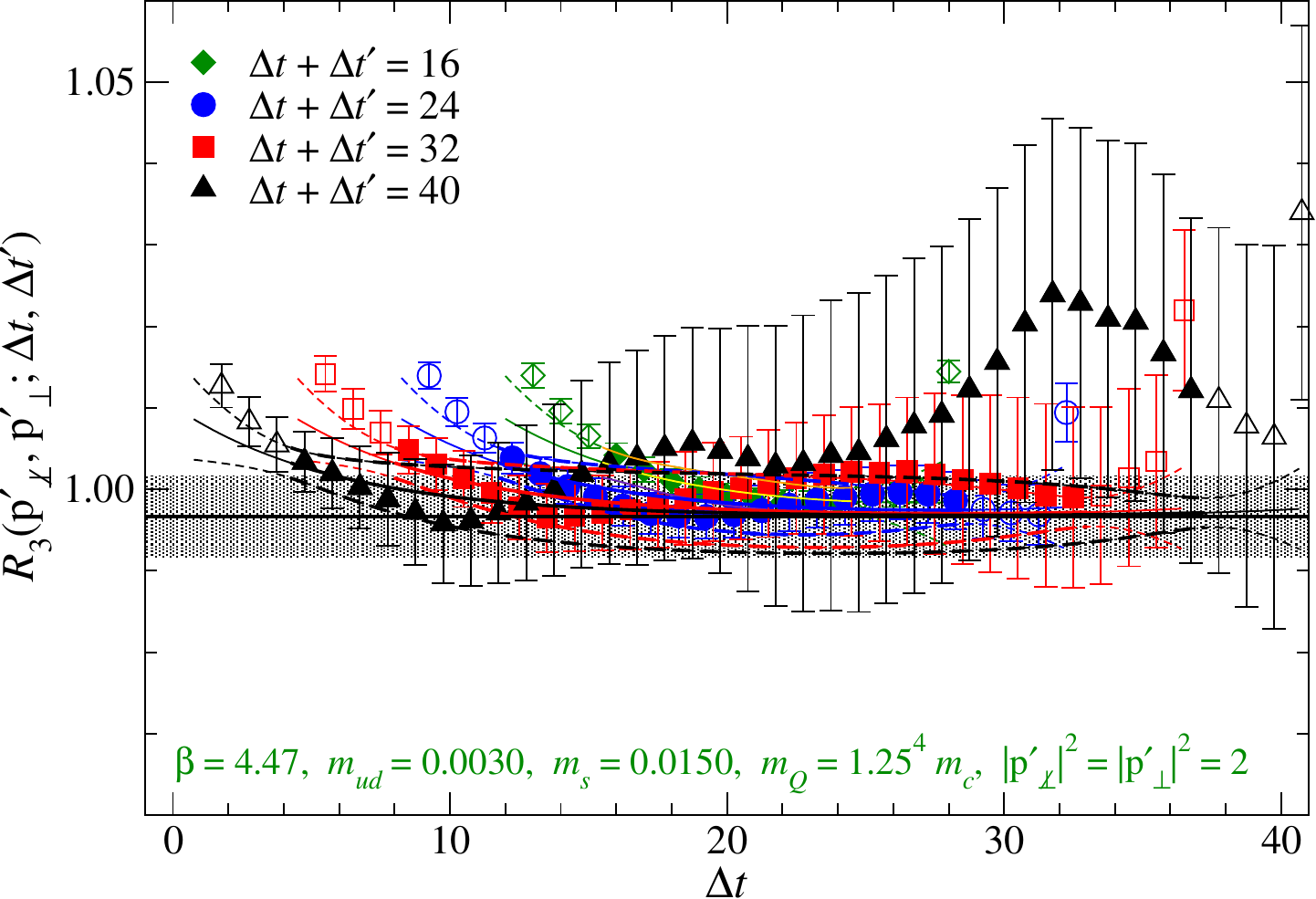}
  \hspace{3mm}
  \includegraphics[angle=0,width=0.48\linewidth,clip]{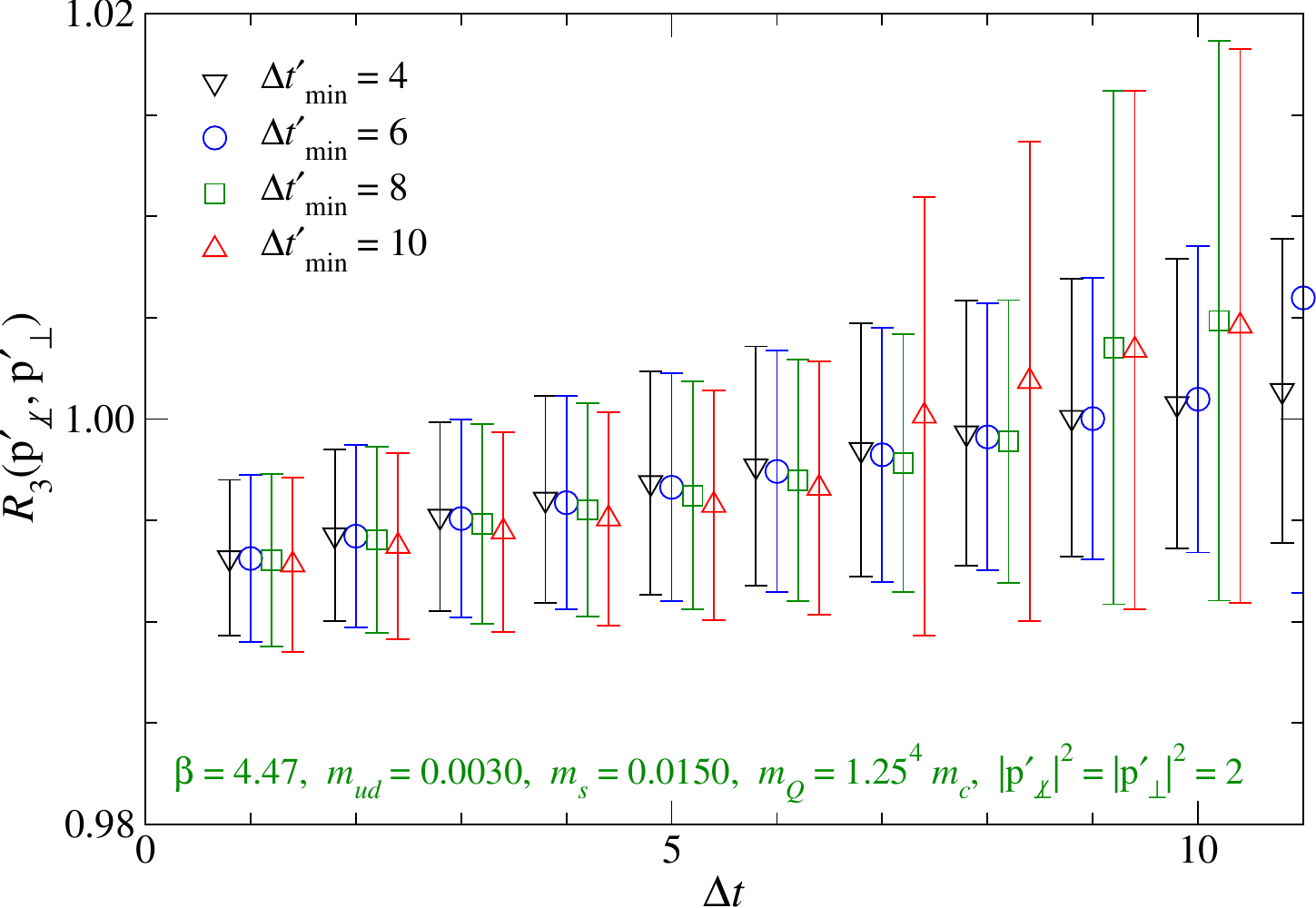}
  \vspace{-3mm}
  \caption{
    Same as Fig.~\protect\ref{fig:ff:drat24}
    but for ratio $R_3(\bfpnperp, \bfpperp; \dt, \dtp)$ (\ref{eqn:ff:ratio:r3})
    to study $h_{A_3}(w)$.
  }
  \label{fig:ff:drat31}
\end{center}
\end{figure}

\begin{figure}[t]
\begin{center}
  \includegraphics[angle=0,width=0.48\linewidth,clip]{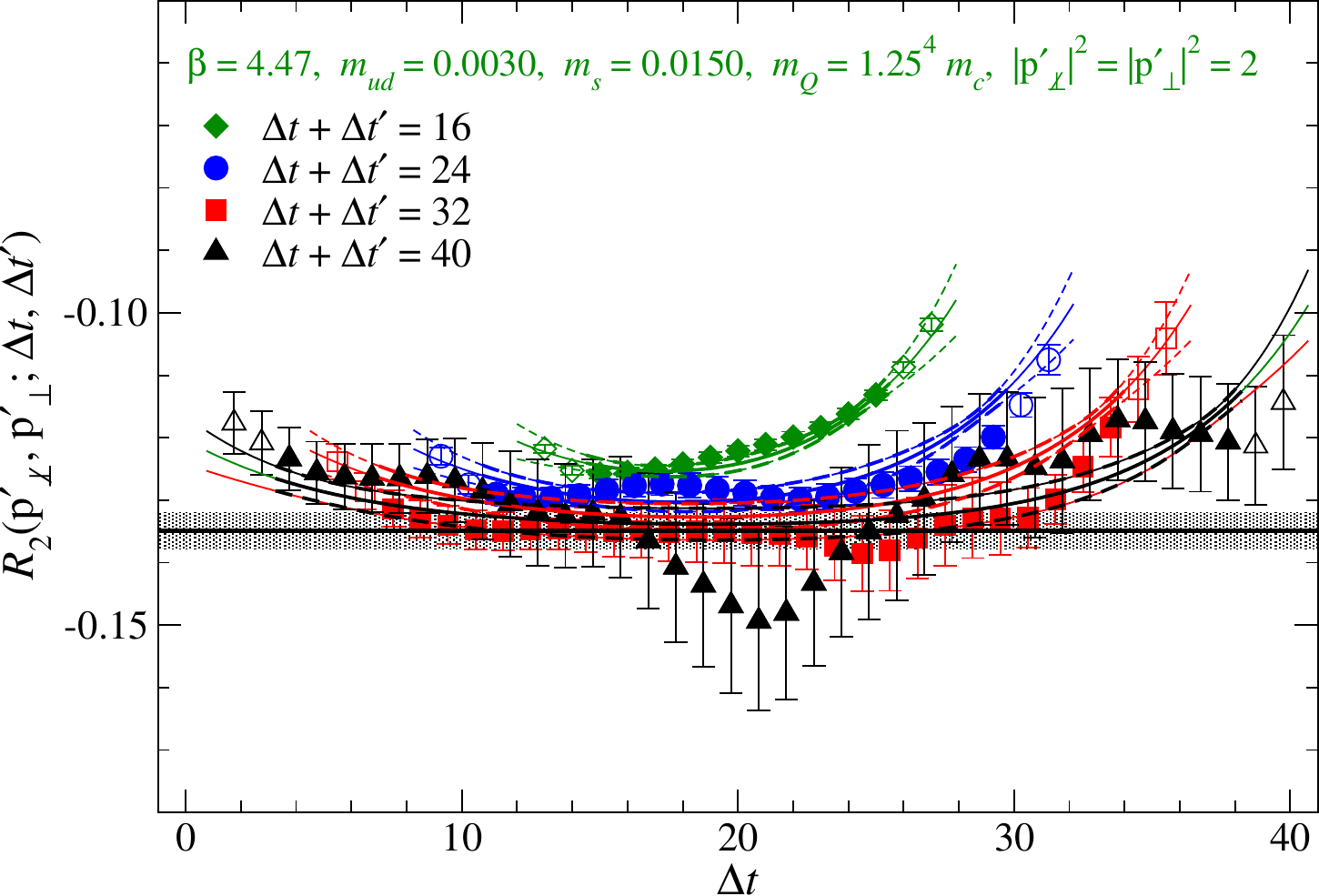}
  \hspace{3mm}
  \includegraphics[angle=0,width=0.48\linewidth,clip]{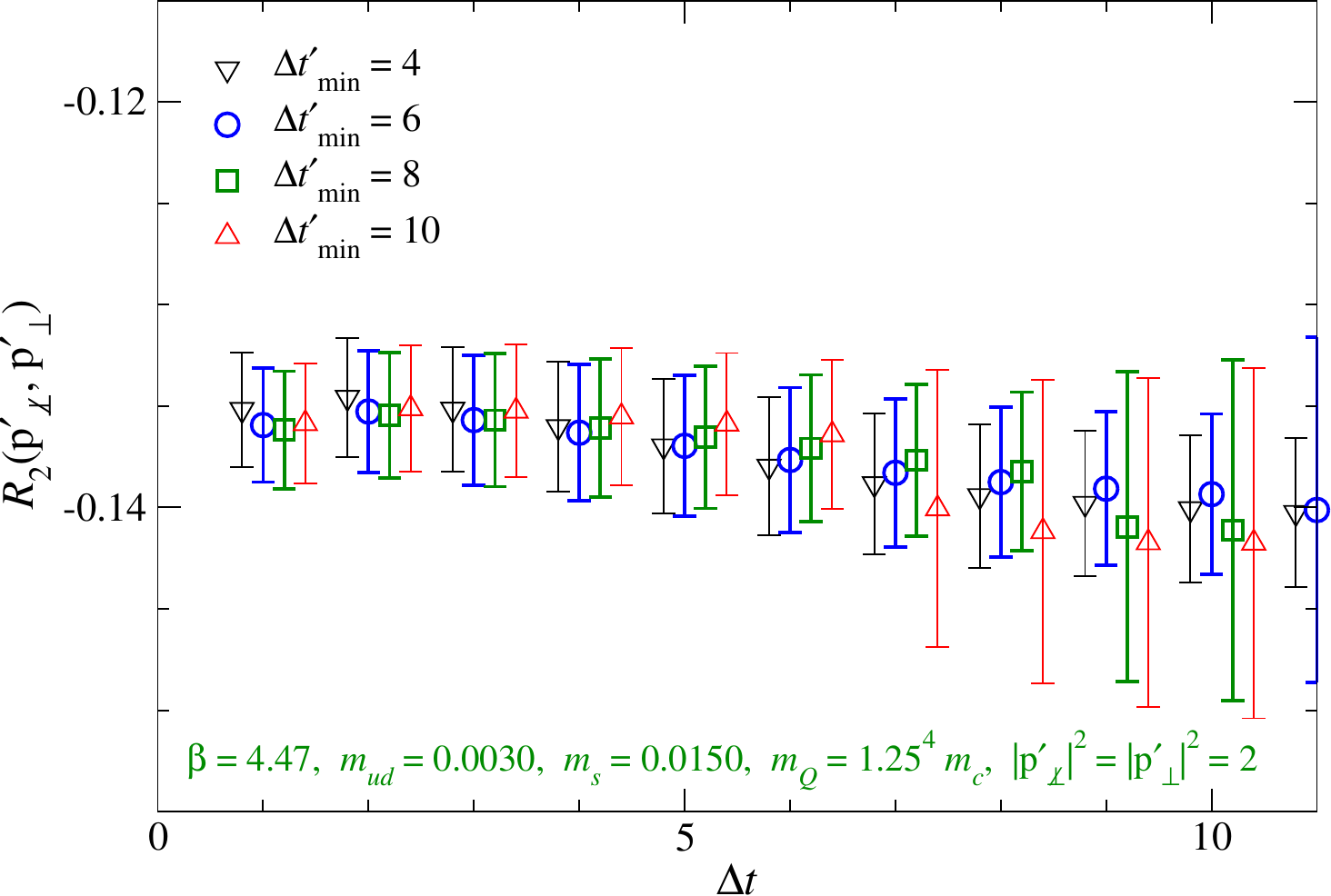}
  \vspace{-3mm}
  \caption{
    Same as Fig.~\protect\ref{fig:ff:drat24}
    but for ratio $R_2(\bfpnperp, \bfpperp; \dt, \dtp)$ (\ref{eqn:ff:ratio:r3})
    to study $h_{A_2}(w)$.
  }
  \label{fig:ff:drat34}
\end{center}
\end{figure}

There is a 2\,$\sigma$ bump of $R_1^\prime$ for $\dt\!+\!\dtp\!=\!40$
(black triangles) around $\dt\!=\!28$ ($\dtp\!=\!12$)
in Fig.~\ref{fig:ff:drat24}.
Since the global topological charge does not fluctuate very much
at the corresponding cutoff,
one may expect bump(s) in the exponential decay of relevant correlators
due to instantons frozen at the temporal separation $\dt^{(\prime)}$
for a certain period of our simulation time.
However, Fig.~\ref{fig:ff:bump} shows that
there is no statistically significant bump of the relevant correlators
at $\dt\!\sim\!28$ and $\dtp\!\sim\!12$.
As mentioned above, the local topological fluctuation is active even
when the global topological charge is fixed.
We attribute 1\,--\,2\,$\sigma$ bumps of correlator ratios
in Figs.~\ref{fig:ff:drat14},
\ref{fig:ff:drat24}\,--\,\ref{fig:ff:ratv},
\ref{fig:ff:drat31}\,--\,\ref{fig:ff:drat34}
to large statistical fluctuations of correlators
at the largest temporal separation $\dt\!+\!\dtp\!=\!40$,
and accidental anti correlation of the statistical fluctuations.
Since these data have larger statistical error than those at smaller 
$\dt\!+\!\dtp$, these bumps do not change results of the fits
(\ref{eqn:ff:ratio:r1:fit}) and (\ref{eqn:ff:ratio:relse:fit})
to extract the ground state contribution.


The axial vector form factors $h_{A_2}$ and $h_{A_3}$ can be extracted from
the axial vector matrix element (\ref{eqn:intro:ff:A:hqet})
with the $D^*$ spatial momentum $\bfppnperp$
not perpendicular to the spatial polarization vector.
The matrix element of $A_1$, for instance, has non-zero sensitivity
to $h_{A_1}$ and $h_{A_3}$.
We use the following ratio to extract $h_{A_3}$
\bea
   R_3(\bfppnperp, \bfppperp; \dt, \dtp)
   & = &
   r_Z 
   \frac{ C_{A_1}^{BD^*}(\dt,\dtp;\bfz,\bfppnperp,\epspp) }
        { C_{A_1}^{BD^*}(\dt,\dtp;\bfz,\bfppperp, \epsp)  }
   \nn \\
   & \xrightarrow[\dt,\dtp\to\infty]{} &
   \epspp^*_1
   -
   \frac{ \epspp^*_4  \vp_{\not\perp,1} }{w+1} \frac{h_{A_3}(w)}{h_{A_1}(w)},
   \label{eqn:ff:ratio:r3}
\eea
where ${\boldsymbol \epsilon}^\prime\!=\!(1,0,0)$,
${\boldsymbol \epsilon}^{\prime *} \bfppperp\!=0$,
${\boldsymbol \epsilon}^{\prime\prime *} \bfppnperp\!\ne\!0$,
and $|\bfppnperp|\!=\!|\bfppperp|$.
The factor
$r_Z\!=\!Z_{D^*}(\bfppperp,\epsp)/Z_{D^*}(\bfppnperp,\epspp)$
appears, since the overlap factor of $D^*$ depends on
whether the polarization vector is perpendicular to the momentum,
even if $|\bfppnperp|\!=\!|\bfppperp|$.
While $r_Z$ can be estimated from individual fit to the relevant two-point functions,
we employ a ratio
$r_Z\!=\!\sqrt{C^{D^*}(\dt_{\rm ref},\bfppperp,\epsp)/C^{D^*}(\dt_{\rm ref},\bfppnperp,\epspp)}$,
which shows better stability against the temporal separation.
The reference temporal separation is chosen as $\dt_{\rm ref}\!\sim\!T/4$
by inspecting the statistical accuracy and the ground state saturation of $r_Z$.
%
%
A similar ratio but with $A_4$ in the numerator is also sensitive to $h_{A_2}$
\bea
   R_2(\bfppnperp, \bfppperp; \dt, \dtp)
   & = &
   r_Z
   \frac{ C_{A_4}^{BD^*}(\dt,\dtp;\bfz,\bfppnperp,\epspp) }
        { C_{A_1}^{BD^*}(\dt,\dtp;\bfz,\bfppperp,\epsp)  }
   \nn \\
   & \xrightarrow[\dt,\dtp\to\infty]{} &
   \epspp^*_4
   \left[
     1 
   - \frac{1}{w+1}
     \left\{
        \frac{h_{A_2}(w)}{h_{A_1}(w)} + v_{\not\perp,4} \frac{h_{A_3}(w)}{h_{A_1}(w)}
     \right\}
   \right].
   \label{eqn:ff:ratio:r2}
\eea


We plot our results for $R_3$ and $R_2$ on our finest lattice
in Figs.~\ref{fig:ff:drat31} and \ref{fig:ff:drat34}, respectively.
Similar to other ratios,
the excited state contamination is not large.
The fit similar to (\ref{eqn:ff:ratio:relse:fit})
yields the ground state contributions,
$R_3(\bfppnperp,\bfppperp)$ and $R_2(\bfppnperp,\bfppperp)$,
which are stable against the choice of the fit range:
we do not observe $\dt^{(\prime)}_{\rm min}$ dependence
beyond 1\,$\sigma$ level
when $\dtmin+\dtpmin\!\lesssim\!16$, that is our smallest value of 
the source-sink separation $\dt+\dtp$.
The error rapidly increases at larger values of $\dt^{(\prime)}_{\rm min}$,
because less data are available for the fit.
From $R_2$, $R_3$ and $h_{A_1}$ extracted above, 
we can calculate $h_{A_2}$ and $h_{A_3}$.


For one of the $D^*$ momenta $\bfpp\!=\!(1,1,1)$,
it is not straightforward to use the above-mentioned correlator ratios,
which use
$C_{A_1}^{BD^*}(\dt,\dtp;\bfz,\bfpp,\epsp)$
with ${\boldsymbol \epsilon}^{\prime *}{\bfpp}\!=\!0$
as a correlator sensitive only to $h_{A_1}(w)$.
In this case, we use a linear combination 
\bea
   C_{A_1}^{BD^*}(\dt,\dtp;\bfz,\bfpp,\epsp)
 - C_{A_1}^{BD^*}(\dt,\dtp;\bfz,\bfpp,\epspp)
   & \propto &
   (w+1) h_{A_1}(w)
\eea
which only involves $h_{A_1}$
with ${\boldsymbol \epsilon}^\prime\!=\!(1,0,0)$
and ${\boldsymbol \epsilon}^{\prime\prime}\!=\!(0,1,0)$.


The double ratio $R_1$ enables us to calculate $h_{A_1}(1)$
with 1\,\% statistical error or better,
as it only involves correlators with zero momentum
and the statistical fluctuation partially cancels
between the numerator and denominator.
At non-zero recoil,
the numerator of $R_1^\prime$ ($R_V$) is exclusively sensitive to $h_{A_1}$ ($h_V$),
and the statistical error of $h_{A_1}$ and $h_V$ is typically a few \%.
This is not the case for $h_{A_2}$ and $h_{A_3}$
with our setup where the $B$ meson is at rest.
The numerator of $R_3$ involves $h_{A_1}$ and $h_{A_3}$,
and the typical precision for $h_{A_3}$ is 10\,--\,20\,\%.
The statistical error increases to $\gtrsim\!40$\,\% for $h_{A_2}$,
since the central value is suppressed by heavy quark symmetry
and $R_2$ is a linear combination of all axial form factors.


We note that, thanks to chiral symmetry,
finite renormalization factors for the weak vector and axial currents
cancel in the above correlator ratios (\ref{eqn:ff:ratio:r1}),
(\ref{eqn:ff:ratio:r1p}), (\ref{eqn:ff:ratio:rv}), (\ref{eqn:ff:ratio:r3})
and (\ref{eqn:ff:ratio:r2}).
While discretization effects to the wave function renormalization
are not small~\cite{fB:Nf3:Moebius:JLQCD:lat16},
these also cancel in the ratios.

\section{Extrapolation to continuum limit and physical quark masses}
\label{sec:cont+chiral_fit}

\subsection{Choice of fitting form}
\label{subsec:cont+chiral_fit:form}


\begin{figure}[t]
\begin{center}
  \includegraphics[angle=0,width=0.48\linewidth,clip]{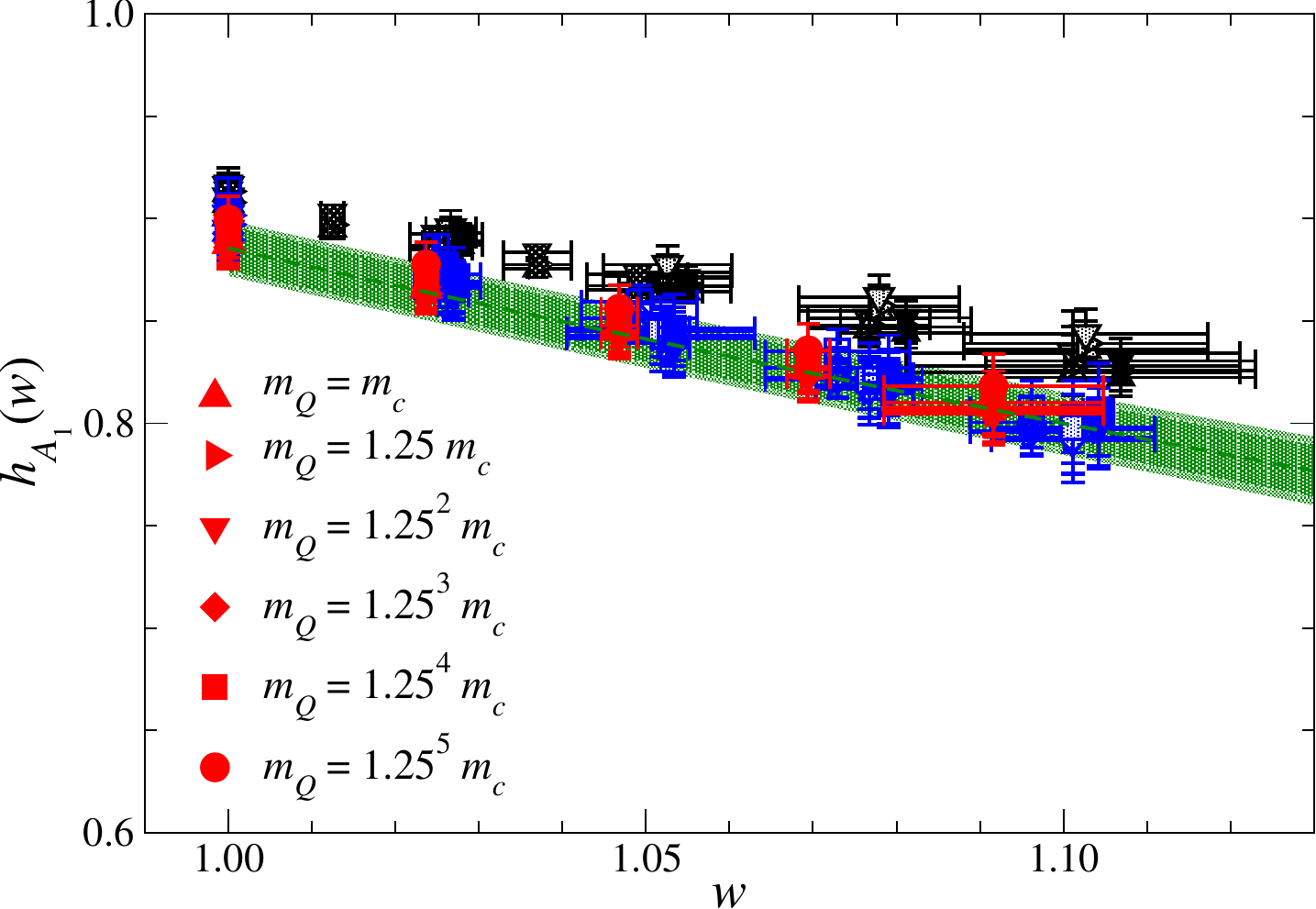}
  \hspace{3mm}
  \includegraphics[angle=0,width=0.48\linewidth,clip]{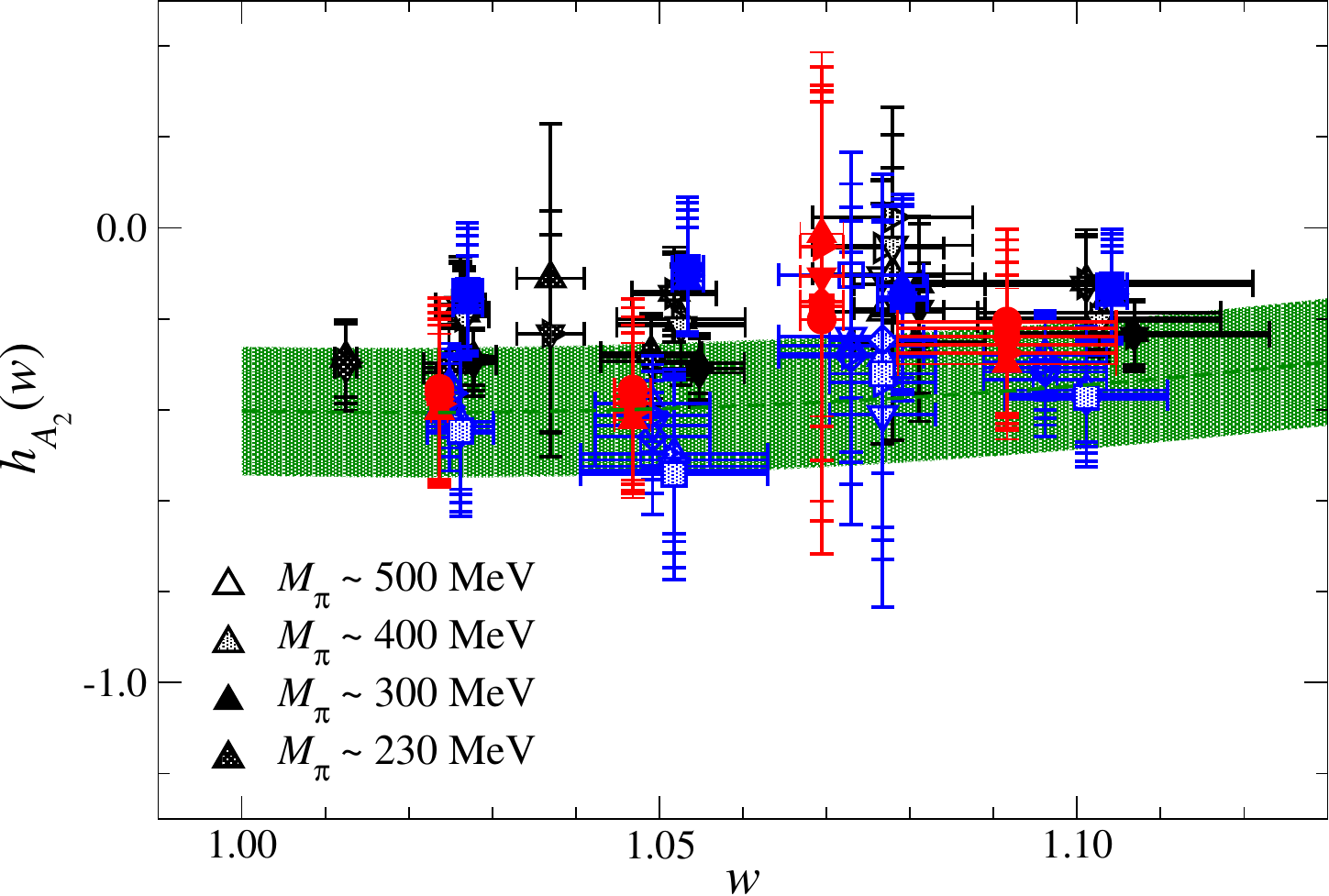}

  \includegraphics[angle=0,width=0.48\linewidth,clip]{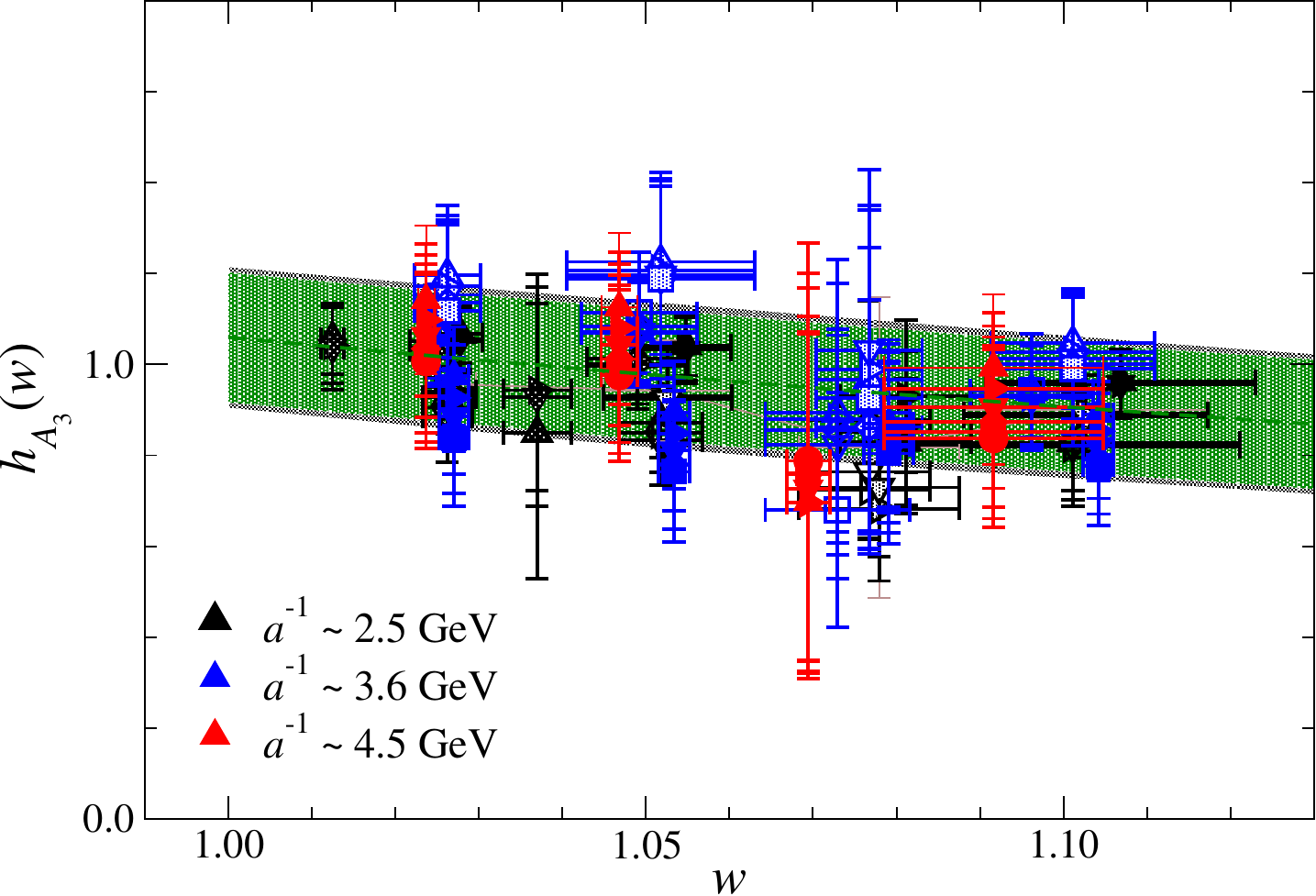}
  \hspace{3mm}
  \includegraphics[angle=0,width=0.48\linewidth,clip]{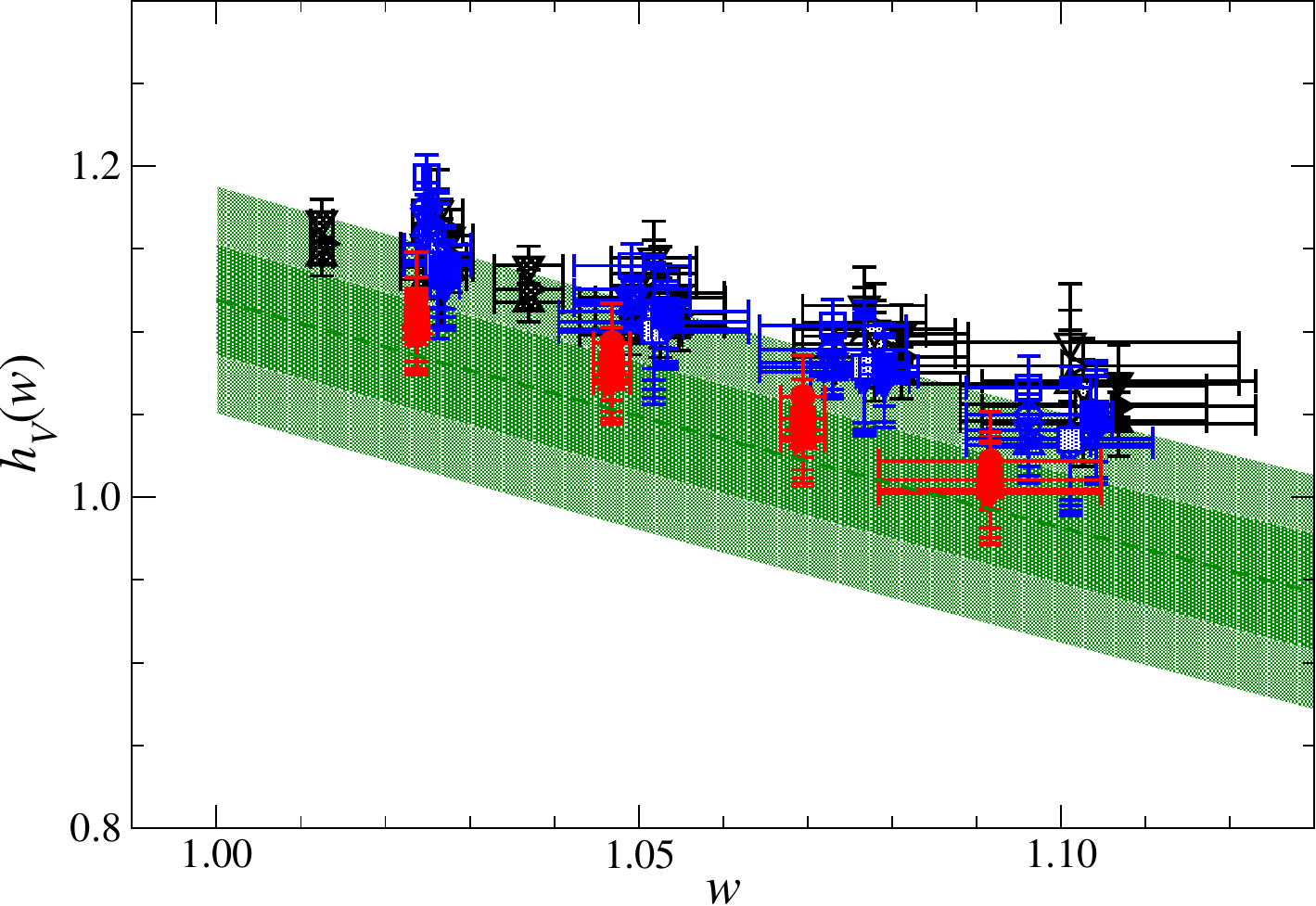}

  \vspace{-3mm}
  \caption{
    Form factors as a function of recoil parameter $w$.
    The top-left, top-right, bottom-left and bottom-right panels show
    results for $h_{A_1}$, $h_{A_2}$, $h_{A_3}$ and $h_V$, respectively.
    The black, blue and red symbols show data at $a^{-1}\!\sim\!2.5$, 3.6 and 4.5~GeV,
    whereas the open, pale-shaded, filled and dark-shaded ones are at
    $M_\pi\!\sim\!500$, 400, 300 and 230~MeV, respectively.
    Symbols with different shapes are obtained with different bottom quark masses.
    (See the legend in the plots.)
    The green bands show the form factors extrapolated to the continuum limit
    and physical quark masses. The dark and pale shaded bands represent
    their statistical and total uncertainties, respectively.
  }
  \label{fig:chiralfit:ff}
\end{center}
\end{figure}

Our results for the form factors obtained at various simulation parameters
are plotted
as a function of the recoil parameter $w$ in Fig.~\ref{fig:chiralfit:ff}.
We find that the results with different heavy and light quark masses are
all close to each other.
It also turns out that the discretization effects are not large.

We extrapolate these results to the continuum limit and physical quark masses
based on next-to-leading order heavy meson chiral perturbation theory
(NLO HMChPT)~\cite{HMChPT:RW,HMChPT:S}.
The extrapolation form is generically written as 
\bea
   h_X/\eta_X
   & = &
   c
 + \frac{\gVPpi^2}{2 \Lambda_\chi^2} 
   \bar{F}_{\rm log}\!\left(\xipi,\Delta_D,\Lambda_\chi\right)
 + c_\pi \xipi + c_{\eta_s} \xietas
 + c_Q \epsQ
 \nn \\
   & &
 + c_a \xia + c_{am_Q} \xiamQ
 + c_w (w-1) + d_w (w-1)^2
 \hspace{5mm}
 (X\!=\!A_1,A_2,A_3 {\rm \ and\ } V),
 \hspace{5mm}
 \label{eqn:chiral_fit:form}
\eea
which comprises the constant term, chiral logarithm at $w\!=\!1$
and leading corrections in small expansion parameters
\bea
   &&
   \xipi   = \frac{M_\pi^2}{\Lambda_\chi^2},
   \hspace{2mm}
   \xietas = \frac{M_{\eta_s}^2}{\Lambda_\chi^2},
   \hspace{2mm}
   \epsQ   = \frac{\bar{\Lambda}}{2m_Q},
   \hspace{2mm}
   \xia    = (\Lambda a)^2,
   \hspace{2mm}
   \xiamQ  = (am_Q)^2,
 \label{eqn:chiral_fit:exparam}
\eea
with $M_{\eta_s}^2\!=\!2M_K^2-M_\pi^2$,
$\Lambda_\chi\!=\!4\pi f_\pi$,
and nominal values of $\bar{\Lambda}\!=\!\Lambda\!=\!500$~MeV.


We denote the one-loop integral function for the chiral logarithm by
$\bar{F}_{\rm log}\!\left(\xipi,\Delta_D,\Lambda_\chi\right)$,
where $\Delta_D\!=\!M_{D^*}-M_D$ is the $D^*$-$D$ mass splitting.
While the loop function can be approximated as 
$\bar{F}_{\rm log}\!\left(\xipi,\Delta_D,\Lambda_\chi\right)
\!=\!\Delta_D^2 \ln[\xipi]+O(\Delta_D^3)$,
we use its explicit form in Ref.~\cite{HMChPT:S}.
For the $D^*D\pi$ coupling,
we take a value of $\gVPpi\!=\!0.53(8)$,
which was employed in the Fermilab/MILC paper~\cite{B2Dstar:Nf3:hA11:Fermilab/MILC2}
and covers previous lattice estimates~\cite{gDDstarpi:DLM12,gDDstarpi:Nf2:BS12,gDDstarpi:Nf3:CEOOT12,gDDstarpi:Nf2:ALPHA14,gDDstarpi:Nf3:RBCUKQCD15}.
This choice does not lead to a large systematic uncertainty,
because the chiral logarithm is suppressed
by the small mass splitting squared $\Delta_D^2$.


\begin{figure}[t]
\begin{center}
  \includegraphics[angle=0,width=0.48\linewidth,clip]{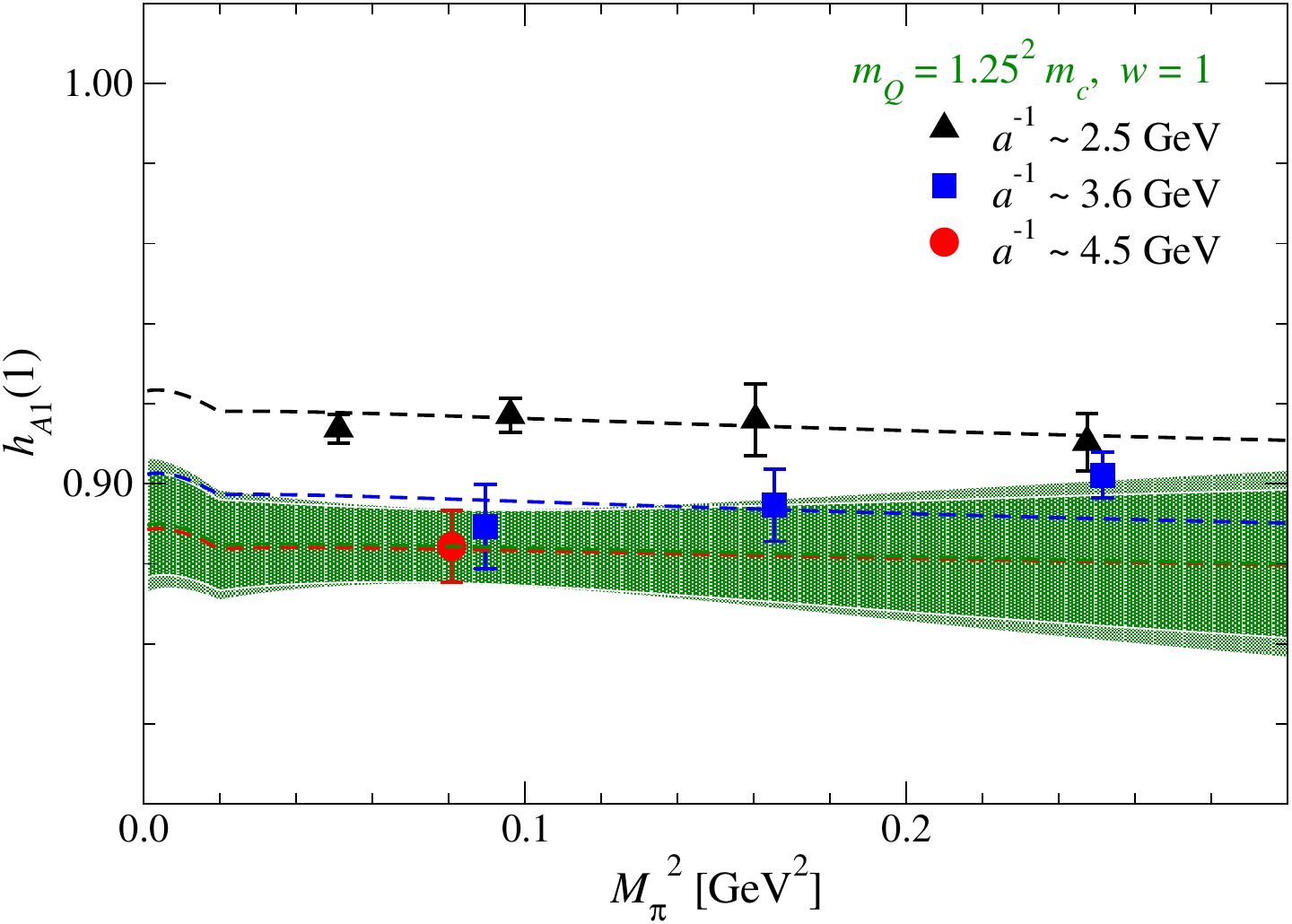}
  \hspace{3mm}
  \includegraphics[angle=0,width=0.48\linewidth,clip]{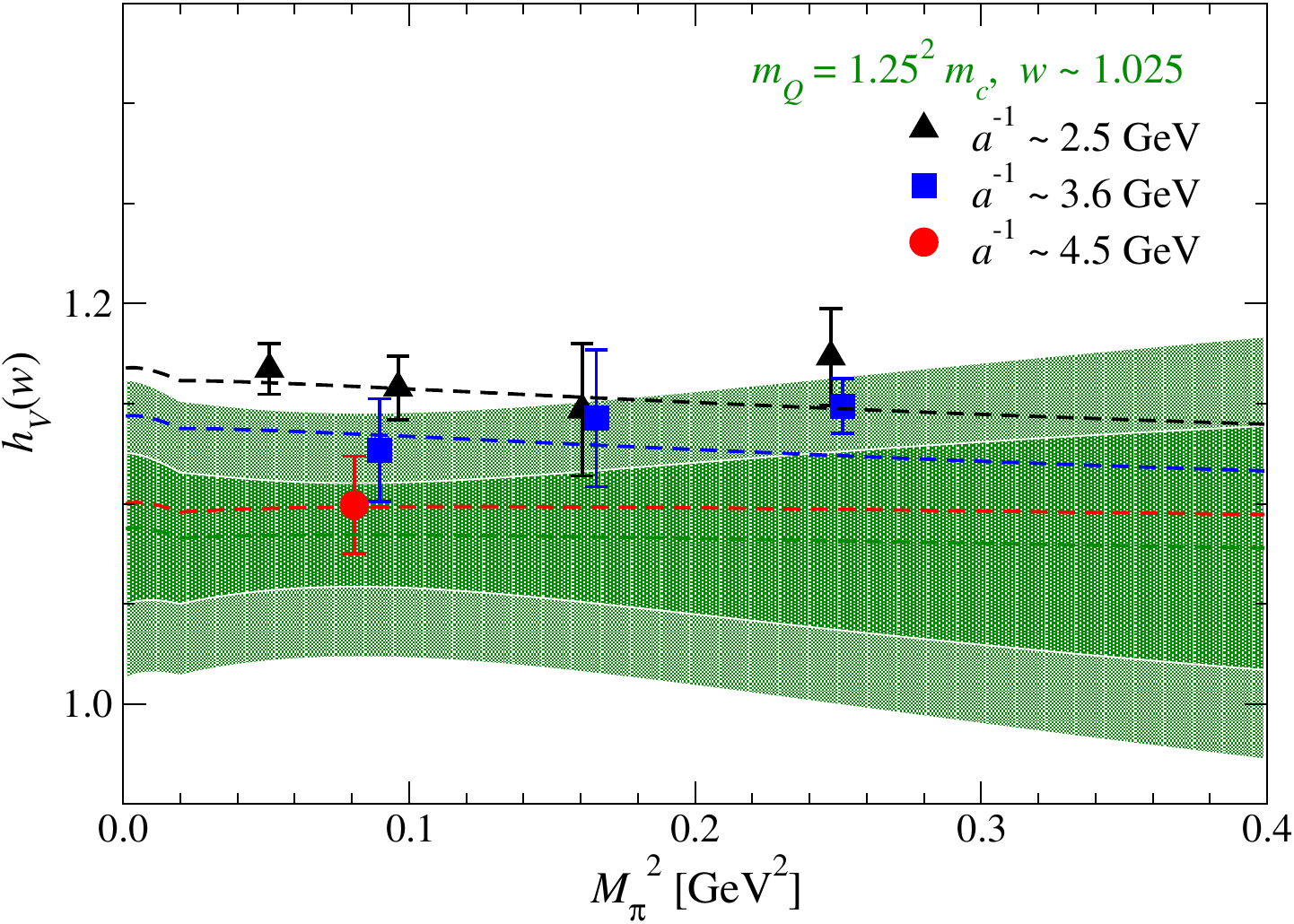}

  \vspace{-3mm}
  \caption{
    Axial vector form factor $h_{A_1}$ (left panel) and 
    vector form factor $h_v$ (right panel) as a function of $M_\pi^2$.
    Both panels show data at $m_Q\!=\!1.25^2 m_c$ and the smallest value of $w$:
    namely, $w\!=\!1.0$ for $h_{A_1}$
    and $\sim\!$~1.205 for $h_V$, respectively.
    For the latter, $w$ is not exactly equal to 1.025
    due to the slight difference in $L$ among simulated lattice spacings
    as well as discretization effects to $w\!=\!E_D^*/M_D^*$ itself.
    The black, blue and red symbols and lines show
    data and fit curve at $a^{-1}\!\sim\!2.5$, 3.6 and 4.5~GeV, respectively.
    The green dashed line shows the form factors extrapolated
    to the continuum limit and physical strange and bottom quark masses.
    The dark and pale shaded green bands represent the statistical and total error,
    respectively.
  }
  \label{fig:chiralfit:vs_Mpi2}
\end{center}
\end{figure}

In Fig.~\ref{fig:chiralfit:vs_Mpi2},
we plot $h_{A_1}$ and $h_V$ at our smallest value of $w$
as a function of $M_\pi^2$.
The $M_\pi^2$ dependence is rather mild,
possibly because there is no valence pion in this decay.
We note that there is a non-trivial concave structure at small $M_\pi^2$
due to the opening of the $D^*\!\to\!D\pi$ decay.
The effect estimated in NLO HMChPT is, however,
not large compared to our statistical accuracy.


For our chiral extrapolation,
we employ the so-called $\xi$ expansion in $\xipi$ and $\xietas$:
namely, we use the measured meson masses squared, $M_\pi^2$ and $M_{\eta_s}^2$,
instead of quark masses $m_{ud}$ and $m_s$, respectively,
as well as quark-mass dependent $f_\pi$ for $\Lambda_\chi$
instead of the decay constant $f$ in the chiral limit.
This is motivated by our observation~\cite{vsChPT:Nf2:JLQCD}
that the $\xi$-expansion shows better convergence of the chiral expansion
for light meson observables compared to the $x$-expansion
with $x\!=\!2Bm_q/(4\pi f)^2$,
where $B$ gives the lowest order relation $M_\pi^2\!=\!2Bm_{ud}$.


\begin{figure}[t]
\begin{center}
  \includegraphics[angle=0,width=0.48\linewidth,clip]{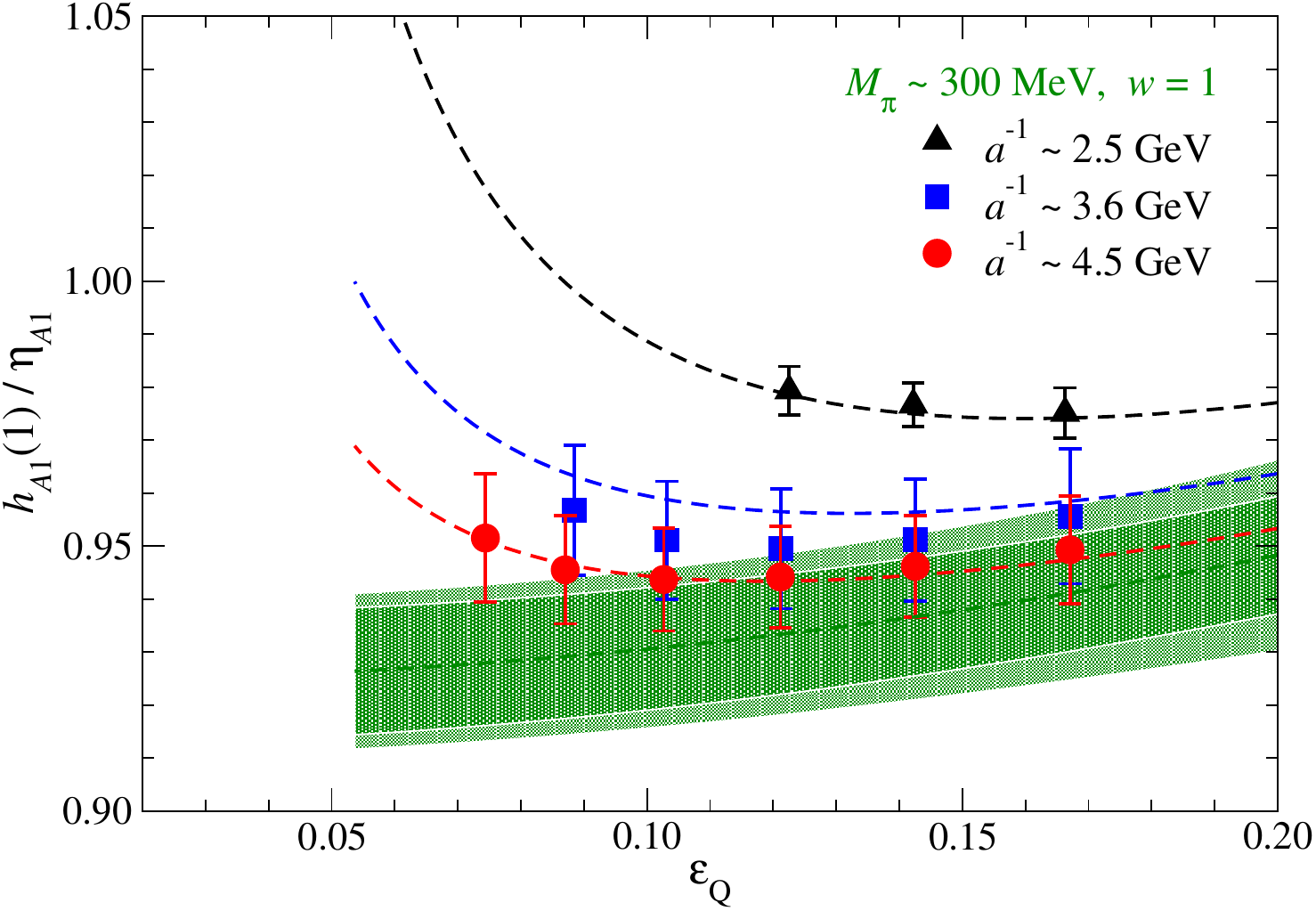}
  \hspace{3mm}
  \includegraphics[angle=0,width=0.48\linewidth,clip]{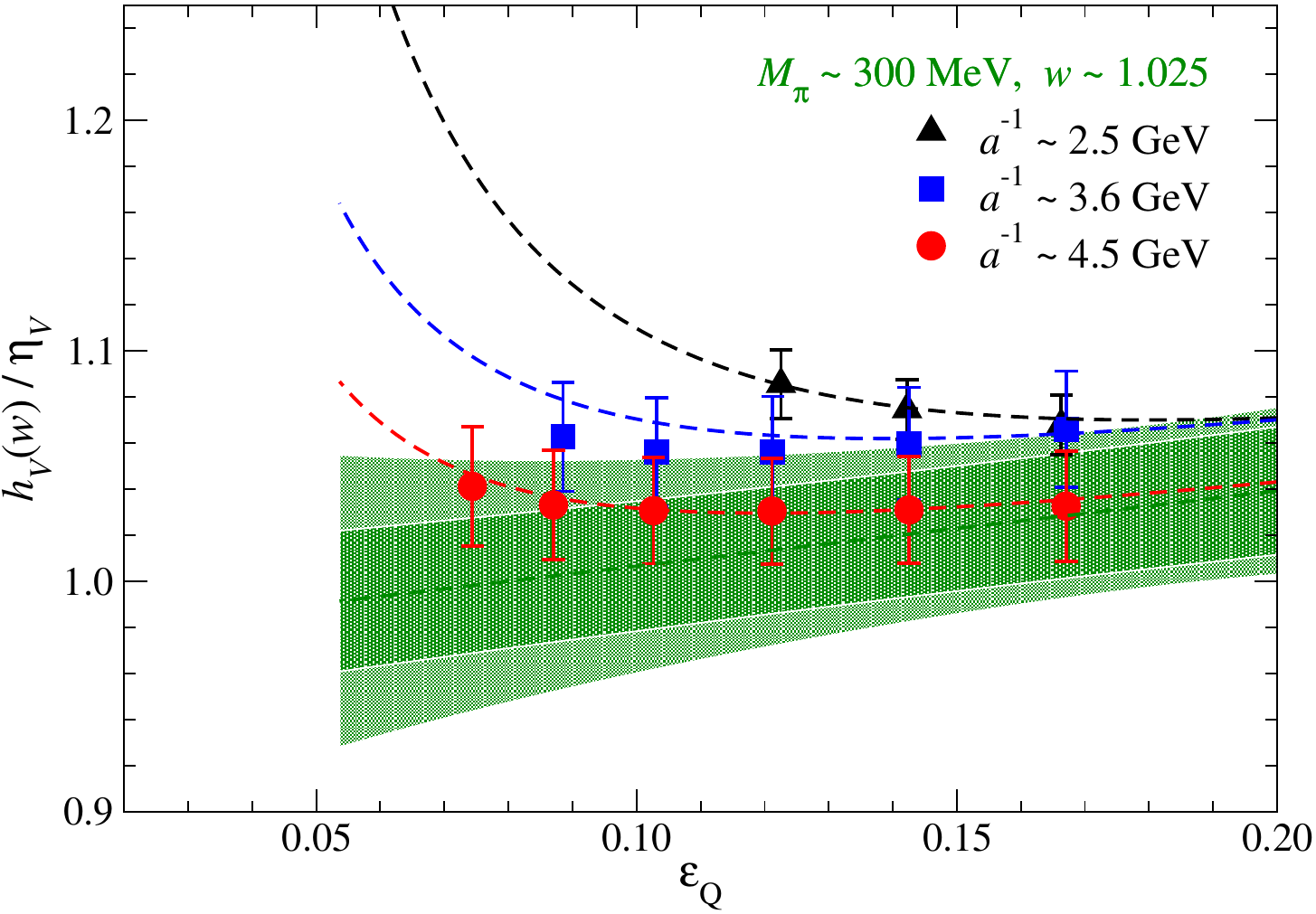}

  \vspace{-3mm}
  \caption{
    Same as Fig.~\ref{fig:chiralfit:vs_Mpi2}
    but as a function of HQET expansion parameter $\epsQ$.
    To cancel non-trivial $m_Q$ dependence
    due to the matching factor $\eta_{\{A_1,V\}}$,
    we plot a ratio to this factor,
    which is fitted to the HMChPT-based polynomial form (\ref{eqn:chiral_fit:form})
    in $\epsQ$ and $\xiamQ$.
  }
  \label{fig:chiralfit:vs_epsb}
\end{center}
\end{figure}

The extrapolation form explicitly includes the one-loop radiative correction $\eta_X$
for the matching of the heavy-heavy currents between QCD and HQET~\cite{matching:HQET2QCD:FGGW,matching:HQET2QCD:FG,matching:HQET2QCD:N}.
Remaining heavy quark mass dependence of the form factors is then parametrized
by polynomials of $\epsQ\!\propto\!1/m_Q$ and $\xiamQ\!\propto\!m_Q$.
The dependence is mild as shown in Fig.~\ref{fig:chiralfit:vs_epsb}.
Our results are well described by the linear terms $c_Q \epsQ$
and $c_{am_Q} \xiamQ$ in Eq.~(\ref{eqn:chiral_fit:form}).
We note that, for $h_{A_1}$,
the term $c_Q \epsQ$ is modified as $c_Q (w-1) \epsQ$
to be consistent with Luke's theorem~\cite{Luke},
and we add the quadratic term $d_Q \epsQ^2$ to take account of
possible $m_Q$ dependence at $w\!=\!1$.


The discretization effects in the form factors also turn out to be mild
with our choice of parameters, $a^{-1}\!\gtrsim\!2.5$~GeV and $m_Q\!\leq 0.7 a^{-1}$.
These are described by an expansion,
{\it i.e.} linear
in the $m_Q$ independent and dependent parameters $\xia$ and $\xiamQ$,
respectively.
Figure~\ref{fig:chiralfit:vs_epsb} shows that
the $m_Q$ dependent discretization error is significant in our results.
However, the net $m_Q$ dependence is described reasonably well
with our fitting form (\ref{eqn:chiral_fit:form}).
We note that
the $m_Q$ dependence for $h_{A_2}$ and $h_{A_3}$ is insignificant
due to their large relative uncertainty.
As we will mention later,
fit results do not change significantly if we exclude data at
$0.5\!\leq\!am_Q\!\leq\!0.7$ in our continuum and chiral extrapolation.


In this study, we first extrapolate the form factors
to the continuum limit and physical quark masses,
and then parametrize the $w$ dependence
by using the model independent BGL parametrization,
because it may not be simply combined
with the chiral and heavy quark expansions.
Concerning the $w$ dependence, therefore,
the fit form (\ref{eqn:chiral_fit:form}) with an expansion in $w-1$
is used to interpolate our results.
Figure~\ref{fig:chiralfit:ff} shows that
our data do not show strong curvature 
in the simulation region of $1\!\leq\!w\!\lesssim\!1.1$,
and can be well described with only up to the quadratic term.


\begin{table}[t]
\centering
\small
\caption{
  Fit parameters from continuum and chiral extrapolation of form factors
  (\ref{eqn:chiral_fit:form}).
}
\vspace{-3mm}
\label{tbl:chiralfit:fit}.
\begin{tabular}{l|llllllllll}
   \hline 
   $h_X$ &
   $\chi^2$/d.o.f &
   $c$    & $c_\pi$   & $c_{\eta_s}$ & $c_Q$     & $d_Q$    &
   $c_a$  & $c_{am_Q}$ &
   $c_w$  & $d_w$  
   \\ \hline
   $h_{A_1}$    &
   0.14(11)    &
   0.898(54)   & 0.68(19)   & 0.08(16)   & -0.18(32)  & 0.59(17)  &
   0.37(26)    & 0.0376(89) & -0.953(70) & 0.67(22)
   \\
   $h_{A_2}$    &
   0.18(13)    &
   -0.5(1.0)   & 0.9(3.6)   & -0.0(2.6)  & 0.13(49)   & --        &
   3.3(3.0)    & -0.022(67) & -0.41(86)  & 10.3(7.1)
   \\
   $h_{A_3}$    &
   0.18(13)    &
   0.91(90)    & -0.4(3.3)  & 0.3(2.4)   & 1.25(47)   & --        &
   -3.4(2.7)   & 0.117(62)  & -1.72(75)  & 1.9(6.1)   
   \\
   $h_V$       &
   0.15(14)    &
   0.87(13)    & 1.47(45)   & 0.39(37)   & 0.33(13)   & --        &
   -0.51(63)   & 0.092(18)  & -1.32(14)  & 0.6(1.3)
\\ \hline
\end{tabular}
\vspace{3mm}
\end{table}

Numerical results of our continuum and chiral extrapolation are summarized
in Table~\ref{tbl:chiralfit:fit}.
The polynomial coefficients turn out to be $O(1)$ or less
with our choice of
the dimensionless expansion parameters~(\ref{eqn:chiral_fit:exparam}),
which are normalized by appropriate scales.
Many of them are consistent with zero reflecting the mild dependence
on the lattice spacing and quark masses mentioned above.
As a result,
our choice of the fitting forms~(\ref{eqn:chiral_fit:form})
results in small values of $\chi^2/{\rm d.o.f.}\!\lesssim\!0.2$,
which, however, are only a rough measure of the goodness of the fits,
because we employ approximated estimators of the covariance matrix
as discussed in the next subsection.

\subsection{Systematic uncertainties}
\label{subsec:cont+chiral_fit:err}


\subsubsection{Correlated fit}
\label{subsubsec:cont+chiral_fiterr:err:cmat}

With limited statistics, 
the covariance matrix $C$ of our form factor data
has exceptionally small eigenvalues with large statistical error.
In this study, we test two methods to take the correlation into account
in the combined continuum and chiral extrapolation.
First, as in the study of $B\!\to\!\pi\ell\nuell$~\cite{B2pi:Nf3:JLQCD},
we introduce a lower cut of the eigenvalue (singular value) $\lambda_{\rm cut}$.
Effects of the exceptionally small eigenvalues are suppressed
by replacing eigenvalues smaller than $\lambda_{\rm cut}$
by $\lambda_{\rm cut}$ in the singular value decomposition of $C$ as
\bea
   C 
   & \sim &
   \sum_{k \leq n_{\rm cut}} \lambda_{\rm cut} \, u_k u_k^T
 + \sum_{k > n_{\rm cut}} \lambda_k  u_k u_k^T,
\eea   
where $u_k$ represents the eigenvector corresponding to the eigenvalue $\lambda_k$.
Here eigenvalues are sorted in ascending order ($\lambda_k \!<\! \lambda_{k+1}$)
and $n_{\rm cut}$ satisfies
$\lambda_{n_{\rm cut}} \!<\! \lambda_{\rm cut} \!<\! \lambda_{n_{\rm cut}+1}$.
We choose $\lambda_{\rm cut}$ such that all eigenvalues below $\lambda_{\rm cut}$
have statistical error larger than 100\,\%.
The second method is the so-called shrinkage estimate of
the covariance matrix~\cite{shrinkage:S,shrinkage:LW},
which amounts to replacing $C_{ij}$ by
\bea
   \rho \delta_{ij} C_{ii} + (1-\rho) C_{ij}.
   \label{eqn:chiral_fit:shrinkage}
\eea
With $0\!\leq\!\rho\!\leq\!1$,
the shrinkage is an interpolation
between the uncorrelated ($\rho\!=\!1$) and correlated ($\rho\!=\!0$) fits.
An optimal value of $\rho$ can be estimated
by minimizing a loss function~\cite{shrinkage:LW}.
We observed, however, that results of the continuum and chiral extrapolation
do not strongly depend on the choice of $\lambda_{\rm cut}$ and $\rho$.
In this study,
we therefore employ the singular value cut for our main results,
since this is a conservative approach,
which only tends to increase the final error.
The shift of the fit results obtained with the shrinkage estimator
is treated as a systematic error due to the estimate of $C$.


\begin{figure}[t]
\begin{center}
  \includegraphics[angle=0,width=0.48\linewidth,clip]{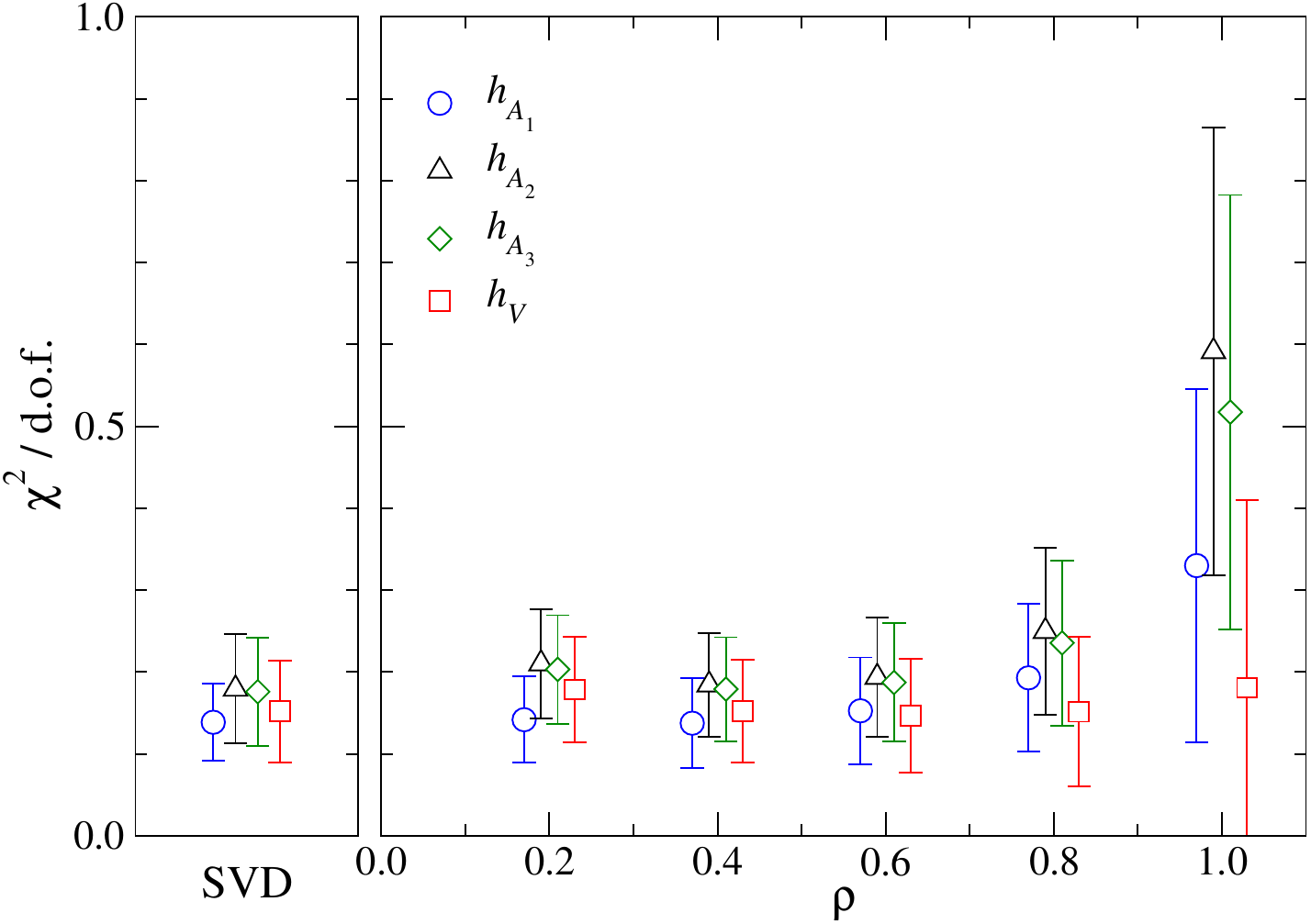}

  \vspace{-3mm}
  \caption{
    Values of $\chi^2/{\rm d.o.f.}$ for our continuum and chiral extrapolation
    of form factors.
    The left panel shows $\chi^2/{\rm d.o.f.}$ with the singular value cut
    for the covariance matrix, whereas the right panel shows
    that with the shrinkage estimator as a function of the parameter $\rho$
    in Eq.~(\ref{eqn:chiral_fit:shrinkage}).
    The error is estimated by the bootstrap method.
  }
  \label{fig:chiralfit:chi2}
\end{center}
\end{figure}

With the approximated estimators of $C$,
$\chi^2/{\rm d.o.f.}$ is only a rough measure of the goodness of the fit.
The impact is shown in Fig.~\ref{fig:chiralfit:chi2}.
We observe that, in the case of our analysis,
$\chi^2/{\rm d.o.f.}$ tends to be slightly smaller
than that for the unquenched fit,
and rather stable against the choice of the estimator
(singular value cut, or shrinkage)
and parameter $\rho$. 
While $C$ is not invertible for the correlated fit
(with $\rho\!=\!0$ for the shrinkage estimator),
we may expect that $\chi^2/{\rm d.o.f.}$ does not change drastically
and serves as a ``rough'' measure of the goodness of the fit
even with the approximated estimators of $C$.
We also note that, in this paper,
the approximated estimators are used only for
the continuum and chiral extrapolation of form factors
due to the large size of $C$ with different choices of
the quark masses (including $m_Q$ for the relativistic approach),
lattice cutoff and recoil parameter.

\subsubsection{Fitting form}
\label{subsubsec:cont+chiral_fiterr:err:form}


The highest order is quadratic in
the recoil parameter $w-1$ expansion and heavy quark expansion for $h_{A_1}$,
and linear for other expansions including chiral and finite $a$ expansions.
The systematic error due to this choice is estimated
by repeating the continuum and chiral extrapolation
without one of the highest order terms,
if the corresponding coefficient is poorly determined,
otherwise by adding a yet higher order term.


Possible higher order corrections in the chiral expansion
are also examined by testing the $x$-expansion.
We also test a multiplicative form
\bea
  h_X/\eta_X
  & = &
  c
+ \left(
    1 + \frac{\gVPpi^2}{2 \Lambda_\chi^2} 
        \bar{F}_{\rm log}\!\left(\xipi,\Delta_D,\Lambda_\chi\right)
      + c_\pi \xipi + c_{\eta_s} \xietas
  \right)
  \left( 1 + c_Q \epsQ \right)
  \nn \\
  & &
  \times \left( 1 + c_a \xia + c_{am_Q} \xiamQ \right)
  \left\{ 1 + c_w (w-1) + d_w (w-1)^2 \right\}
 \hspace{5mm}
 \label{eqn:chiral_fit:form:multi}
\eea
to examine the potential effect of cross terms. We find that these alternative fits
yield a change of the fit results well below the statistical error.

\subsubsection{Inputs}
\label{subsubsec:cont+chiral_fiterr:err:input}


The lattice cutoff $a^{-1}$, physical meson masses $M_\pi$, $M_K$, $M_{\eta_b}$,
and the $D^*D\pi$ coupling $\gVPpi$ are inputs to the continuum and chiral extrapolation.
Since the masses have been measured very precisely,
we take account of the uncertainty due to $a^{-1}$ and $\gVPpi$
by repeating the extrapolation with each input shifted by $\pm$ 1 $\sigma$.


We set $M_\pi$ and $M_K$ to those in the isospin limit
as in our determination of $a$~\cite{B2pi:Nf3:JLQCD}.
To examine the isospin breaking effects to the form factors,
we test the continuum and chiral extrapolation to $M_\pi\!=\!M_{\pi^0}$ and $M_{\pi^\pm}$
as well as $M_K\!=\!M_{K^0}$ and $M_{K^\pm}$.
We also test the heavy quark expansion parameter
replaced with $\epsQ\!=\!\bar{\Lambda}/2M_B$
and compare the extrapolations using $M_B=M_{B^0}$ and $M_{B^\pm}$.
The difference in the form factor can be considered
as an estimate of the strong isospin breaking effect.
It turns out to be small
as suggested by the mild quark mass dependence of the form factors.
We include this in the systematic uncertainty
so that our synthetic data for the form factors
can be used for both the neutral and charged $B$ meson decays.

\begin{figure}[t]
\begin{center}
  \includegraphics[angle=0,width=0.48\linewidth,clip]{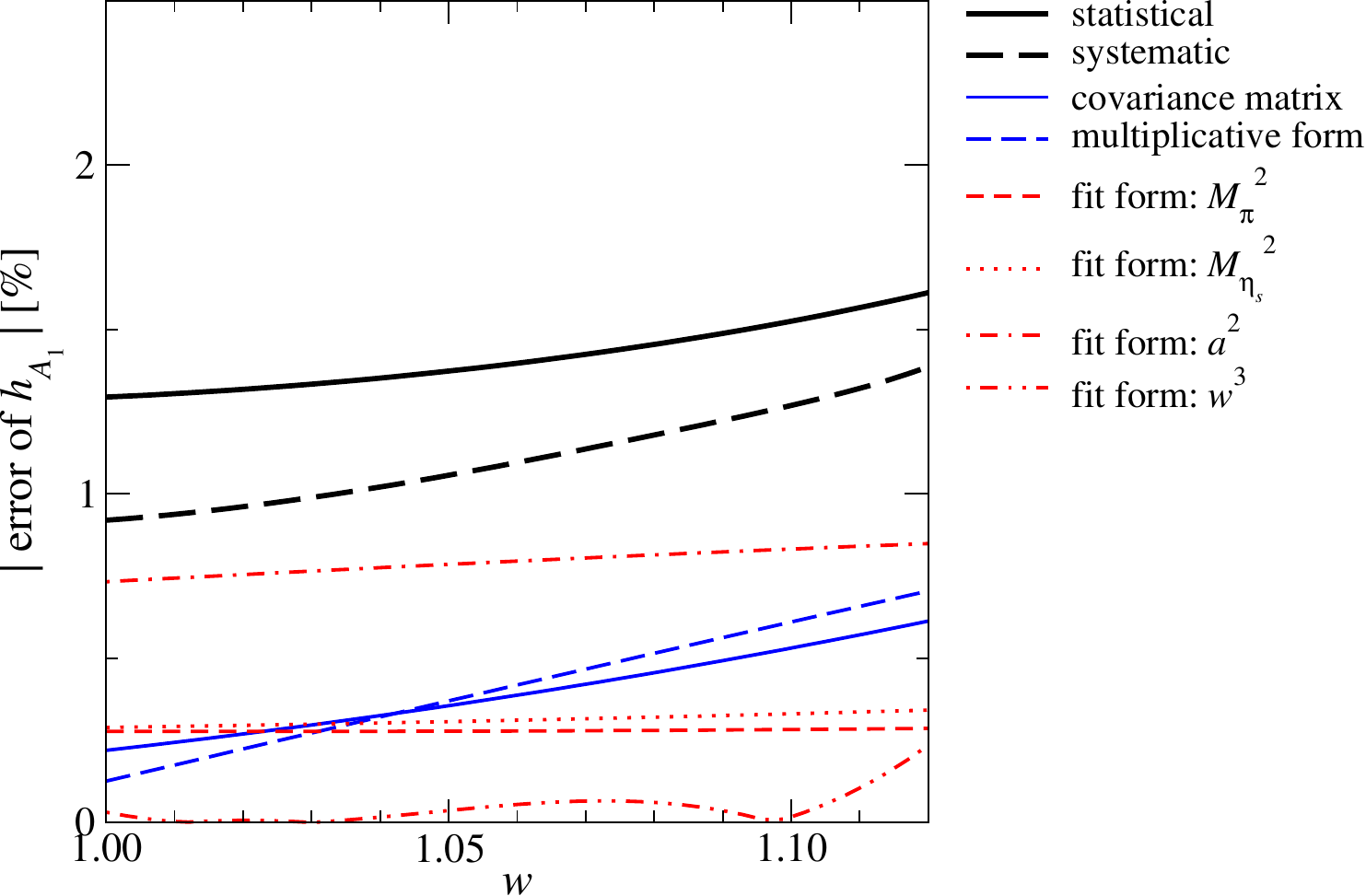}
  \hspace{3mm}
  \includegraphics[angle=0,width=0.48\linewidth,clip]{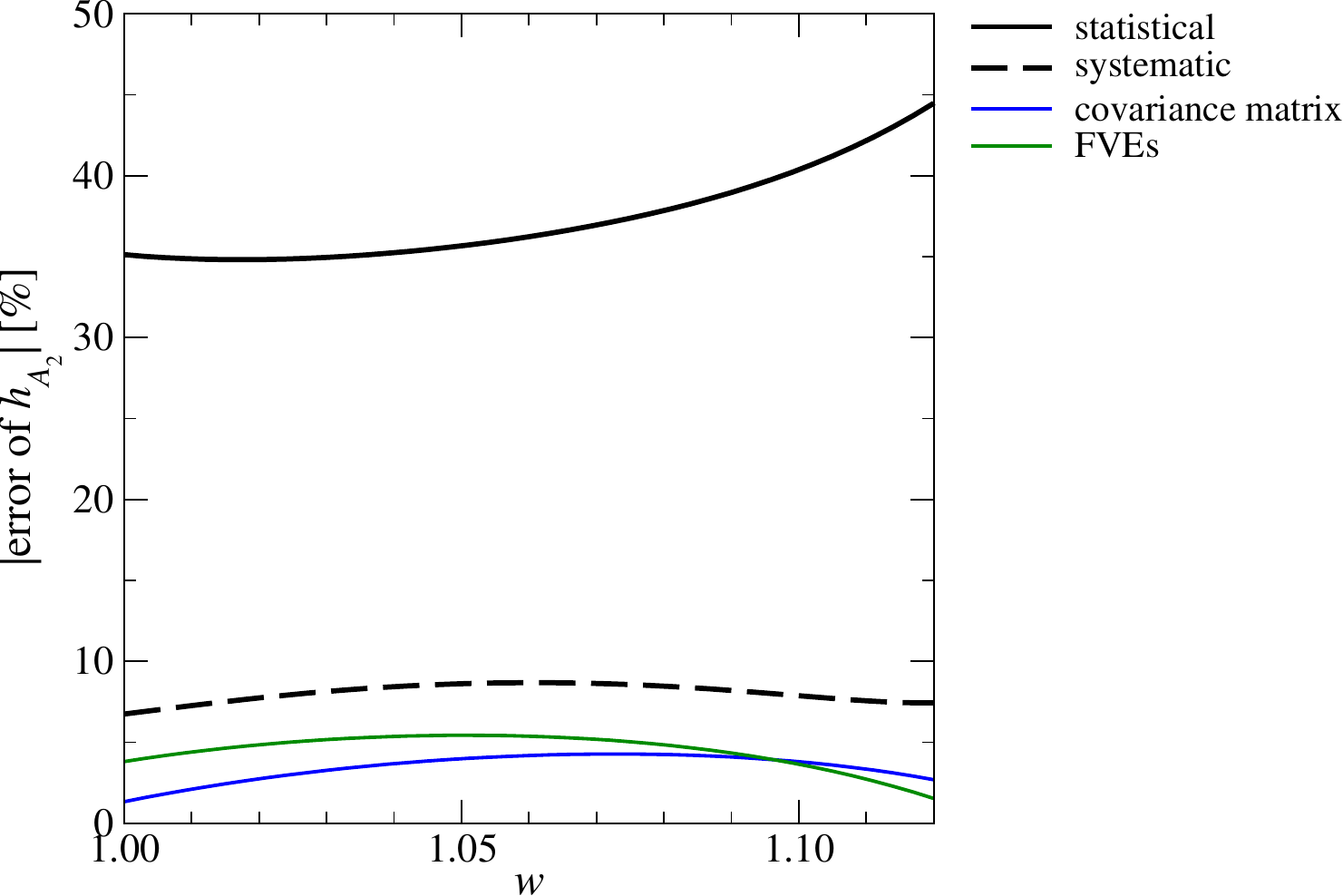}
  \vspace{3mm}

  \includegraphics[angle=0,width=0.48\linewidth,clip]{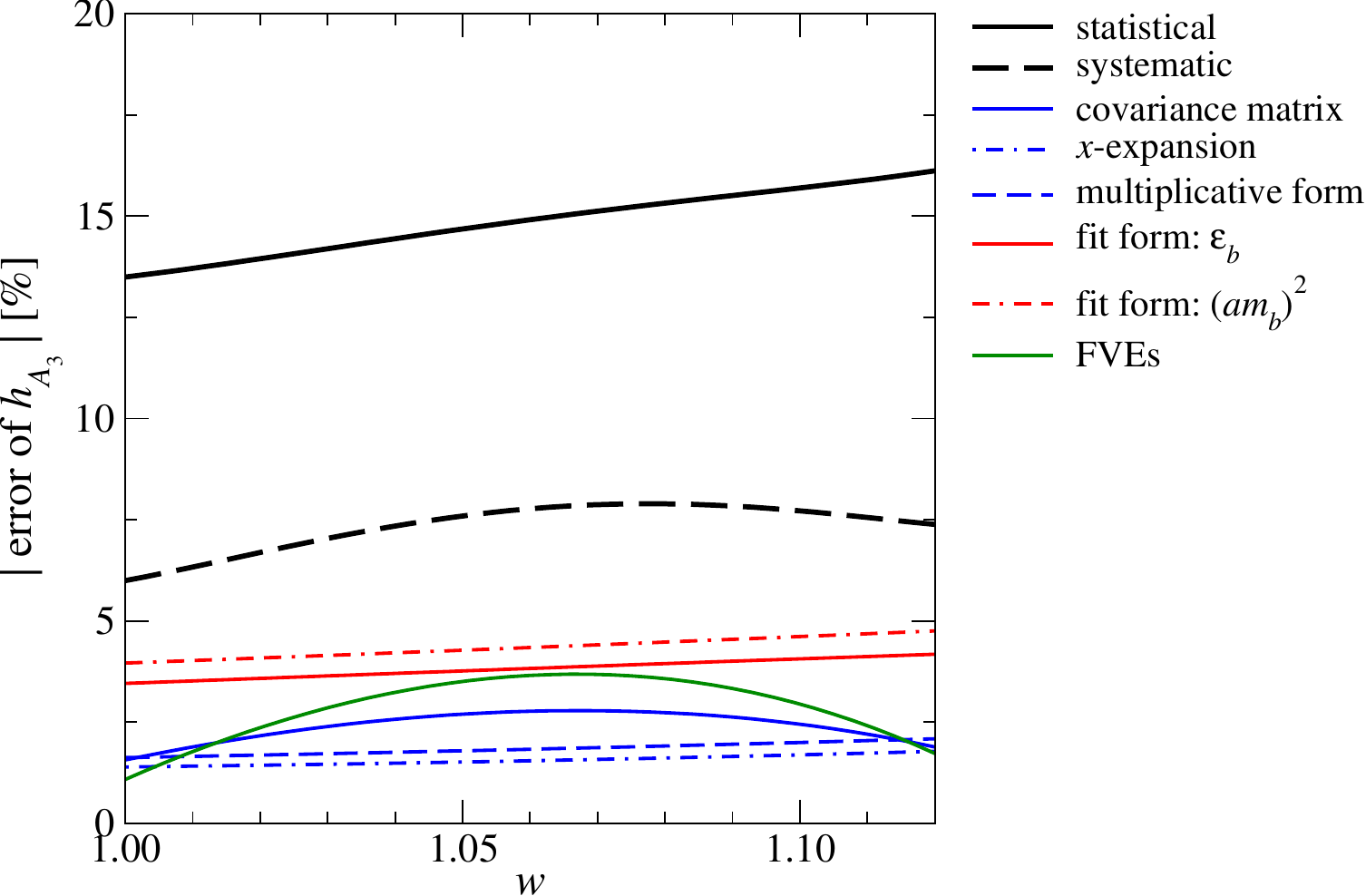}
  \hspace{3mm}
  \includegraphics[angle=0,width=0.48\linewidth,clip]{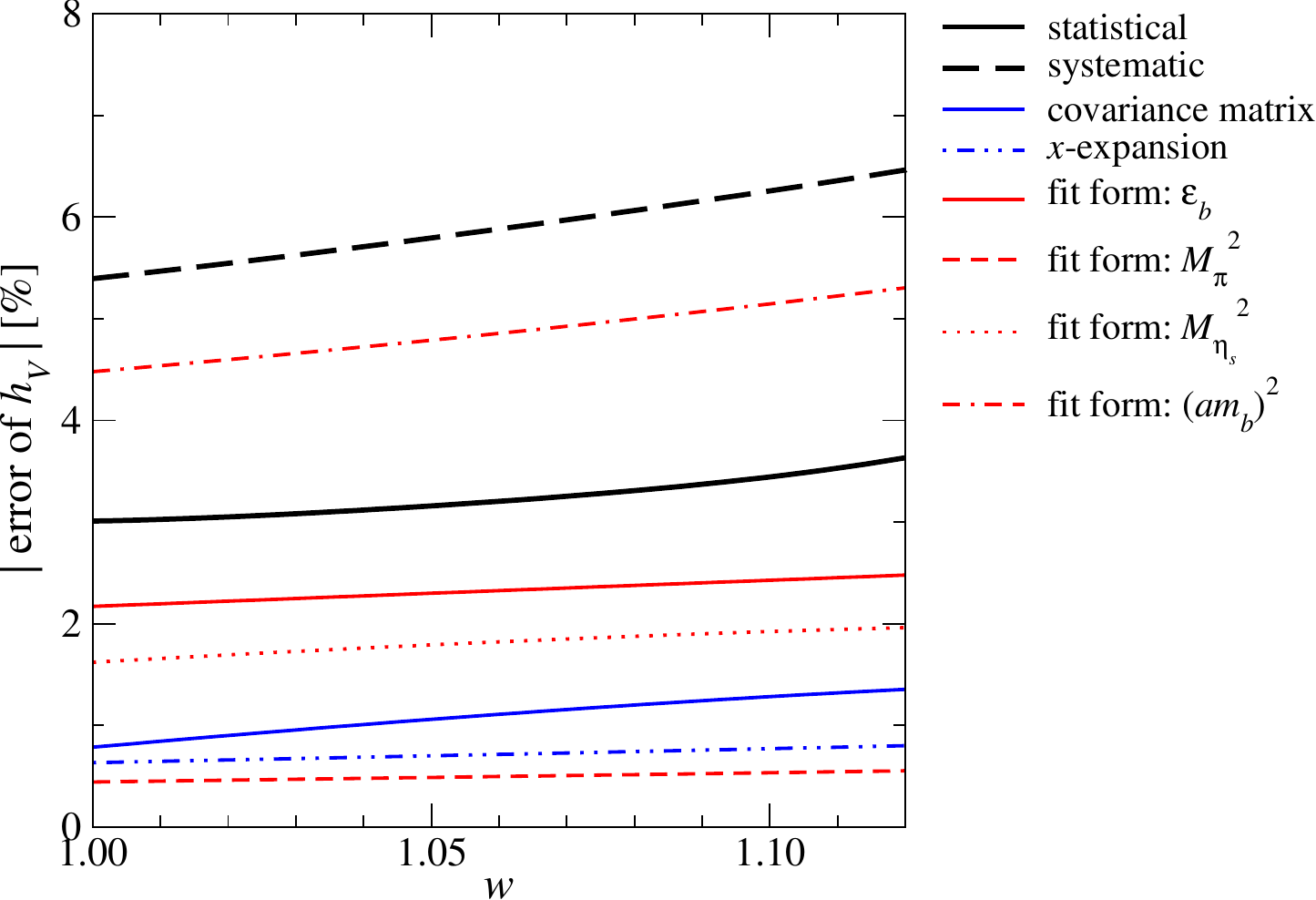}

  \vspace{-3mm}
  \caption{
    Uncertainties of form factors as a function of $w$.
    The top-left, top-right, bottom-left and bottom-right panels show
    uncertainties of $h_{A_1}$, $h_{A_2}$, $h_{A_3}$ and $h_V$, respectively.
    The thick solid and dashed lines show the statistical and total systematic
    uncertainties, while the thin lines show a breakdown of the latter.
    Just for clarity,
    we omit small systematic errors below 0.2\,\% ($h_{A_1}$), 4\,\% ($h_{A_2}$),
    1\,\% ($h_{A_3}$) and 0.5\,\% ($h_V$) in the whole simulated region of $w$.
  }
  \label{fig:chiralfit:syserr}
\end{center}
\end{figure}

\subsubsection{Finite volume effects (FVEs)}
\label{subsubsec:cont+chiral_fiterr:err:fve}


Our spatial lattice size satisfies $M_\pi L\!\geq\! 4$.
At our smallest pion mass on the coarsest lattice,
we also simulate a smaller size $L\!\simeq\!3.0 M_\pi^{-1}$
to directly examine FVEs.
Comparison between the two volumes at that simulation point
does not show any significant deviation:
for instance,
$h_{A_1}$ at $w\!=\!1$ is consistent within statistical error 
of $\lesssim\,1$\,\%.
We also repeat the continuum and chiral extrapolation
by replacing the form factors at the smallest $M_\pi$
with those on the smaller $L$.
The change in the fit results is included in the systematic uncertainty,
but turns out to be well below the statistical uncertainty.
We conclude that finite volume effects are not significant
in our simulation region of $M_\pi\!\gtrsim\!230$~MeV.

\subsubsection{Accuracy of continuum and chiral extrapolation}
\label{subsubsec:cont+chiral_fiterr:err:comments}


Figure~\ref{fig:chiralfit:syserr} shows 
the estimate of various uncertainties for the form factors
in the continuum limit and at the physical quark masses as a function of $w$.
As discussed in Sec.~\ref{sec:ff},
the axial (vector) matrix element
is exclusively sensitive to $h_{A_1}(w)$ ($h_V(w)$)
with an appropriate choice of the $D^*$ meson polarization vector and momentum.
Combined with a precise determination at zero-recoil,
the total uncertainty of $h_{A_1}$ is about 2\,\% in our simulation region of $w$,
and the largest uncertainty comes from statistics and discretization effects.
On the other hand,
either $D^*$ or $B$ needs to have non-zero momentum to determine $h_V$,
which is, therefore, less accurate.
The total accuracy is $\lesssim\!7$\,\%
with 3\,--4\,\% statistical and $\sim\!5$\,\% discretization errors.


Since the simulated bottom quark masses are not far below the lattice cut-off
($m_Q\!\leq\!0.7a^{-1}$),
we repeat the continuum and chiral extrapolation
only using the data with $m_Q\!\leq\!0.5\,a^{-1}$.
The change in the fit results is well below the statistical uncertainty.
We therefore conclude that
our continuum extrapolation in the relativistic approach is controlled.


On the other hand,
the axial form factors, $h_{A_2}$ and $h_{A_3}$, have larger uncertainty,
{\it i.e.} 40\,\% and 15\,\%, respectively, dominated by the statistical error.
Towards a more precise determination,
it would be advantageous to simulate moving $B$ mesons
so as to have matrix elements more sensitive to these form factors.


\begin{figure}[t]
\begin{center}
  \includegraphics[angle=0,width=0.48\linewidth,clip]{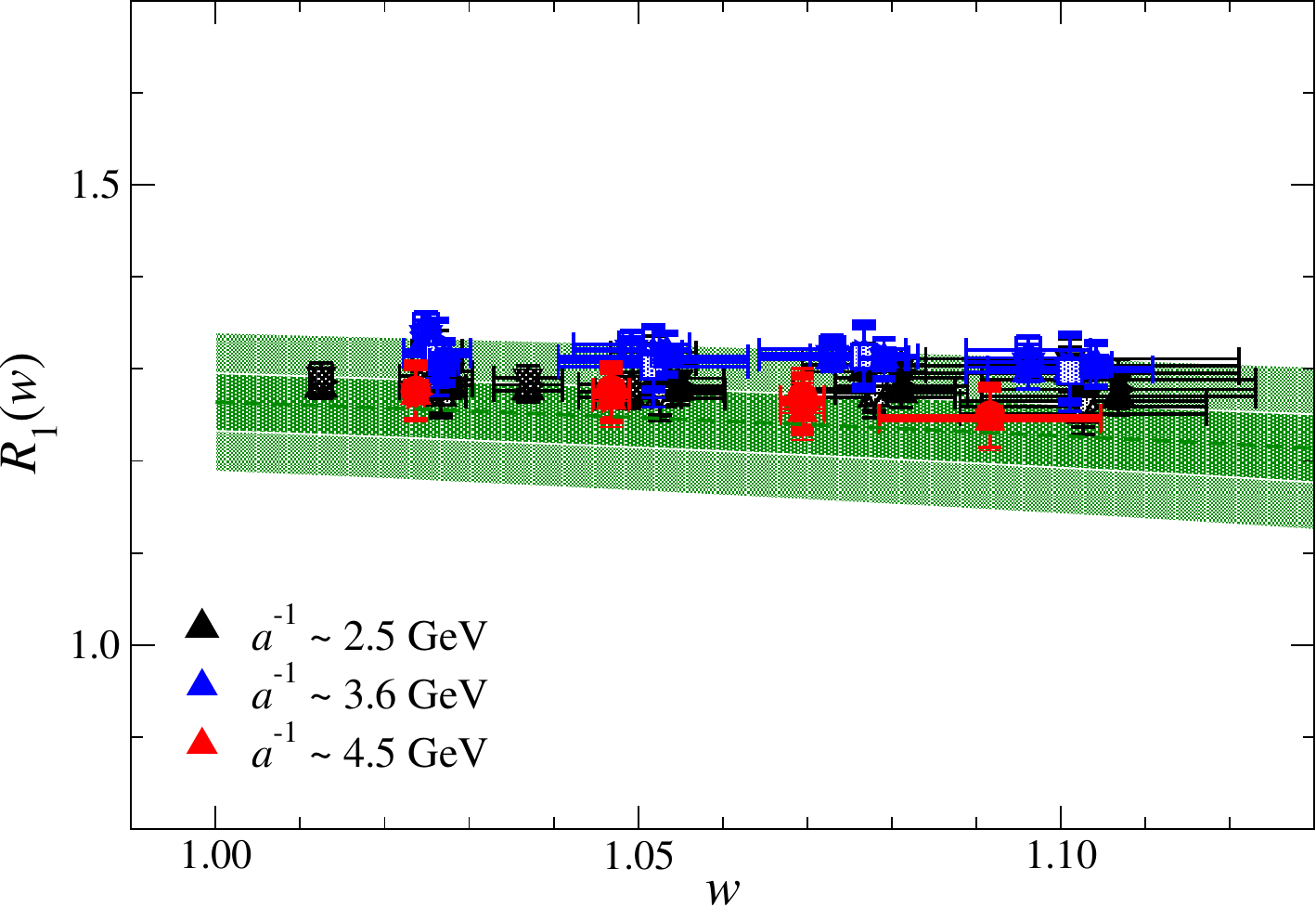}
  \hspace{3mm}
  \includegraphics[angle=0,width=0.48\linewidth,clip]{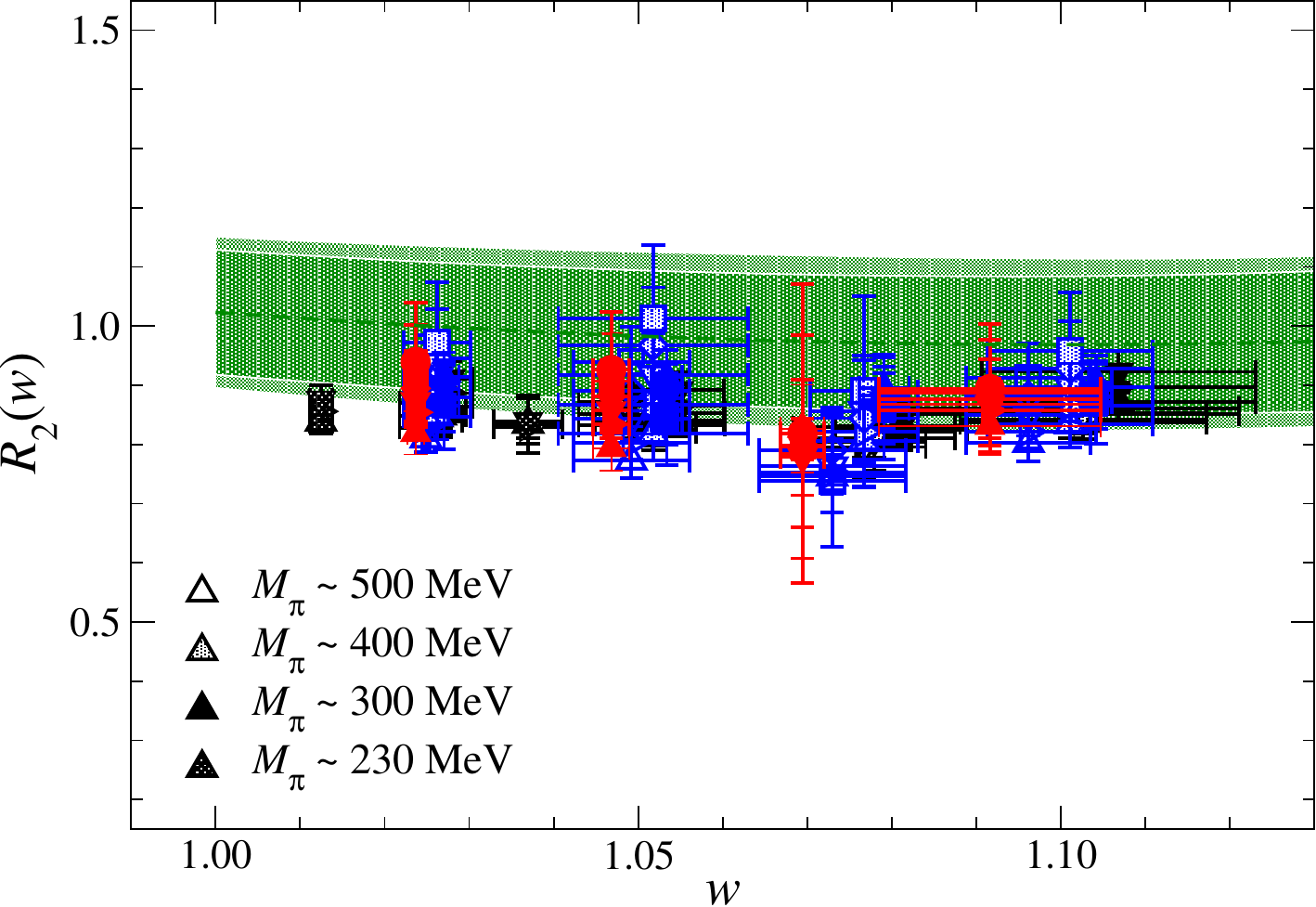}

  \vspace{-3mm}
  \caption{
    Form factor ratios $R_1(w)$ (left panel) and $R_2(w)$ (right panel)
    as a function of recoil parameter $w$.
    Results at simulated parameters are plotted in symbols,
    which have the same meaning as in Fig.~\ref{fig:chiralfit:ff}.
    The green bands show these ratios in the continuum limit
    and at physical quark masses 
    reproduced by the continuum and chiral extrapolation
    of form factors~(\ref{eqn:chiral_fit:form}).
  }
  \label{fig:chiralfit:r1+2}
\end{center}
\end{figure}

Previous determinations of $\Vcb$ often employed
a HQET-based parametrization by Caprini, Lellouch and Neubert
of $h_{A_1}$ and form factor ratios
\bea
   R_1(w) \!=\! \frac{h_V(w)}{h_{A_1}(w)}, 
   \hspace{5mm}
   R_2(w) \!=\! \frac{r h_{A_2}(w) + h_{A_3}(w)}{h_{A_1}(w)},
\eea
where higher order HQET corrections are expected to be partially canceled~\cite{CLN}.
Figure~\ref{fig:chiralfit:r1+2} shows that
these ratios also have mild parameter dependences
and can be safely extrapolated to the continuum limit and physical quark masses.
We, however, do not employ the model-dependent CLN parametrization
in the following analysis.

\section{Form factors as a function of the recoil-parameter}
\label{sec:q2-param}


\subsection{Fit to lattice data}
\label{subsec:q2-param:lat}


From the continuum and chiral extrapolation in the previous section,
we generate synthetic data of the form factors in the relativistic convention,
\bea
   \langle D^*(\epsp,\pp) | V_\mu | B(p) \rangle
   & = &
   i \varepsilon_{\mu\nu\rho\sigma} \, \epsps^{\nu} p^{\prime\rho} p^\sigma \, g(w),
   \label{eqn:q2-param:ff:V:rel}
   \\[2mm]
   \langle D^*(\epsp,\pp) | A_\mu | B(p) \rangle
   & = &
   \epsilon_\mu^{\prime *} \,  f(w)
 - (\epsps p)
   \left\{
      (p+\pp)_\mu \, a_+(w) + (p-\pp)_\mu \, a_-(w)
   \right\},
   \hspace{10mm}
   \label{eqn:q2-param:ff:A:rel}
\eea   
with which the BGL parametrization was originally proposed.
A kinematic invariant
$k\!=\!\sqrt{ \left(t_+-q^2\right)\left(t_--q^2\right)/4t}$
is defined with the maximum value of the momentum transfer
$q^2_{\rm max}\!=\!t_-\!=\!\left(M_B-M_{D^*}\right)^2$,
which corresponds to the zero recoil limit,
and the threshold $t_+\!=\!\left(M_B+M_{D^*}\right)^2$ 
for the $W\!\to\!BD^*$ production channel.
Instead of $a_+$ and $a_-$,
it is common to use linear combinations
\bea
   \Fo(w)
   & = &
   \frac{1}{M_{D^*}}
   \left\{
      \frac{1}{2} \left( M_B^2 - M_{D^*}^2 - q^2 \right) f(w)
      + 2 k^2 q^2 a_+(w) 
   \right\},
   \label{eqn:q2-param:f1}
   \\
   \Ft(w)
   & = &
   \frac{1}{M_{D^*}}
   \left\{
      f(w) + \left( M_B^2 - M_{D^*}^2 \right) a_+(w)
           + q^2 a_-(w)
   \right\}.
   \label{eqn:q2-param:f2}
\eea
These are related to the form factors in the HQET
convention as
\bea
   g(w)
   & = &
   \label{eqn:ff:ff:V:rel}
   \frac{1}{M_B\sqrt{r}} h_V(w),
   \label{eqn:q2-param:ff_conv:g}
   \\
   f(w)
   & = &
   M_B\sqrt{r} (w+1) h_{A_1}(w),
   \label{eqn:q2-param:ff_conv:f}
   \\
   \Fo(w)
   & = &
   M_B^2 \sqrt{r} (w+1)
   \left[
     (w-r) h_{A_1}(w) - (w-1)\left\{ r h_{A_2}(w) + h_{A_3}(w) \right\}
   \right],
   \label{eqn:q2-param:ff_conv:f1}
   \\
   \Ft(w)
   & = &
   \frac{1}{\sqrt{r}}
   \left\{
   (w+1) h_{A_1}(w) + (rw - 1) h_{A_2}(w) + (r-w)h_{A_3}
   \right\}.
   \label{eqn:q2-param:ff_conv:f2}
\eea
We note that there are two kinematical constraints to be satisfied
at the minimum ($w\!=\!1$) and maximum values of $w$
($\wmax\!=\!(1+r^2)/2r$ with $m_\ell\!=\!0$)
\bea
   \Fo(1) & = & M_B(1-r)f(1),
   \label{eqn:q2-param:const:w1}
   \\
   \Ft(\wmax) & = & \frac{1+r}{M_B^2r(1-r)(\wmax+1)} \Fo(\wmax).
   \label{eqn:q2-param:const:wmax}
\eea

   
With these form factors, we can write
the helicity amplitudes for a given helicity of the intermediate $W$ boson,
and the differential decay rate with respect to $w$
in simple forms
\bea
   H_\pm(w)
   & = &
   f(w) \mp M_B^2 r \sqrt{w^2-1} g(w),
   \label{eqn:q2-param:Hamp:+-}
   \\
   H_0(w)
   & = &
   \frac{1}{\sqrt{q^2}}\Fo(w),
   \label{eqn:q2-param:Hamp0}
   \\
   H_S(w)
   & = &
   M_B r \sqrt{\frac{w^2-1}{1-2rw+r^2}} \Ft(w),
   \label{eqn:q2-param:Hamps}
\eea
and
\bea
   \frac{d\Gamma}{dw}
   & = &
   \frac{G_F^2}{16\pi^3} |V_{cb}|^2 |\eta_{EW}|^2
   M_B r^2 \sqrt{ w^2 - 1 } \left(1-\frac{m_\ell^2}{q^2}\right)^2
   \nn \\[3mm]
   &&
   \hspace{5mm}
   \times
   \left[
      \frac{q^2}{3}
      \left( 1 + \frac{m_\ell^2}{2q^2} \right)
      \left\{ |H_+(w)|^2 + |H_-(w)|^2 + |H_0(w)|^2 \right\}
      +\frac{m_\ell^2}{2} |H_S(w)|^2
      \right].
   \hspace{10mm}
   \label{eqn:q2-param:ddr:ddr}
\eea


From the continuum and chiral extrapolation presented
in Sec.~\ref{sec:cont+chiral_fit},
synthetic data of $f$, $g$ and $\Fot$ are calculated
at reference values of the recoil parameter $\wref$ through
the relations~(\ref{eqn:q2-param:ff_conv:g})\,--\,(\ref{eqn:q2-param:ff_conv:f2}).
We take $\wref\!=\!1.025$, 1.060 and 1.100,
which span the whole simulated region of $w$,
since our continuum and chiral extrapolation
interpolates $h_{A_{\{1,2,3\}}}$ and $h_V$ in $w$.
{\it Three} values are chosen,
because the statistical covariance matrix of the synthetic data
develops ill-determined eigenvalues with four or more values of $\wref$.
This might be partly because, in our continuum and chiral extrapolation,
we parametrize the $w$ dependence of each form factor
using a quadratic form with three independent parameters. 
The numerical values together with their covariance matrix are
presented in Table~\ref{tbl:q2-param:ff:synthed}.


\begin{sidewaystable}[h]
\centering
\small
\caption{
  Synthetic data of form factors $f$, $\Fot$ and $g$
  in the relativistic convention.
  The central value and total error are in the second column,
  and the correlation matrix is in the third to fourteenth columns.
}
\vspace{3mm}
\label{tbl:q2-param:ff:synthed}.
\begin{tabular}{l|l|lll|lll|lll|lll}
\hline
& &
$f(1.025)$   &
$f(1.060)$   &
$f(1.100)$   &
$\Fo(1.025)$  &
$\Fo(1.060)$  &
$\Fo(1.100)$  &
$\Ft(1.025)$  &
$\Ft(1.060)$  &
$\Ft(1.100)$  &
$g(1.025)$   &
$g(1.060)$   &
$g(1.100)$   
\\ \hline
$f(1.025)$   &  5.6928(936)   &  1.00000  &  0.98385  &  0.93944  &  0.94354  &  0.71378  &  0.49823  &  0.24518  &  0.22123  &  0.21755  &  0.22911  &  0.22641  &  0.22699
\\
$f(1.060)$   &  5.5879(992)   &  0.98385  &  1.00000  &  0.98338  &  0.92328  &  0.70780  &  0.49984  &  0.22221  &  0.20401  &  0.20620  &  0.22112  &  0.22194  &  0.22694
\\
$f(1.100)$   &  5.473(109)    &  0.93944  &  0.98338  &  1.00000  &  0.87531  &  0.68095  &  0.48820  &  0.19269  &  0.18105  &  0.18953  &  0.20747  &  0.21181  &  0.22233
\\ \hline
$\Fo(1.025)$ &  18.612(324)   &  0.94354  &  0.92328  &  0.87531  &  1.00000  &  0.88532  &  0.70711  &  0.36140  &  0.35455  &  0.37432  &  0.20808  &  0.20305  &  0.20067
\\
$\Fo(1.060)$ &  18.308(410)   &  0.71378  &  0.70780  &  0.68095  &  0.88532  &  1.00000  &  0.94630  &  0.58476  &  0.60631  &  0.65922  &  0.15762  &  0.15344  &  0.15170
\\
$\Fo(1.100)$ &  17.983(545)   &  0.49823  &  0.49984  &  0.48820  &  0.70711  &  0.94630  &  1.00000  &  0.66784  &  0.70838  &  0.78248  &  0.11626  &  0.11328  &  0.11262
\\ \hline
$\Ft(1.025)$ &  2.1704(962)   &  0.24518  &  0.22221  &  0.19269  &  0.36140  &  0.58476  &  0.66784  &  1.00000  &  0.98622  &  0.96645  &  0.05204  &  0.04696  &  0.04049
\\
$\Ft(1.060)$ &  2.097(106)    &  0.22123  &  0.20401  &  0.18105  &  0.35455  &  0.60631  &  0.70838  &  0.98622  &  1.00000  &  0.98356  &  0.04418  &  0.03974  &  0.03438
\\
$\Ft(1.100)$ &  1.992(109)    &  0.21755  &  0.20620  &  0.18953  &  0.37432  &  0.65922  &  0.78248  &  0.96645  &  0.98356  &  1.00000  &  0.04246  &  0.03872  &  0.03398
\\ \hline
$g(1.025)$   &  0.3326(212)   &  0.22911  &  0.22112  &  0.20747  &  0.20808  &  0.15762  &  0.11626  &  0.05204  &  0.04418  &  0.04246  &  1.00000  &  0.99764  &  0.98991
\\
$g(1.060)$   &  0.3178(213)   &  0.22641  &  0.22194  &  0.21181  &  0.20305  &  0.15344  &  0.11328  &  0.04696  &  0.03974  &  0.03872  &  0.99764  &  1.00000  &  0.99622
\\
$g(1.100)$   &  0.3013(215)   &  0.22699  &  0.22694  &  0.22233  &  0.20067  &  0.15170  &  0.11262  &  0.04049  &  0.03438  &  0.03398  &  0.98991  &  0.99622  &  1.00000
\\ \hline
\end{tabular}
\vspace{3mm}
\end{sidewaystable}




The synthetic data are fitted to the model-independent BGL parametrization
\bea
   F(z)
   & = &
   \frac{1}{P_F(z)\phi_F(z)}
   \sum_{k=0}^{N_F} a_{F,k} z^k
   \hspace{5mm}
   (F=f,g,\Fo,\Ft),
   \label{eqn:q2-param:bgl}
\eea
where the $z$ parameter
\bea
   z(w) 
   & = &
   \frac{\sqrt{w+1}-\sqrt{2}}{\sqrt{w+1}+\sqrt{2}}
\eea
maps the whole semileptonic kinematical region
$0 \!\leq\! q^2[{\rm GeV}^2] \!\lesssim\!11$
($1\!\leq\!w\!\lesssim\!1.5$) ($m_\ell\!=\!0$)
into a small parameter region of $z$, $0\!\leq\!z\!\lesssim\!0.06$.


\begin{table}[t]
\centering
\small
\caption{
  Two sets of input resonance masses with quantum number $J^P$.
  All values are in GeV units.
  The input-A is a compilation of 
  slightly old experimental measurement~\cite{PDG16}
  and lattice QCD~\cite{Mreso:lat:DDHH12,Mreso:lat:HPQCD15}
  and phenomenological~\cite{Mreso:ph:G04,Mreso:ph:RD13} studies.
  A different phenomenological estimate~\cite{Mreso:ph:EQ94}
  for the (axial) vector resonances is used for input-B,
  where the pseudo-scalar masses are set
  from a recent edition of the Particle Data Book~\cite{PDG22} ($n\!=\!0,1$)
  and another phenomenological study~\cite{Mreso:ph:LLWL23} ($n\!=\!2$).
  We employ input-A in our main analysis, whereas input-B is used
  to test the stability of the BGL parametrization against the choice of the input.
  See the text for more details.
}
\vspace{3mm}
\label{tbl:q2-param:reso}.
\begin{tabular}{l|ll|ll}
\hline
       &  \multicolumn{2}{|c}{input-A}     & \multicolumn{2}{|c}{input-B}
\\ \hline
$J^P$  &  $n_{\rm pole}$ & $M_{{\rm pole},n}$ & $n_{\rm pole}$ & $M_{{\rm pole},n}$
\\ \hline
$1^+$  &  4  &  6.739, 6.750, 7.145, 7.150 & 4 & 6.730, 6.736, 7.135, 7.142
\\
$1^-$  &  3  &  6.329, 6.920, 7.020        & 4 & 6.337, 6.899, 7.012, 7.280
\\
$0^-$  &  3  &  6.275, 6.842, 7.250        & 3 & 6.274, 6,871, 7220
\\ \hline
\end{tabular}
\vspace{3mm}
\end{table}

The Blaschke factors
\bea
   P_f(z) = P_\Fo(z) = P_{1^+}(z), \hspace{3mm}
   P_g(z) = P_{1^-}(z), \hspace{3mm}
   P_\Ft(z) = P_{0^-}(z)
\eea   
factor out the pole singularities
outside the semileptonic region but below the threshold $t_+$
due to the resonances for a given channel $J^P$ of $b\bar{c}$ mesons.
The explicit form is given as
\bea
   P_{J^P}(z)
   & = &
   \prod_k^{n_{\rm pole}} \frac{z-z_{{\rm pole},k}}{1-zz_{{\rm pole},k}},
\eea   
where 
\bea
   z_{{\rm pole},k}
   & = &
   \frac{\sqrt{t_+-M_{{\rm pole},k}^2}-\sqrt{t_+ - t_-}}
        {\sqrt{t_+-M_{{\rm pole},k}^2}+\sqrt{t_+ - t_-}}
\eea   
is the $z$-parameter corresponding to
the $k$-th resonance (in term of $q^2$ rather than $w$).
In our main analysis,
we employ resonance masses denoted as ``input-A'' in Table~\ref{tbl:q2-param:reso}.
This choice consists of
slightly old experimental measurement~\cite{PDG16}
($M_{{\rm pole},\{0,1\}}$ for $J^{P}\!=\!0^-$)
and estimates based on
lattice QCD~\cite{Mreso:lat:DDHH12,Mreso:lat:HPQCD15}
($M_{{\rm pole},0}$ for $1^+$ and $M_{{\rm pole},\{0,1\}}$ for $1^-$)
and quark potential models~\cite{Mreso:ph:G04,Mreso:ph:RD13}
(other $M_{{\rm pole},n}$'s).
This input has been used in 
some of previous phenomenological analysis using the BGL parametrization~\cite{Vcb:anly:BGS17,Vcb:BGL:GJS19}
and lattice calculations of form factors~\cite{B2Dstar:Nf3:Fermilab/MILC,B2Dstar:Nf4:HPQCD}.
Another choice (``input-B'' in Table~\ref{tbl:q2-param:reso}) in the literature 
is used to estimate the systematics uncertainty
due to the use of input-A (see below).


The outer function $\phi_F(z)$ is chosen
such that a constraint on the expansion coefficients
derived from unitarity of the theory has a simple form
\bea
   \sum_{k=0}^{\infty} |a_{g,k}|^2 \leq 1,
   \hspace{5mm}
   \sum_{k=0}^{\infty} |a_{f,k}|^2 + \sum_{k=0}^{\infty} |a_{\Fo,k}|^2 \leq 1,
   \hspace{5mm}
   \sum_{k=0}^{\infty} |a_{\Ft,k}|^2 \leq 1.
   \label{eqn:q2-param:unitary_bound}
\eea   
The explicit form is given as
\bea
   \phi_g(z)
   & = &
   16r^2
   \sqrt{ \frac{n_I}{3\pi \tilde{\chi}^T_{1^-}(0)} }
   \frac{ (1+z)^2 (1-z)^{-1/2} }
        { \left\{ (1+r)(1-z) + 2\sqrt{r}(1+z) \right\}^4 },
   \label{eqn:q2-param:ofunc:g}
   \\
   \phi_f(z)
   & = &
   \frac{4r}{ M_B^2}
   \sqrt{ \frac{n_I}{3\pi \chi^T_{1^+}(0)} }
   \frac{ (1+z) (1-z)^{3/2} }
        { \left\{ (1+r)(1-z) + 2\sqrt{r}(1+z) \right\}^4 },
   \label{eqn:q2-param:ofunc:f}
   \\
   \phi_\Fo(z)
   & = &
   \frac{4r}{ M_B^3}
   \sqrt{ \frac{n_I}{6\pi \chi^T_{1^+}(0)} }
   \frac{ (1+z) (1-z)^{5/2} }
        { \left\{ (1+r)(1-z) + 2\sqrt{r}(1+z) \right\}^5 },
   \label{eqn:q2-param:ofunc:F1}
   \\
   \phi_\Ft(z)
   & = &
   8\sqrt{2}r^2
   \sqrt{ \frac{n_I}{\pi \chi^L_{1^+}(0)} }
   \frac{ (1+z)^2 (1-z)^{-1/2} }
        { \left\{ (1+r)(1-z) + 2\sqrt{r}(1+z) \right\}^4 },
   \label{eqn:q2-param:ofunc:F2}
\eea
where $n_I\!=\!2.6$ is the number of the spectator quarks
with a correction due to SU(3) breaking.
By $\chi_{J^P}^{\{T,L\}}(0)$,
we denote the derivative of the vacuum polarization function of the weak current
at zero momentum $q^2\!=\!0$ for a given channel $J^P$ and polarization $\{T,L\}$.
As in the literature,
we replace $\chi^T_{1^-}$ by $\tilde{\chi}^T_{1^-}$,
from which the one-particle state contribution due to $B_c^*$ is subtracted
for a better saturation of the unitarity bound (\ref{eqn:q2-param:unitary_bound}).
While a lattice QCD estimate is available~\cite{B2Dstar:chi:lat:MSV21},
we employ a perturbative estimate~\cite{Vcb:HQS:BGS17,B2Dstar:chi:pQCD:GHMS12,B2Dstar:chi:pQCD:BG16} summarized in Table~\ref{tbl:q2-param:suscep}.
We note that our choice of the resonance masses (input-A) and derivatives are 
the same as previous lattice studies
to allow a direct comparison of the BGL parametrization with them.
It is known that the unitarity bound (\ref{eqn:q2-param:unitary_bound})
is not well saturated by the $B\!\to\!D^*\ell\nuell$ channel,
and hence is rather weak.
In this study, we do not impose this bound,
but confirm that fit results satisfy it within the error.

\begin{table}[t]
\centering
\small
\caption{
  Derivatives of vacuum polarization functions
  entering other functions
  (\ref{eqn:q2-param:ofunc:g})\,--\,(\ref{eqn:q2-param:ofunc:F2}).
  Perturbative estimate (upper line) is employed in our main analysis.
}
\vspace{2mm}
\label{tbl:q2-param:suscep}.
\begin{tabular}{l|lll}
\hline
&
$\chi^T_{1^+}(0)$ [GeV$^{-2}$]  &
$\chi^T_{1^-}(0)$ [GeV$^{-2}$]  &
$\chi^L_{1^+}(0)$ 
\\ \hline
perturbation~\cite{Vcb:HQS:BGS17,B2Dstar:chi:pQCD:GHMS12,B2Dstar:chi:pQCD:BG16}
\hspace{2mm} &
$3.894 \times 10^{-4}$  &  $5.131 \times 10^{-4}$  &  $1.9421 \times 10^{-2}$
\\ \hline
lattice QCD~\cite{B2Dstar:chi:lat:MSV21} &
$4.69(20) \times 10^{-4}$  &  $5.84(44) \times 10^{-4}$  &  $2.19(19) \times 10^{-2}$
\\ \hline
\end{tabular}
\vspace{3mm}
\end{table}


\begin{figure}[t]
\begin{center}
  \includegraphics[angle=0,width=0.48\linewidth,clip]{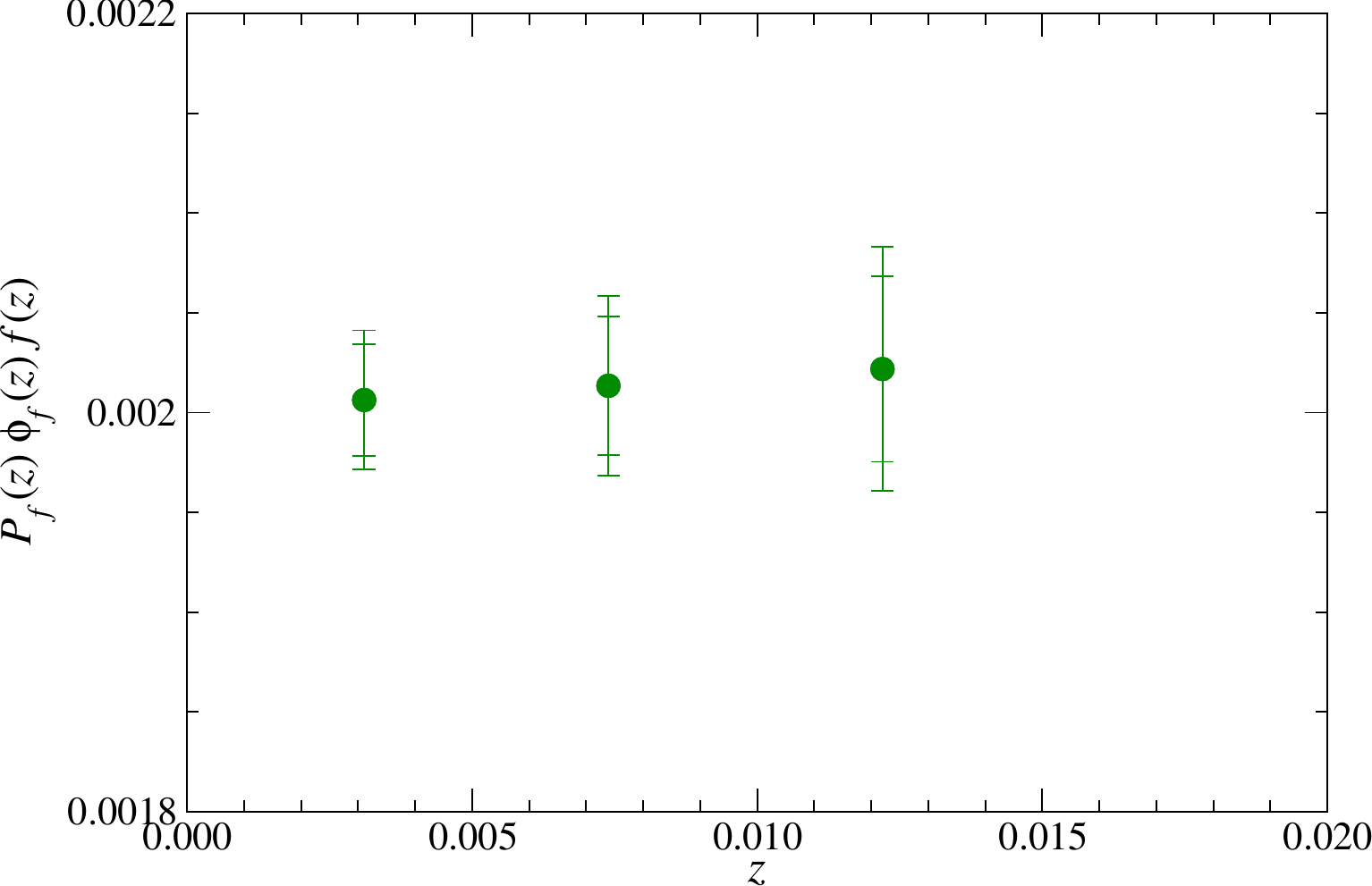}

  \vspace{-3mm}
  \caption{
    Regular part of axial form factor $P_f(z)\phi_f(z)f(z)$ as a function of $z$.
    The two error bars in this figure and Fig.~\ref{fig:q2-param:fx_vs_w}
    show the statistical and total errors, respectively.
  }
  \label{fig:q2-param:ff-reg}
\end{center}
\end{figure}

Figure~\ref{fig:q2-param:ff-reg} shows the regular part of the axial form factor
$P_f(z) \phi_f(z) f(z)$, which is to be expanded in $z$.
We observe rather mild $z$ dependence of all form factors.
While there is no clear sign of curvature in our simulation region,
we employ a quadratic form $N_f\!=\!N_g\!=\!N_\Fo\!=\!N_\Ft\!=\!2$
to safely suppress the truncation error.
Among twelve coefficients,
constants $a_{\Fo,0}$ and $a_{\Ft,0}$ are fixed to fulfill 
the kinematical constraints~(\ref{eqn:q2-param:const:w1}) at $w=1$,
where $z\!=\!0$ and the linear and quadratic terms
in the $z$-parameter expansion (\ref{eqn:q2-param:bgl}) vanish,
and (\ref{eqn:q2-param:const:wmax}) at $\wmax$,
where $z\!\sim\!0.05$ is well below unity,
and hence the constant term gives rise to a large contribution to $\Ft$.
To be consistent with the unitarity bound (\ref{eqn:q2-param:unitary_bound}),
we impose the bound $|a_{F,k}|\!\leq\!1$ ($k\!=\!0,1,2$) for all form factors
$F\!=\!g,f,\Fo$ and $\Ft$.
Table~\ref{tbl:q2-param:bgl} shows fit results,
which describe the synthetic data 
with an acceptable value of $\chi^2/{\rm d.o.f.}\!=\!0.51$.


\begin{sidewaystable}[h]
\centering
\small
\caption{
  Expansion coefficients and their correlation
  for BGL parametrization (\ref{eqn:q2-param:bgl}) of
  form factors $g$, $f$, $\Fo$ and $\Ft$.
  The parameters $a_{\Fo,0}$ and $a_{\Ft,0}$ are fixed from other parameters
  to satisfy kinematical constraints (\ref{eqn:q2-param:const:w1}) and
  (\ref{eqn:q2-param:const:wmax}), respectively.
}
\vspace{3mm}
\label{tbl:q2-param:bgl}.
\begin{tabular}{l|l|rrr|rrr|rrr|rrr}
\hline
& &
$a_{g,0}$   &
$a_{g,1}$   &
$a_{g,2}$   &
$a_{f,0}$   &
$a_{f,1}$   &
$a_{f,2}$   &
$a_{\Fo,0}$ &
$a_{\Fo,1}$ &
$a_{\Fo,2}$ &
$a_{\Ft,0}$ &
$a_{\Ft,1}$ &
$a_{\Ft,2}$
\\ \hline
$a_{g,0}$   & 0.0291(18)   &    1.00000 &    0.46208 &    0.00157 &    0.23345 &    0.01973 & $-$0.00342 &    0.23345 & $-$0.00136 &    0.01362 &    0.08685 & $-$0.00553 &    0.01628
\\
$a_{g,1}$   & $-$0.045(35) &    0.46208 &    1.00000 &    0.01024 &    0.06133 &    0.21056 &    0.06890 &    0.06133 & $-$0.01804 &    0.02956 & $-$0.03921 &    0.00394 &    0.03000
\\
$a_{g,2}$   & $-$1.0(1.7)  &    0.00157 &    0.01024 &    1.00000 &    0.01137 &    0.02396 & $-$0.01746 &    0.01137 & $-$0.00789 &    0.01312 &    0.00205 &    0.01056 &    0.00708
\\ \hline
$a_{f,0}$   & 0.01198(19)  &    0.23345 &    0.06133 &    0.01137 &    1.00000 &    0.06216 &    0.12057 &    1.00000 &    0.07462 & $-$0.09380 &    0.39328 &    0.19208 & $-$0.13641
\\
$a_{f,1}$   & 0.018(11)    &    0.01973 &    0.21056 &    0.02396 &    0.06216 &    1.00000 & $-$0.49732 &    0.06216 &    0.49162 & $-$0.24351 &    0.15088 &    0.45046 & $-$0.25008
\\
$a_{f,2}$   & $-$0.10(45)  & $-$0.00342 &    0.06890 & $-$0.01746 &    0.12057 & $-$0.49732 &    1.00000 &    0.12057 & $-$0.33668 &    0.20338 & $-$0.18421 & $-$0.23045 &    0.19828
\\ \hline
$a_{\Fo,0}$ & 0.002006(31) &    0.23345 &    0.06133 &    0.01137 &    1.00000 &    0.06216 &    0.12057 &    1.00000 &    0.07462 & $-$0.09380 &    0.39328 &    0.19208 & $-$0.13641
\\
$a_{\Fo,1}$ & 0.0013(41)   & $-$0.00136 & $-$0.01804 & $-$0.00789 &    0.07462 &    0.49162 & $-$0.33668 &    0.07462 &    1.00000 & $-$0.51671 &    0.17631 &    0.84705 & $-$0.49244
\\
$a_{\Fo,2}$ & $-$0.03(21)  &    0.01362 &    0.02956 &    0.01312 & $-$0.09380 & $-$0.24351 &    0.20338 & $-$0.09380 & $-$0.51671 &    1.00000 &    0.19200 & $-$0.48706 &    0.98112
\\ \hline
$a_{\Ft,0}$ & 0.0484(16)   &    0.08685 & $-$0.03921 &    0.00205 &    0.39328 &    0.15088 & $-$0.18421 &    0.39328 &    0.17631 &    0.19200 &    1.00000 &    0.07882 &    0.12820
\\
$a_{\Ft,1}$ & $-$0.059(87) & $-$0.00553 &    0.00394 &    0.01056 &    0.19208 &    0.45046 & $-$0.23045 &    0.19208 &    0.84705 & $-$0.48706 &    0.07882 &    1.00000 & $-$0.55376
\\
$a_{\Ft,2}$ & $-$0.9(1.1)  &    0.01628 &    0.03000 &    0.00708 & $-$0.13641 & $-$0.25008 &    0.19828 & $-$0.13641 & $-$0.49244 &    0.98112 &    0.12820 & $-$0.55376 &    1.00000
\\ \hline
\end{tabular}
\vspace{3mm}
\end{sidewaystable}


\begin{table}[t]
\centering
\small
\caption{
  Stability of BGL parametrization coefficients $a_{\{g,f,\Fo,\Ft\},\{0,1\}}$
  given in Table~\ref{tbl:q2-param:bgl}.
  We list their shift in units of their uncertainty
  due to different choices of input and reference values $\wref$.
  We omit $a_{\Fo,0}$, which is proportional to $a_{f,0}$
  due to the kinematical constraint~(\ref{eqn:q2-param:const:w1}),
  as well as $a_{\{g,f,\Fo,\Ft\},2}$ for quadratic terms,
  which show tiny shift due to their large relative uncertainty.
  Note that $a_{g,n}$ is independent of the pseudo-scalar resonance mass input.
  The maximum (minimum) value of $\wref$ is denoted as $w_{\rm ref, max}$
  ($w_{\rm ref, min}$).
  The quadratic fit ($N_{\{f,g,\Fo,\Ft\}}\!=\!2$) is impossible
  with two values of $\wref$,
  with which we compare coefficients of the linear fit ($N_{\{f,g,\Fo,\Ft\}}\!=\!1$).
}
\vspace{3mm}
\label{tbl:q2-param:input-dep}.
\begin{tabular}{l|rrrrrrr}
\hline
choice of input-B, $\chi^{{T,L}}_{J^P}$ or $\wref$
  & $a_{g,0}$  & $a_{g,1}$   & $a_{f,0}$  & $a_{f,1}$   
               & $a_{\Fo,1}$ & $a_{\Ft,0}$ & $a_{\Ft,1}$ 
\\ \hline
i) $M_{{\rm pole},n}$ for $J^P\!=\!1^+, 1^-$ 
  & $-1.62$ &  0.32 & $-2.22$ &  0.04 &  0.03 &  0.01 &  0.00
\\
ii) $M_{{\rm pole},\{0,1\}}$ for $J^P\!=\!0^-$ 
  & --    & --    &  0.00 &  0.00 &  0.00 &  0.57 & $-0.04$
\\
iii) $M_{{\rm pole},2}$ for $J^P\!=\!0^-$ 
  & --    & --    &  0.00 &  0.00 &  0.00 & $-1.59$ & 0.10
\\ \hline
iv) $\chi^{{T,L}}_{J^P}$ from lattice
  & $-0.74$ &  0.08 & $-4.18$ & $-0.11$ &  0.02 & $-1.24$ &  0.03
\\ \hline
v) $w_{\rm ref, max}\!=\!1.080$
  & $-0.00$ & $-0.01$ & $-0.02$ & $-0.05$ & $-0.10$ &  0.12 &  0.31 
\\
vi) $w_{\rm ref, min}\!=\!1.040$
  & $-0.00$ & $-0.01$ & $-0.03$ & $-0.02$ & $-0.04$ &  0.30 & $-0.30$
\\ 
vii) 2 $\wref$'s $=\!1.025, 1.100$
  & $-0.01$ & $-0.09$ & $-0.04$ & $-0.08$ &  0.01 &  0.11 & $-0.04$
\\ \hline
\end{tabular}
\vspace{3mm}  
\end{table}

We test the stability of the fit with different choices of the input
and reference values of $w$.
Input-B in Table~\ref{tbl:q2-param:reso} of the resonance masses
for the Blaschke factors
consists of the pseudo-scalar masses $M_{{\rm pole},{\{0,1\}}}$
from recent experimental measurements~\cite{PDG22},
and other resonance masses from phenomenological studies
different from input-A:
namely,
Ref.~\cite{Mreso:ph:LLWL23} for $J^P\!=\!0^-$ and $n\!=\!2$,
and Ref.~\cite{Mreso:ph:EQ94} for $J^P\!=\!1^{\{+,-\}}$.
We note that these vector and axial masses
have been used in a phenomenological analysis~\cite{Vcb:anly:GK17}
and Belle analyses~\cite{B2Dstar:Belle:tag,B2Dstar:Belle:untag}
of the $B\!\to\!D^*\ell\nu$ decay.
For the derivatives of the vacuum polarization functions $\chi^{{T,L}}_{J^P}$,
we test the lattice QCD estimate in Ref.~\cite{B2Dstar:chi:lat:MSV21}.
Shift of the expansion coefficients $a_{\{g,f,\Fo,\Ft\},\{0,1\}}$
in units of their uncertainty is summarized in Table~\ref{tbl:q2-param:input-dep},
where we test four choices of the input: namely,
i) vector and axial resonance masses from Ref.~\cite{Mreso:ph:EQ94},
ii) ground and first excited pseudo-scalar masses from Ref.~\cite{PDG22},
iii) second excited pseudo-scalar mass from Ref.~\cite{Mreso:ph:LLWL23},
and iv) lattice estimate of $\chi^{{T,L}}_{J^P}$.
The shift of $a_{\{g,f,\Fo,\Ft\},\{0,1\}}$ is at the 2-$\sigma$ level
or typically well below it.
A large shift ($\sim\!4 \sigma$) in $a_{f,0}$ with the lattice QCD estimate
of $\chi^{{T,L}}_{J^P}$ can be attributed to about 20\% shift of $\chi^T_{1^+}$
for the axial channel. As suggested in Eq.~(\ref{eqn:q2-param:ofunc:f}),
this simply leads to rescaling of all $a_{\{f,\Fo\},\{0,1,2\}}$
for the corresponding channel.
Also for other coefficients,
the changes in the Blaschke factors and outer functions
due to the different inputs are largely absorbed into
the shift in $a_{\{g,f,\Fo,\Ft\},\{0,1,2\}}$ to reproduce the synthetic data
of the form factors.
As a result,
the shift in physical observables, $R(D^*)$ and $|V_{cb}|$ in this paper,
is well below 1\,$\sigma$ as we will see later.
This also means that our results for $a_{\{g,f,\Fo,\Ft\},\{0,1\}}$
in Table~\ref{tbl:q2-param:bgl}
should be used with our choice of the input,
namely input-A for resonance masses in Table~\ref{tbl:q2-param:reso}
and perturbative estimate of $\chi^{{T,L}}_{J^P}$
in Table~\ref{tbl:q2-param:suscep},
otherwise the uncertainty suggested in Table~\ref{tbl:q2-param:input-dep}
should be taken into account.


In Table~\ref{tbl:q2-param:input-dep},
we also test three different choices of the reference values $\wref$:
namely, v) decreased maximum value $w_{\rm ref, max}\!=\!1.080$,
vi) increased minimum value $w_{\rm ref, min}\!=\!1.040$
and vii) two values of $\wref\!=\!1.025, 1.100$
against our main choice of $\wref\!=\!1.025, 1.060, 1.100$.
Table~\ref{tbl:q2-param:input-dep} shows that the shift of the coefficients
$a_{\{g,f,\Fo,\Ft\},\{0,1\}}$ is well below 1\,$\sigma$.
Their relative accuracy is slightly worse (typically by 5\,--\,10\,\%)
with the above choices of $\wref$
suggesting that our main choice is reasonably good.


We confirm the unitarity bound for the axial vector channel as 
$\sum_{k=0}^{N_f} |a_{f,k}|^2 + \sum_{k=0}^{N_\Fo} |a_{\Fo,k}|^2 = 0.012(96)$,
which is poorly saturated by $B\!\to\!D^*\ell\nuell$.
The bounds for other channels are satisfied as
$\sum_{k=0}^{N_g} |a_{g,k}|^2 = 1.0(3.5)$
and $\sum_{k=0}^{N_\Ft} |a_{\Ft,k}|^2 = 0.9(2.1)$.
The large uncertainty
comes from the poorly-determined quadratic coefficients.
The sums up to the linear term,
$\sum_{k=0}^{1} |a_{g,k}|^2 = 0.0029(31)$ and
$\sum_{k=0}^{1} |a_{\Ft,k}|^2 = 0.006(10)$,
suggest that the unitarity bound is rather weak as mentioned above.


\begin{figure}[t]
\begin{center}
  \includegraphics[angle=0,width=0.47\linewidth,clip]{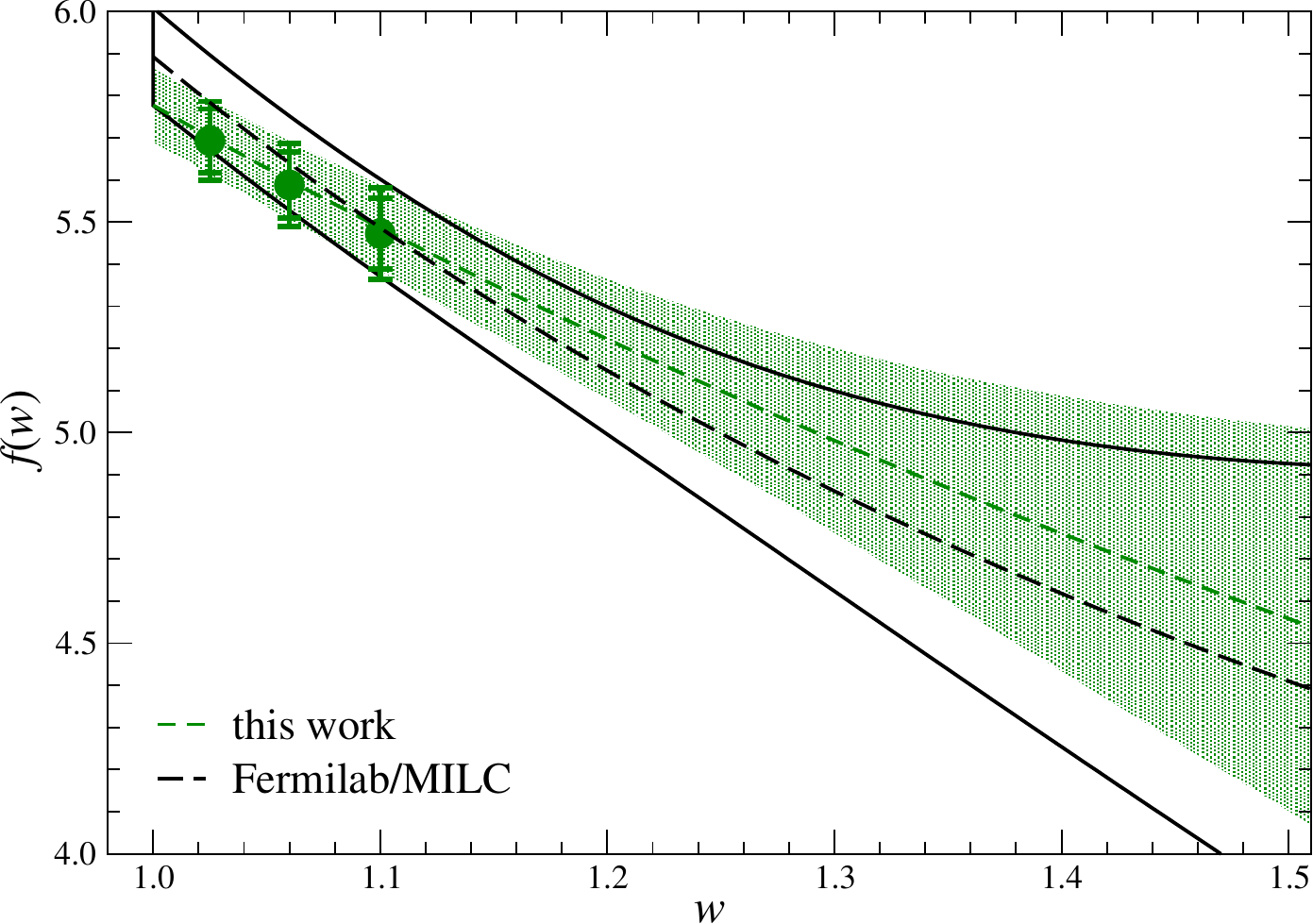}
  \hspace{5mm}
  \includegraphics[angle=0,width=0.47\linewidth,clip]{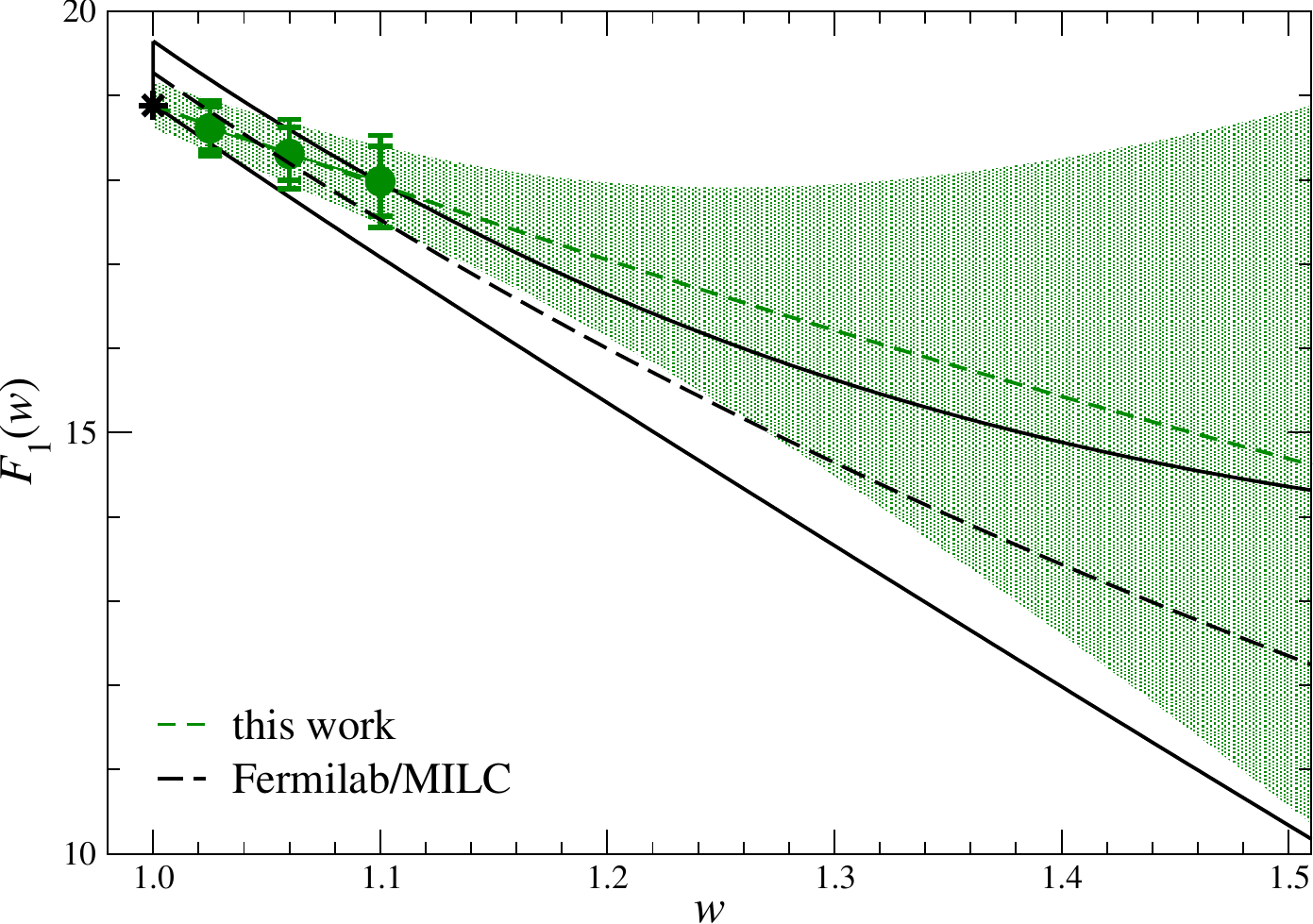}
  \vspace{0mm}

  \includegraphics[angle=0,width=0.47\linewidth,clip]{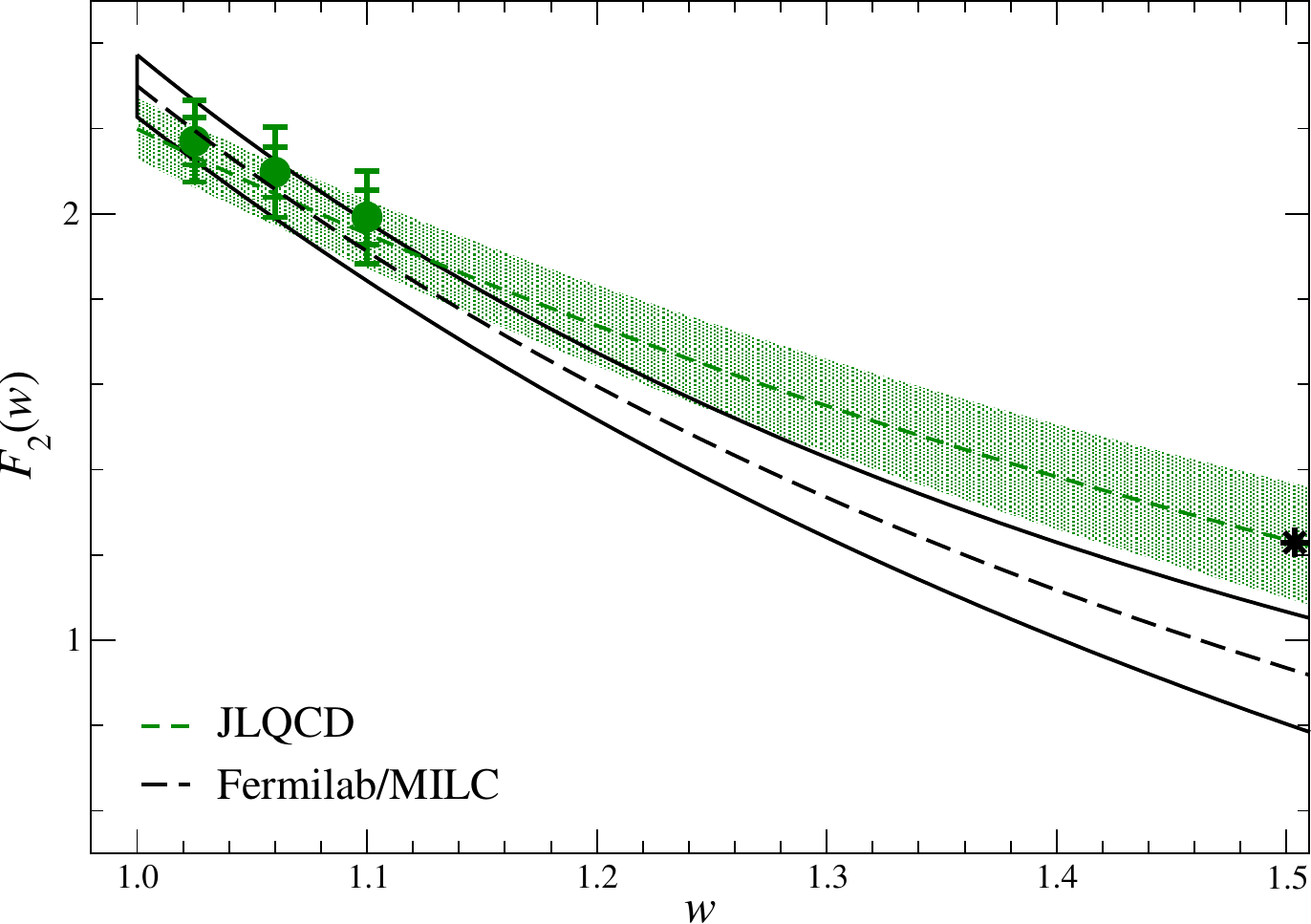}
  \hspace{5mm}
  \includegraphics[angle=0,width=0.47\linewidth,clip]{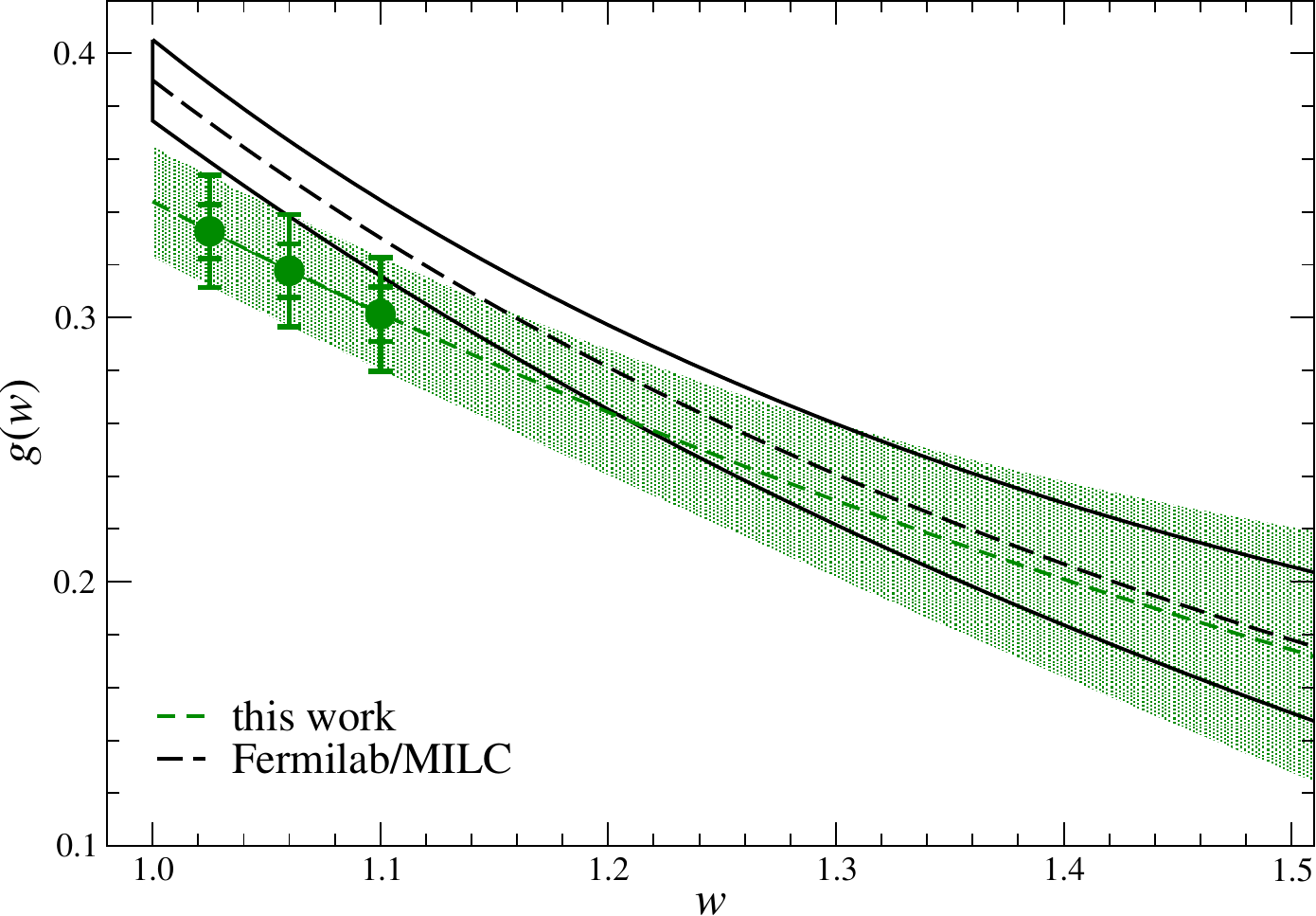}

  \vspace{-3mm}
  \caption{
    BGL parametrization (green band) of the synthetic data (green circles)
    in the whole semileptonic region $1\!\leq\!w\!\lesssim\!1.5$.
    The top-left, top-right, bottom-left and bottom-right panels
    show $f(w)$, $\Fo(w)$, $\Ft(w)$ and $g(w)$, respectively.
    We also plot the parametrization by Fermilab/MILC~\cite{B2Dstar:Nf3:Fermilab/MILC}
    by the open black bands.
    The stars in the panels for $\Fo$ and $\Ft$ represent the values
    fixed from the kinematical constraints (\ref{eqn:q2-param:const:w1}) and
    (\ref{eqn:q2-param:const:wmax}), respectively.
  }
  \label{fig:q2-param:fx_vs_w}
\end{center}
\end{figure}

\begin{figure}[t]
\begin{center}
  \includegraphics[angle=0,width=0.305\linewidth,clip]{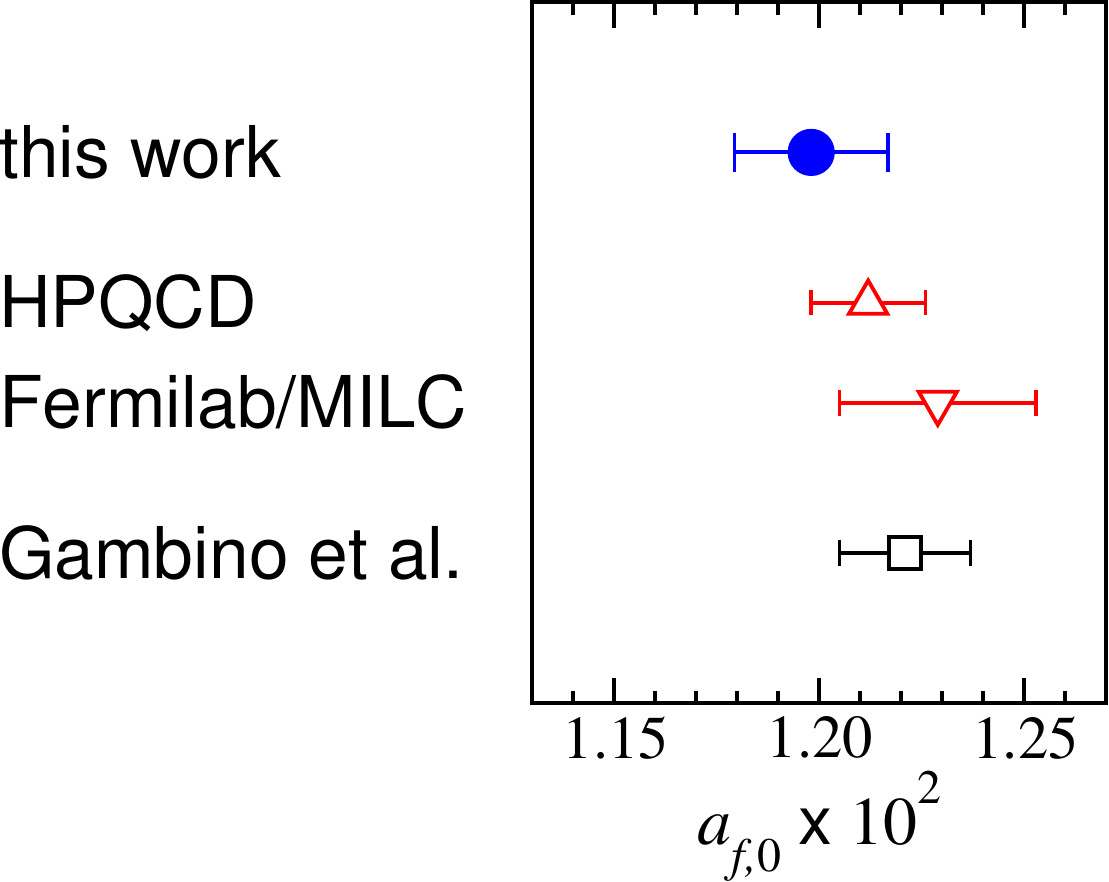}
  \hspace{6mm}
  \includegraphics[angle=0,width=0.17\linewidth,clip]{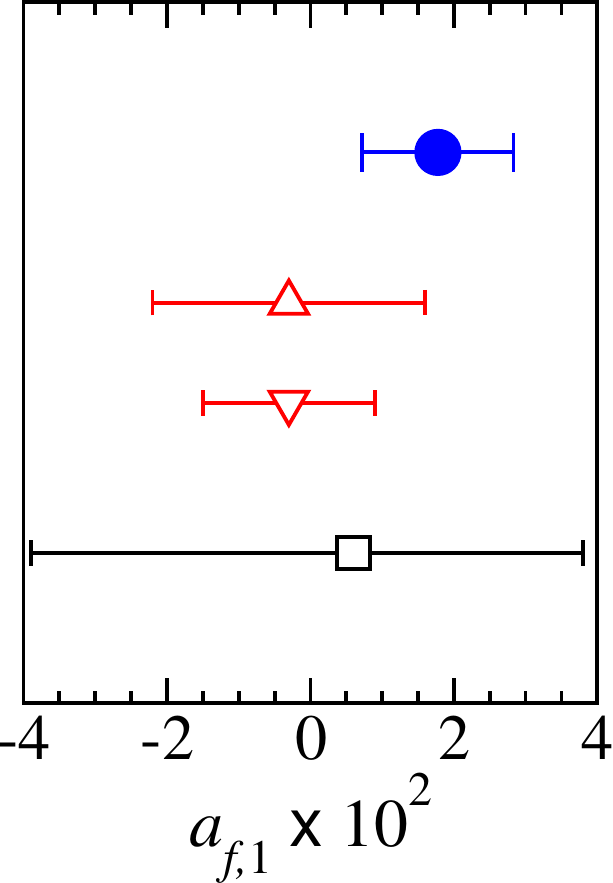}
  \hspace{6mm}
  \includegraphics[angle=0,width=0.17\linewidth,clip]{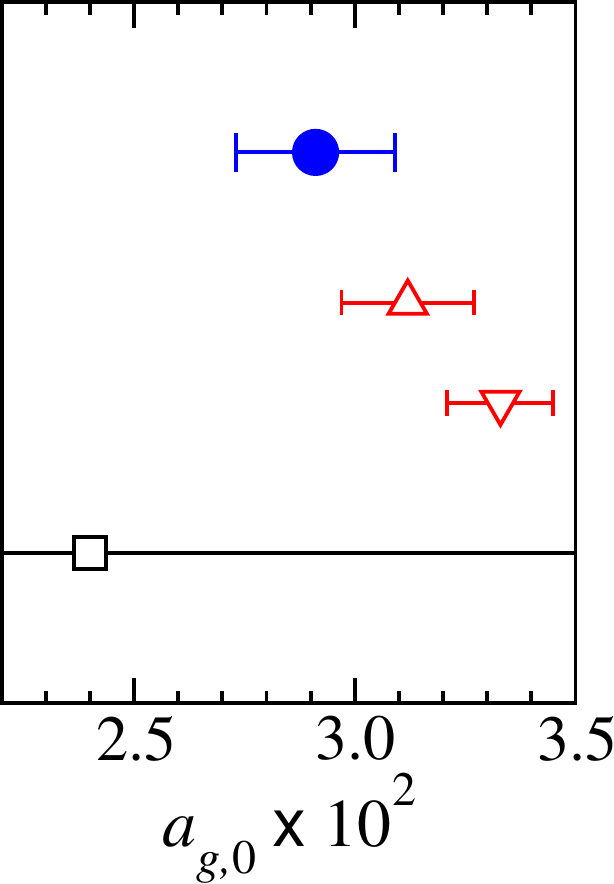}
  \hspace{6mm}
  \includegraphics[angle=0,width=0.17\linewidth,clip]{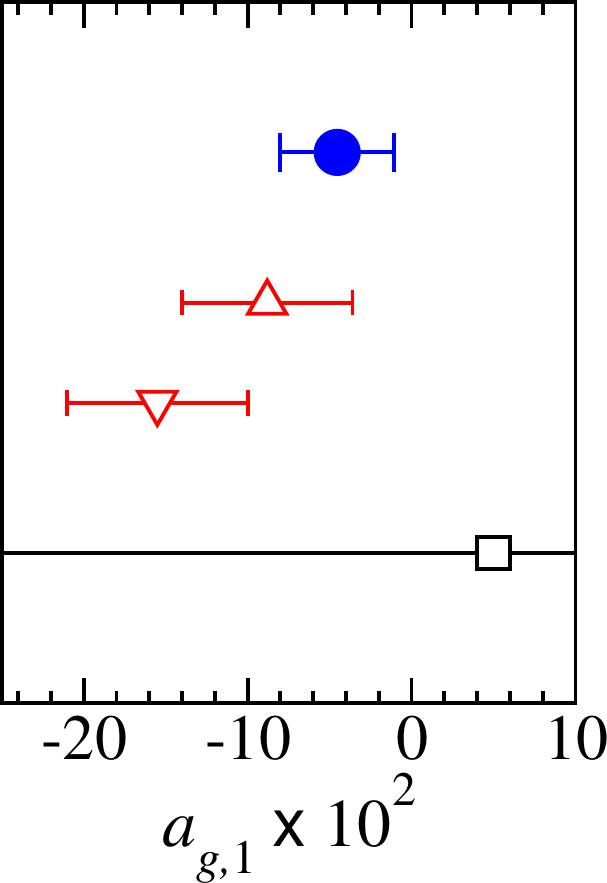}

  \includegraphics[angle=0,width=0.17\linewidth,clip]{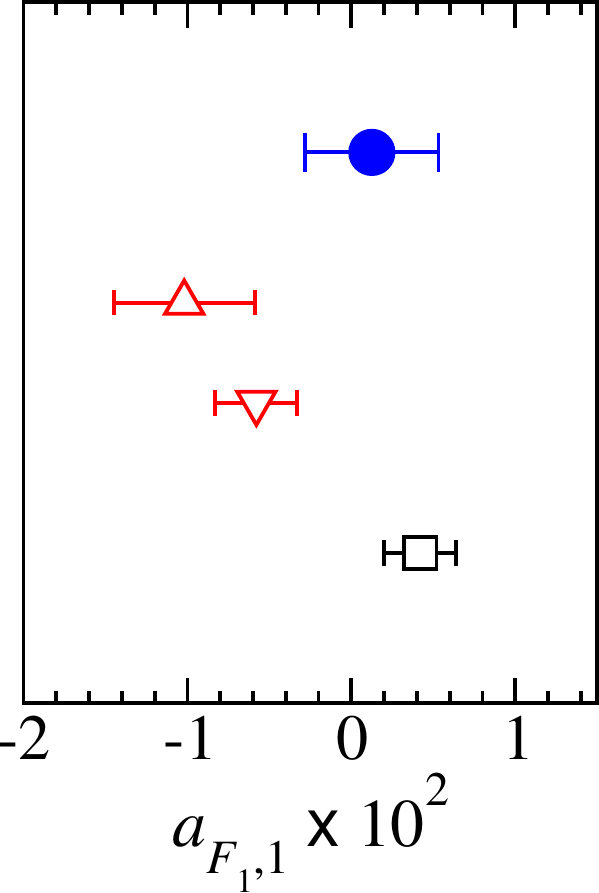}
  \hspace{6mm}
  \includegraphics[angle=0,width=0.17\linewidth,clip]{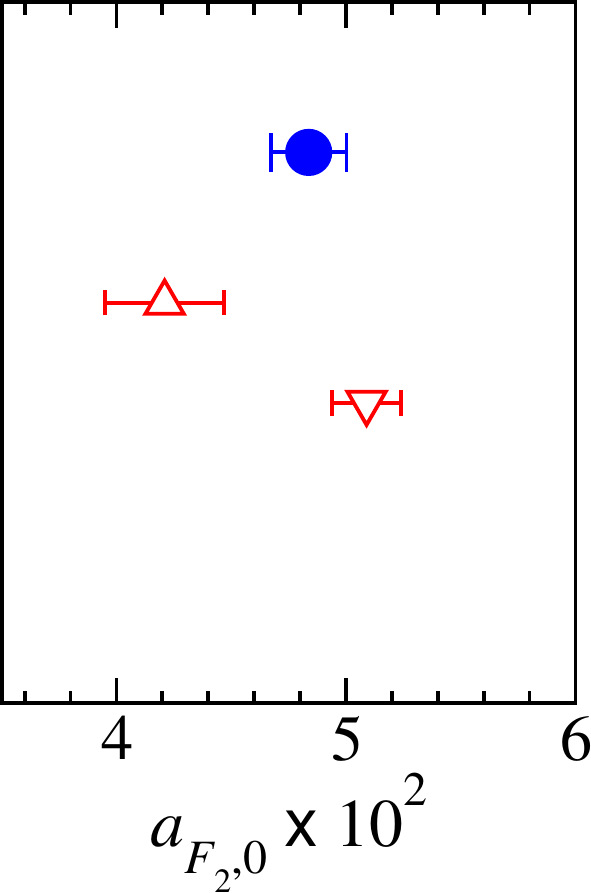}
  \hspace{6mm}
  \includegraphics[angle=0,width=0.17\linewidth,clip]{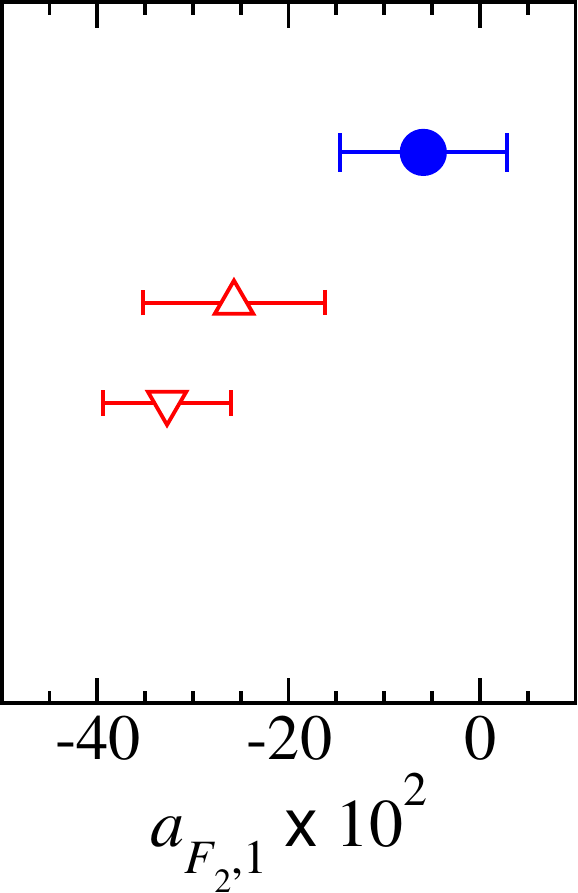}

  \vspace{-3mm}
  \caption{
    Comparison of expansion coefficients for BGL parametrization
    (\ref{eqn:q2-param:bgl}).
    Our results are plotted as filled blue circles.
    Open red triangles are obtained by HPQCD (triangles up)~\cite{B2Dstar:Nf4:HPQCD}
    and Fermilab/MILC (triangles down)~\cite{B2Dstar:Nf3:Fermilab/MILC}.
    We also plot results from a recent phenomenological analysis of
    the Belle data~\cite{B2Dstar:Belle:untag}
    as open black squares~\cite{Vcb:BGL:GJS19}.
    Since all these studies impose the kinematical constraint (\ref{eqn:q2-param:const:w1})
    leading to $a_{\Fo,0}\!\propto\!a_{f,0}$,
    we omit a panel for $a_{\Fo,0}$.
  }
  \label{fig:q2-param:afx}
\end{center}
\end{figure}

The BGL parametrization and synthetic data of the form factors are plotted 
in Fig.~\ref{fig:q2-param:fx_vs_w}.
We observe reasonable consistency
with the Fermilab/MILC results~\cite{B2Dstar:Nf3:Fermilab/MILC}
except the vector form factor $g(w)$ near zero-recoil
and the axial form factor $\Ft$ extrapolated to large recoils.
This can be seen in a more detailed comparison of the expansion coefficients
in Fig.~\ref{fig:q2-param:afx},
where there is slight tension in $a_{g,0}$, $a_{g,1}$
and $a_{\Fo,1}$.
Recent HPQCD results show better consistency with ours,
except for $a_{\Fo,1}$ and $a_{\Ft,0}$.

In the same figure, 
we also plot a phenomenological estimate~\cite{Vcb:BGL:GJS19}
obtained from the Belle data~\cite{B2Dstar:Belle:untag}
combined with a lattice input of the FLAG average of
$h_{A_1}(1)\!\propto\!f(1)$~\cite{FLAG19}.
Note that this analysis employs the same BGL parametrization
with the input ($M_{{\rm pole},n}$ and $\chi_{J^P}^{\{T,L\}}$) and
orders ($N_f$, $N_g$, $N_\Fo$),
which are the same as our main analysis.
The phenomenological result of $a_{f,0}$ is well determined
and shows good consistency with lattice results,
simply because it is mainly fixed from the lattice input.
Other poorly determined coefficients suggest the importance of lattice calculations
to obtain controlled BGL parametrization.


On the other hand, the slope $a_{F_1,1}$ is well determined from the phenomenological
analysis of the experimental data, which are precise at large recoils. 
This study observes a reasonable agreement in $a_{F_1,1}$ and, hence,
in the differential decay rate as discussed in the next subsection
(see Fig.~\ref{fig:q2-param:DDR}).
We note, however, that
the Fermilab/MILC and HPQCD studies reported tension in the $w$ dependence of
the differential decay rate with experiment.


Since $\Ft$ describes the contribution suppressed by $m_\ell^2$
to the decay rate (\ref{eqn:q2-param:ddr:ddr}),
its expansion coefficients are poorly constrained by the experimental data
for the light lepton channels $B\!\to\!D^*\ell\nuell$ ($\ell\!=\!e,\mu$).
The lattice determination of $\Ft$ is, therefore, helpful towards
a precision new physics search using the $\tau$ channel.
Through numerical integration of Eq.~(\ref{eqn:q2-param:ddr:ddr})
for $\ell\!=\!e,\mu$ and $\tau$ with fit results in Table~\ref{tbl:q2-param:bgl},
we obtain a pure SM value 
\bea
   R(D^*)
   & = &
   0.252(22)(4),
   \label{eqn:q2-param:rdstar}
\eea
which does not resort to phenomenological models nor experimental data.
The second error comes from the test of the stability of the BGL parametrization
summarized in Table~\ref{tbl:q2-param:input-dep}.
It is mainly comes from the shift in $\wref_{\rm min}$ and $\wref_{\rm max}$,
but smaller than statistical and other systematic errors of the BGL parametrization.
This estimate can be improved as $R(D^*)\!=\!0.252(+0.009/\!\!-\!\!0.016)(4)$
by imposing the unitarity bound (\ref{eqn:q2-param:unitary_bound})
on the expansion coefficients.
These are consistent with Fermilab/MILC and HPQCD estimates,
0.265(13) and 0.279(13), respectively,
and to be compared with the experimental average 0.295(14)~\cite{HFLAV22}.
For a more precise and reliable estimate of $R(D^*)$,
it is helpful to simulate larger recoils
as well as resolving the tensions in $a_{\Ft,\{0,1\}}$ shown
in Fig.~\ref{fig:q2-param:afx}.


\subsection{Fit including Belle data}
\label{subsec:q2-param:lat+belle}


In this subsection,
we carry out a simultaneous fit to our lattice and Belle data to estimate $\Vcb$.
The differential decay rate with respect to $w$ and three decay angles $\theta_l$,
$\theta_v$ and $\chi$ is given as
\bea
   \frac{d\Gamma}{dw d\costhl d\costhv d\chi}
   & = &
   \frac{3G_F^2}{1024\pi^4} |V_{cb}|^2 \eta_{\rm EW}^2
   M_B r^2 \sqrt{w^2-1}\, q^2
   \nn \\
   & &
   \hspace{4mm}
   \times
   \left\{
      (1-\costhl)^2 \sinthvt H_+^2(w)
    + (1+\costhl)^2 \sinthvt H_-^2(w)
   \right.
   \nn \\
   & &
   \hspace{10mm}
   \left.
    + 4 \sinthlt \costhvt H_0^2(w)
    - 2 \sinthlt \sinthvt \cos[2\chi] H_+(w) H_-(w)
   \right.
   \nn \\
   & &
   \hspace{10mm}
   \left.
    - 4 \sinthl (1-\costhl) \sinthv \costhv \cos[\chi] H_+(w) H_0(w)
   \right.
   \nn \\
   & &
   \hspace{10mm}
   \left.
    + 4 \sinthl (1+\costhl) \sinthv \costhv \cos[\chi] H_-(w) H_0(w)
   \right\}
\eea
with the helicity amplitudes
given in (\ref{eqn:q2-param:Hamp:+-}) and (\ref{eqn:q2-param:Hamp0}).
For a simultaneous fit with the Belle data for $B^0\!\to\!D^{*-}\ell^+\nuell$
$(\ell\!=\!e,\mu)$ in Ref.~\cite{B2Dstar:Belle:untag},
we set $m_\ell\!=\!0$, and include the Coulomb factor $(1+\alpha\pi)$
in the right-hand side.
The decay angles are chosen the same as for the Belle paper,
where the full kinematical distribution
is described by four differential decay rates
$d\Gamma/dw$, $d\Gamma/d\costhl$, $d\Gamma/d\costhv$ and $d\Gamma/d\chi$,
which are integrated over the other three variables.
The differential decay rates are provided
at ten equal-size bins in the range of each variable,
$w\!\in\![1.0,1.5]$, $\cos[\theta_{\{\ell,v\}}]\!\in\![-1,1]$ and $\chi\!\in\![-\pi,\pi]$.


To estimate $\Vcb$,
we fit our synthetic data of the form factors
to the BGL parametrization~(\ref{eqn:q2-param:bgl})
alongside the Belle data of the calculated differential decay rates
where the expansion coefficients $a_{\{g,f,\Fo,\Ft\},\{0,1,2\}}$
are shared parameters
and $\Vcb$ is an additional parameter determined from the fit.
The expression~(\ref{eqn:q2-param:ddr:ddr}) is used for $d\Gamma/dw$.
We calculate $d\Gamma/d\cos[\theta_{\{l,v\}}]$ and $d\Gamma/d\chi$
by analytically integrating over two other decay angles
and by numerically integrating over $w$
via Simpson's rule with 800 steps within each bin
as in the Belle paper~\cite{B2Dstar:Belle:untag}.


\begin{table}[t]
\centering
\small
\caption{
  Fit results of simultaneous fit to our lattice and Belle data.
}
\vspace{0mm}
\label{tbl:q2-param:bgl:lat+belle}.
\begin{tabular}{lllll}
\hline
$F$        & $g$          & $f$          & $\Fo$        & $\Ft$
\\ \hline
$a_{F,0}$   & 0.02896(84)  & 0.01196(18)  & 0.002003(30) & 0.0489(14)
\\
$a_{F,1}$   & $-$0.050(23) & 0.0227(92)   & 0.0040(14)   & $-$0.018(24)
\\
$a_{F,2}$   & $-$1.0(1.9)  & $-$0.46(24)  & $-$0.040(27)  & $-$1.0(2.0)
\\ \hline
\end{tabular}
\vspace{3mm}  
\end{table}

\begin{figure}[t]
\begin{center}
  \includegraphics[angle=0,width=0.48\linewidth,clip]{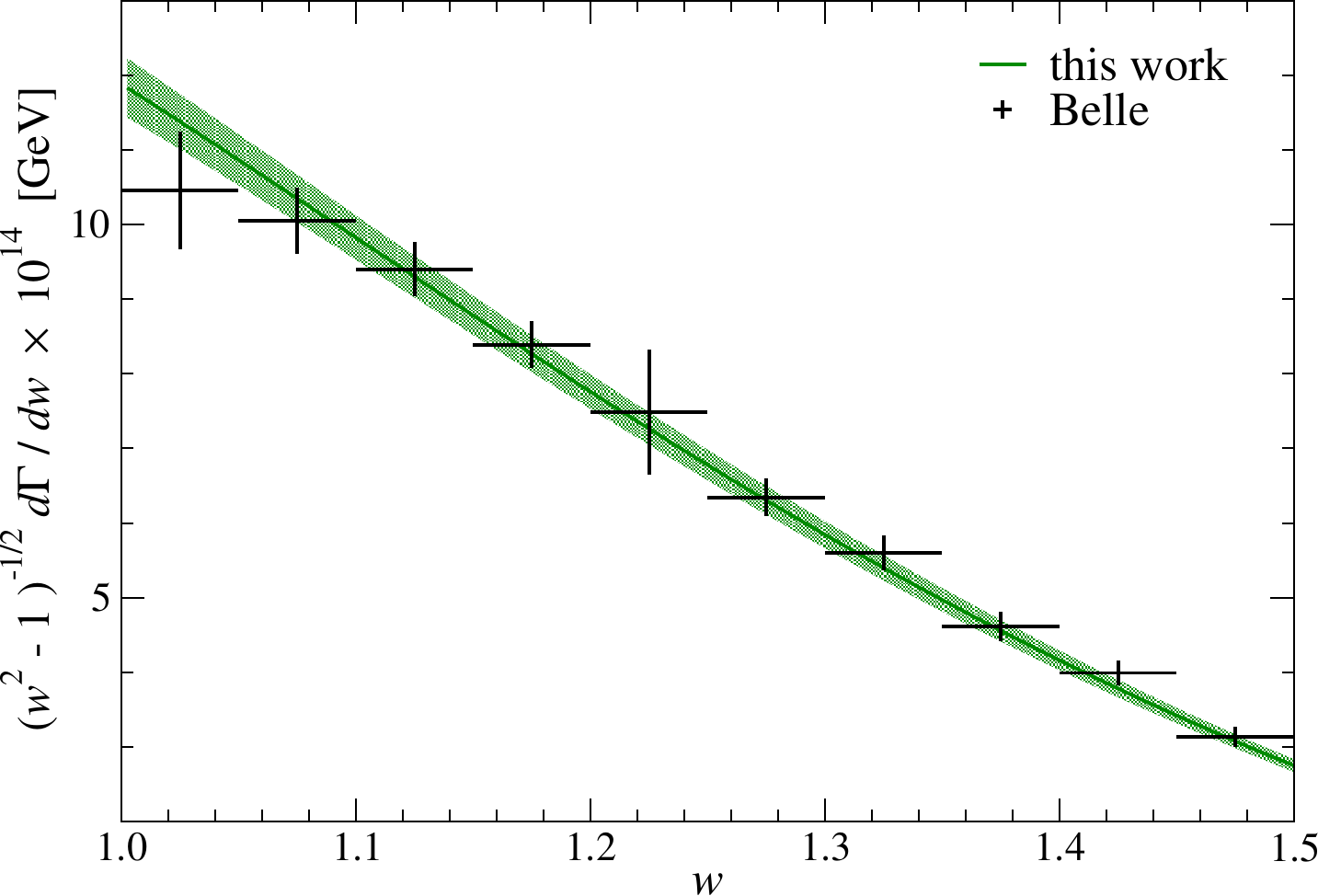}
  \hspace{1mm}
  \includegraphics[angle=0,width=0.48\linewidth,clip]{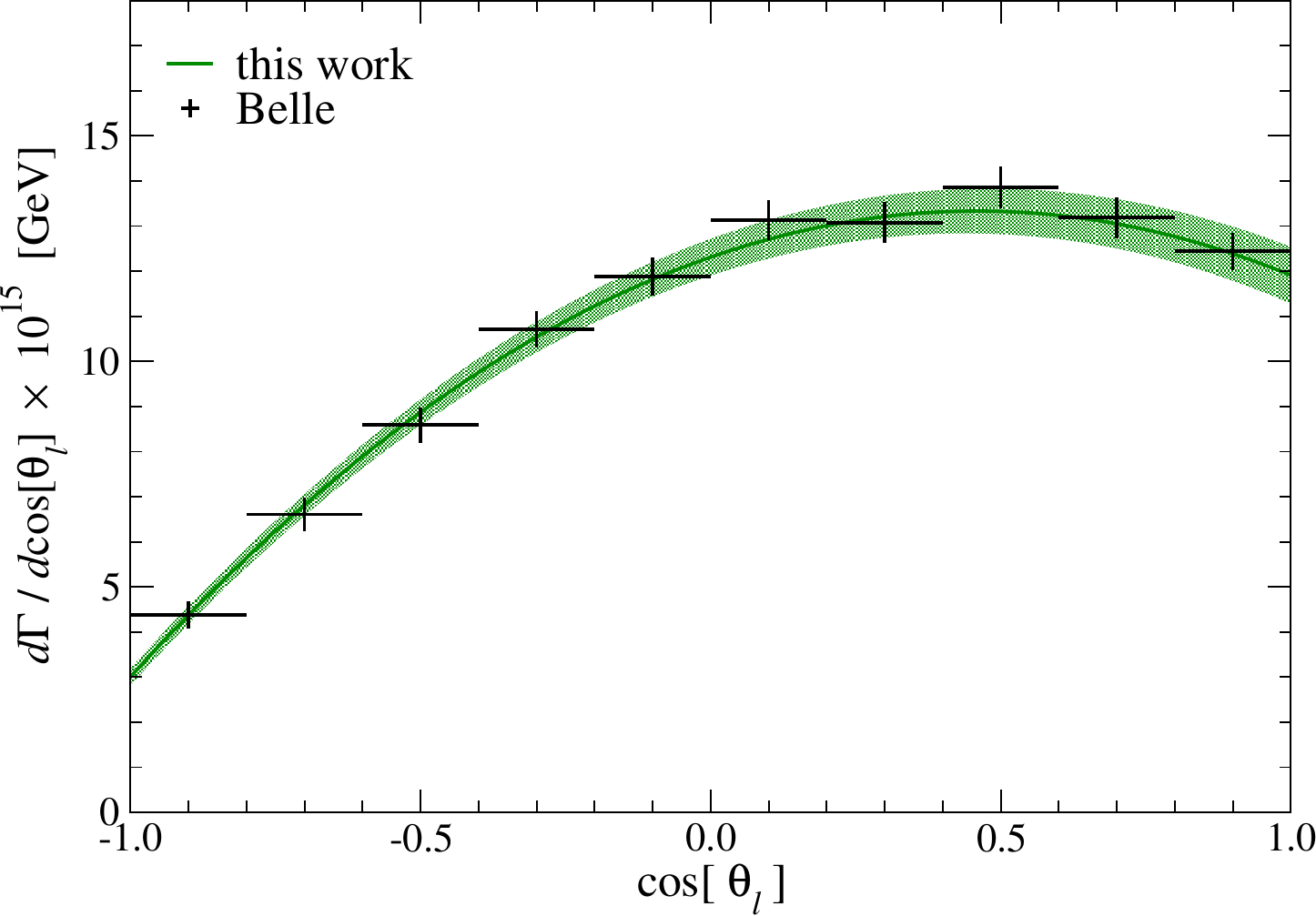}

  \includegraphics[angle=0,width=0.48\linewidth,clip]{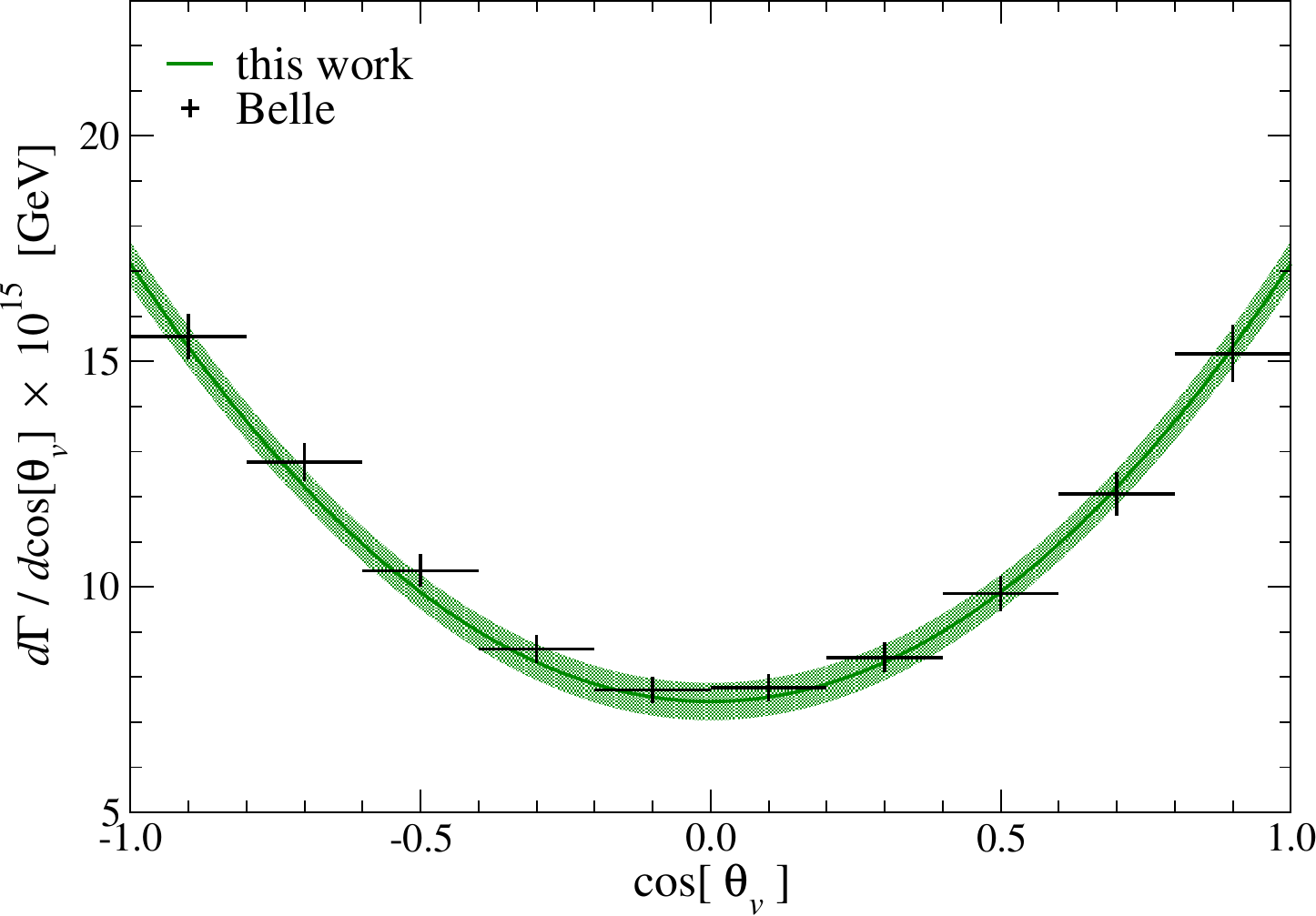}
  \hspace{1mm}
  \includegraphics[angle=0,width=0.48\linewidth,clip]{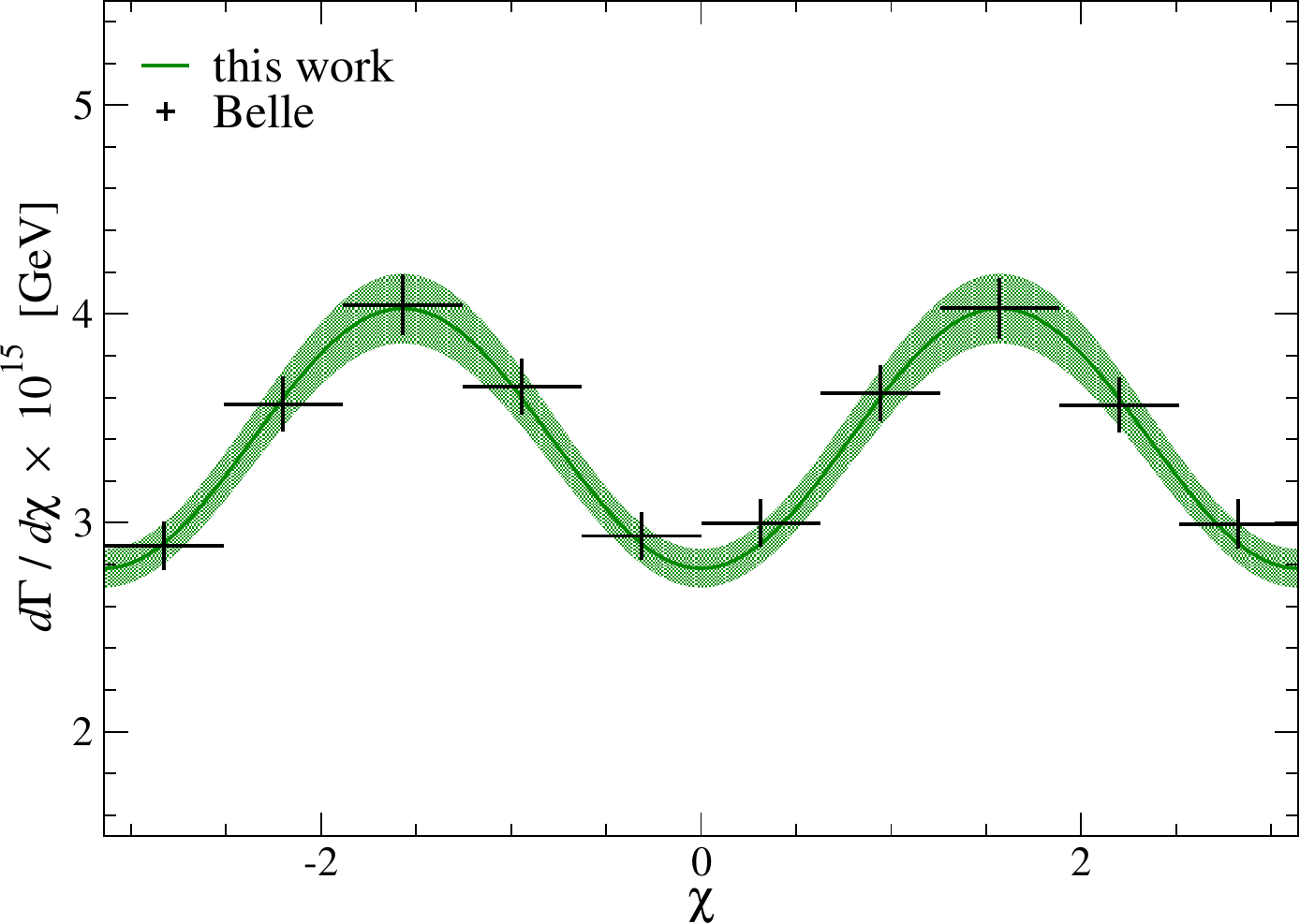}

  \vspace{0mm}
  \caption{
    Differential decay rate with respect to recoil parameter,
    $d\Gamma/dw$ (top-left panel), and three decay angles, 
    $d\Gamma/d\cos[\theta_\ell]$ (top-right panel),
    $d\Gamma/d\cos[\theta_v]$ (bottom-left panel)
    and $d\Gamma/d\chi$ (bottom-right panel).
    We divide $d\Gamma/dw$ by a phase space factor $\sqrt{w^2-1}$
    to make its $w$ dependence milder and monotonous.
    The symbols show Belle results, whereas the green band is reproduced
    from our simultaneous fit to our lattice and Belle data.
  }
  \label{fig:q2-param:DDR}
\end{center}
\end{figure}

Our results for the expansion coefficients are summarized
in Table~\ref{tbl:q2-param:bgl:lat+belle}.
These are consistent with those in Table~\ref{tbl:q2-param:bgl},
which are obtained without the experimental data,
with slightly smaller uncertainty.
We obtain
\bea
   \Vcb = 39.19(90)(12) \times 10^{-3},
\eea
where the first error comes from our main fit 
with input-A in Table~\ref{tbl:q2-param:reso},
perturbative ${\chi}^{\{T,L\}}_{J^P}$ in Table~\ref{tbl:q2-param:suscep},
and $\wref\!=\!1.025, 1.060, 1.100$.
The second error is obtained by testing the stability of the fit
against the different choices of the input and $\wref$
as in Table~\ref{tbl:q2-param:input-dep}.
This is in good agreement with
the previous determination from the exclusive decay~(\ref{eqn:Vcb:excl}).
In contrast to previous lattice studies,
as shown in Fig.~\ref{fig:q2-param:DDR},
we observe consistency between our lattice and Belle data
leading to an acceptable value of $\chi^2/{\rm d.o.f.}\!\sim\!0.90$.
Towards the resolution of the $\Vcb$ tension,
we need a more detailed analysis taking care of, for instance,
bias in $\Vcb$ discussed in Refs.~\cite{Vcb:BGL:GJS19,Vcb:BGLvsCLN:FUW21}
due to D'Agostini effects~\cite{DAgostini}
and the strong systematic correlation of experimental data.
That analysis is left for future work.

\section{Conclusion}
\label{sec:concl}


In this article,
we present our calculation of the $B\!\to\!D^*\ell\nuell$ semileptonic form factors
in 2+1-flavor lattice QCD.
The M\"obius domain-wall action is employed for all relevant quark flavors 
to simulate lattices of cutoffs up to 4.5~GeV with chiral symmetry
preserved to good accuracy.
This removes the leading $O(a)$ discretization errors
and explicit renormalization is not necessary to calculate
the SM form factors through the correlator ratios.
The ground state saturation of the correlator ratios is carefully studied
by simulating four source-sink separations.
The bottom quark mass is limited to $m_b\!\leq\!0.7a^{-1}$
to suppress $O(a^2)$ and higher discretization error.
Note also that the chiral logarithm is suppressed by the small
$D^*$\,--\,$D$ mass splitting squared.
As a result,
the lattice spacing and quark mass dependence turns out to be mild
in our simulation region
leading to good control of the continuum and chiral extrapolation.
We find finite size effects to be small by direct simulations of
two volumes at our smallest pion mass.


One of the main results is the synthetic data of the form factors
$g$, $f$, $\Fo$ and $\Ft$ in the relativistic convention
at three reference values of the recoil parameter $w\!=\!1.025$, 1.060 and 1.100,
which can be used in future analyses to determine $\Vcb$ and
to search for new physics.
Through the BGL parametrization, we obtain $R(D^*)\!=\!0.252(22)$ in the SM.
A combined analysis with the Belle data yields $\Vcb\!=\!39.19(91) \times 10^{-3}$.
These are consistent with previous lattice studies.
However, 
the expansion coefficients for $g$, $\Fo$ and $\Ft$
show significant tension among lattice studies. 
In particular, our slope for $\Fo$ is slightly larger
and consistent with a phenomenological analysis of the Belle data.
As a result, our data show better consistency with the Belle data
in the differential decay rates.


High statistics simulations on finer lattices would be helpful
in resolving this tension.
Simulating larger recoils can lead to a better estimate of $R(D^*)$
as well as more detailed comparison with experiment
in a wide range of the recoil parameter to search for new physics.
%
In order to resolve the $\Vcb$ tension, 
it would also be very helpful to study the inclusive decay on the lattice
for more direct comparison of the exclusive and inclusive analyses
in the same simulations~\cite{Inclusive:H17}.
While it is an ill-posed problem to calculate the relevant hadronic tensor
from discrete lattice data on a finite volume,
a joint project of the JLQCD and RBC/UKQCD Collaborations 
along a strategy proposed in Ref.~\cite{Inclusive:SmrSpecFunc,Inclusive:B2Xc:GH20}
is in progress~\cite{Inclusive:R&D:Nf3:JLQCD+RBC/UKQCD:lat22:B,Inclusive:R&D:Nf3:JLQCD+RBC/UKQCD:lat22:K,Incl:Nf3:JLQCD+UKQCD}
towards a realistic study of $B\!\to\!X_c\ell\nuell$.
\vspace{5mm}

\begin{acknowledgements}

We thank F.~Bernlochner, B.~Dey, D.~Ferlewicz, P.~Gambino, S.~Iguro, M.~Jung,
T.~Kitahara,  P.~Urquijo and E.~Waheed for helpful discussions.
This work used computational resources of supercomputer
Fugaku provided by the RIKEN Center for Computational Science
and Oakforest-PACS provided by JCAHPC
through the HPCI System Research Projects
(Project IDs: hp170106, hp180132, hp190118, hp210146, hp220140 and hp230122)
and Multidisciplinary Cooperative Research Program (MCRP)
in Center for Computational Sciences, University of Tsukuba
(Project ID: xg18i016),
SX-Aurora TSUBASA at the High Energy Accelerator Research
Organization (KEK) under its Particle, Nuclear and Astrophysics Simulation Program
(Project IDs: 2019L003, 2020-006, 2021-007, 2022-006 and 2023-004),
Wisteria/BDEC-01 Odyssey at the University of Tokyo
provided by MCRP (Project ID: Wo22i049), 
and Yukawa Institute Computer Facility.
This work is supported in part by JSPS KAKENHI Grant Numbers
JP18H01216, JP21H01085, JP22H01219, JP22H00138 and JP22K21347,
by ``Program for Promoting Researches on the Supercomputer Fugaku''
(Simulation for basic science: approaching the new quantum era),
by the Toshiko Yuasa France Japan Particle Physics Laboratory
(TYL-FJPPL project FLAV\_03 and  FLAV\_06),
and by Joint Institute for Computational Fundamental Science (JICFuS).
J.K. acknowledges support by the European Research Council
(ERC) under the European Union's Horizon 2020 research and
innovation program through Grant Agreement No. 771971-SIMDAMA.

\end{acknowledgements}


\bibliography{Nf3_BtoDstar}

\end{document}